\documentclass[usenatbib,useAMS]{mnras}
\usepackage{natbib}
\usepackage{graphicx}
\usepackage{rotating}
\usepackage{verbatim}
\usepackage{amssymb}
\usepackage{amsmath}
\usepackage{gensymb}
\usepackage{gensymb}

\newcommand{\HI}{H{\sc i}}

\defcitealias{TC12}{TC12}

\begin{document}
\title[H{\small I} kinematics of Tully-Fisher calibrators]
{Detailed \HI\ kinematics of Tully-Fisher calibrator galaxies} 
\author[Ponomareva et al.]
{Anastasia A. Ponomareva${^1}$\thanks{E-mail:ponomareva@astro.rug.nl}, 
Marc A. W. Verheijen${^1}$ and Albert Bosma${^2}$\vspace*{0.2cm}\\
$^1$Kapteyn Astronomical Institute, University of Groningen, Postbus 800, NL-9700 AV Groningen, The Netherlands\\
$^2$Aix Marseille Universit\'e CNRS, LAM (Laboratoire d'Astrophysique de Marseille) UMR 7326, 13388, Marseille, France\\}


\pagerange{\pageref{firstpage}--\pageref{lastpage}}
\pubyear{2015}

\maketitle

\label{firstpage}

\begin{abstract}

We present spatially--resolved H{\sc i} kinematics of 32 spiral
galaxies which have Cepheid or/and Tip of the Red Giant Branch
distances, and define a calibrator sample for the Tully-Fisher relation. 
The interferometric H{\sc i} data for this sample were collected
from available archives and supplemented with new GMRT observations. 
This paper describes an uniform analysis of the H{\sc i} kinematics of
this inhomogeneous data set.  Our main result is an atlas for our
calibrator sample that presents global H{\sc i} profiles, integrated
H{\sc i} column--density maps, H{\sc i} surface density profiles and,
most importantly, detailed kinematic information in the form of
high--quality rotation curves derived from highly--resolved,
two--dimensional velocity fields and position--velocity diagrams. 
  
\end{abstract}


\section{Introduction}

The Tully-Fisher relation (TFr) is one of the main scaling relations for
rotationally supported galaxies, describing an empirical correlation
between the luminosity or visible mass of a spiral galaxy and its
rotational velocity or dynamical mass \citep{tf77}.  Numerous
observational studies of the statistical properties of the TFr have been
undertaken in the past.  Generally, their aim was to find observables
that reduce the scatter in the relation, and thus improve the distance
measure to spiral galaxies.  An accurate distance measure is necessary
to address some of the main cosmological questions pertaining to the
local Universe, such as the value of the Hubble constant, the local
large--scale structure and the local cosmic flow field, e.g. 
\citealt{TC12}, hereafter \citetalias{TC12}. 

Use of the implied scaling law and understanding its origin is one of
the main challenges for theories of galaxy formation and evolution.  In
particular the detailed statistical properties of the TFr provide
important constraints to semi--analytical models and numerical
simulations of galaxy formation and evolution.  It is an important test
for theoretical studies to reproduce the slope, scatter and the zero
point of the TFr in different photometric bands simultaneously.  Recent
cosmological simulations of galaxy formation have reached sufficient
maturity to construct sufficiently realistic galaxies that follow the
observed TFr \citep{mar14, vog14, schaye15}. 
  
An important outstanding issue, however, is how to connect the results
from simulations to the multitude of observational studies, which often
do not agree with each other \citep{mcg12, zar14}.  Moreover, while
performing these comparison tests, it is important to ensure that the
galaxy parameters derived from simulations and observations have the
same physical meaning.  For example, for simulated galaxies the circular
velocity of a galaxy is usually derived from the potential of the dark
matter halo, while the observed velocity, based on the width of the
global H{\sc i} profile obtained with single--dish telescopes, is a good
representation for the dark matter halo potential only in rather limited
cases.  Note that the rotational velocity of a galaxy, inferred from the
width of its global H{\sc i} profile, can be directly related to the
dark matter potential only when the gas is in co--planar, circular
orbits and the rotation curve has reached the turnover to a constant
velocity at radii properly sampled by the gas disk.  However, this is
often not the case as galaxies display a variety of rotation curves
shapes, which can not be characterised with a single parameter such as
the corrected width of the global H{\sc i} line profile.  Furthermore,
the number of observational studies that take into account the
two--dimensional distribution of H{\sc i} and the detailed geometry and
kinematics of a galaxy's disk is rather limited. 

With the developments of new methods and instruments, the observed
scatter in the Tully-Fisher relation has decreased significantly, mainly
due to more accurate photometric measurements.  Photographic magnitudes
\citep{tf77} were improved upon by CCD photometry.  The advent of
infra--red arrays shifted photometry to the JHK bands, minimizing
extinction corrections, and then to space--based infra--red photometry
with the Spitzer Space Telescope \citep{wer04}, which better samples the
older stellar population and maximizes photometric calibration
stability.  As was shown by \citet{sorce12}, the measurement errors
on the total $3.6 \mu m$ luminosity are reduced to a point where they no
longer contribute significantly to the observed scatter in the
Tully--Fisher relation.  Hence, other measurement errors, such as
uncertainties on a galaxy's inclination and/or rotational velocity,
combined with a certain intrinsic scatter are responsible for the total
observed scatter.

\begin{figure}
\begin{center}
\includegraphics[scale=0.75]{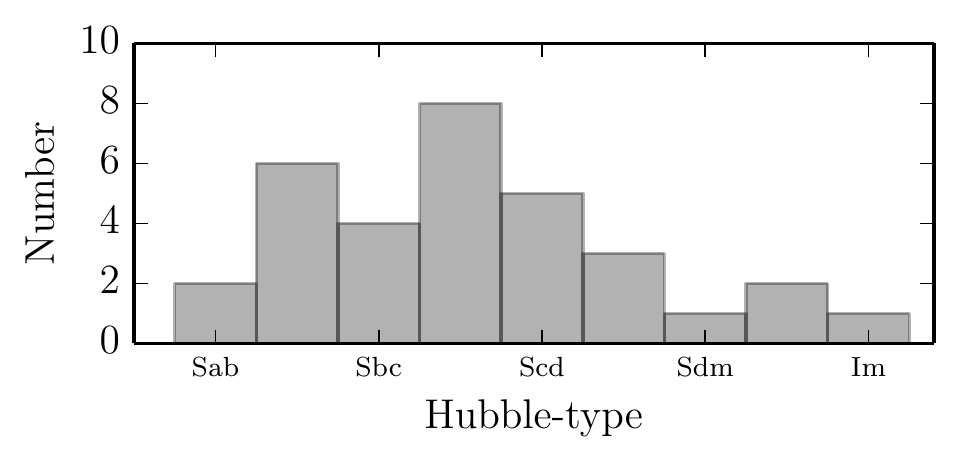}
\caption{
The distribution of morphological types of our sample galaxies.
\label{fig_ttype}}
\end{center}
\end{figure}

It was shown by \citet{v01}, with a study of nearly equidistant spiral
galaxies in the Ursa Major cluster with deep K'--band photometry, that
the observed scatter in the Tully--Fisher relation can be reduced
significantly when the measured velocity of the outer, flat part of
the H{\sc i} rotation curve (V$_{\rm flat}$) is used instead of the
rotational velocity as estimated from the width of the global H{\sc i}
profile.  Indeed, the spatially-resolved, two--dimensional velocity
fields of gas disks, obtained with radio interferometric arrays, often
reveal the presence of warps, streaming motions and kinematic
lopsidedness of the gas disks, while the global H{\sc i} profiles do not
allow the identification of such features.  Therefore, the width of a
global H{\sc i} line profile, albeit easily obtained with a single--dish
telescope, may not be an accurate representation of V$_{\rm flat}$. 
Moreover, the analysis of spatially resolved H{\sc i} kinematics shows
that the rotation curves of spiral galaxies are not always flat and
featureless. Rotation curves may still be rising at the outermost
observable edge of the gas disk, in which case the width of the global
H{\sc i} profile provides a lower limit on V$_{\rm flat}$.  Rotation
curves may also be declining beyond an initial turnover radius, in which
case V$_{\rm flat}$ derived from the rotation curve is systematically
lower than the circular velocity derived from the corrected width of the
global H{\sc i} profile.  In this paper, we investigate the
differences between three measures of the rotational velocity of spiral
galaxies: The rotational velocity from the corrected width of the global
H{\sc i} profile, the maximal rotation velocity of the rotation curve
(V$_{\rm max}$) and the rotational velocity of the outer flat part of
the rotation curve (V$_{\rm flat}$).  The effects of these different
velocity measures on the observed scatter and the intrinsic properties
of the TFr will be discussed in a forthcoming paper.

To derive V$_{\rm max}$ and V$_{\rm flat}$ we analyse in detail the
H{\sc i} synthesis imaging data for a Tully-Fisher calibrator sample of
32 nearby, well--resolved galaxies and present these data in the form
of an Atlas and tables.  We will discuss the identification and
corrections made for warps and streaming motions in the gas disks when
deriving the H{\sc i} rotation curves.  Forthcoming papers will use
these data to investigate the luminosity--based TFr using panchromatic
photometric data, as well as the baryonic TFr for the same calibrator
sample. 

This paper is organised as follows: Section 2 describes the sample of
selected galaxies.  Section 3 presents the data collection and analysis,
including subsections on the comparison between the rotational
velocities obtained from the rotation curves and the corrected width of
the global H{\sc i} profiles.  Section 4 presents the properties of the
gaseous disks of the sample galaxies.  A summary and concluding remarks
are presented in Section 5.  The Atlas is described in the Appendix,
together with notes on individual galaxies.

\begin{table*}
\begin{tabular}{llrlllllrl}
\hline
Name & Hubble type  & P.A. & Incl.  & Log$\,d_{25}$ & $m_{T}^{b,i} [3.6]$ & $\mu_{0}^{i} [3.6]$ & Log$\,h_{r} [3.6]$  & Distance \\
         &         & deg   & deg  & arcsec &  mag   &mag arcsec$^{-2}$& arcsec & Mpc \\
\hline
NGC 0055 & SBm     &  101  &  86  &  2.47  &  5.64  &  17.54 & 2.32 &  2.09 \\  
NGC 0224 & SAb     &   35  &  78  &  3.25  &  0.48  &  17.74 & 2.46 &  0.76 \\  
NGC 0247 & SABd    &  166  &  76  &  2.29  &  6.27  &  18.52 & 2.26 &  3.45 \\  
NGC 0253 & SABc    &   52  &  81  &  2.42  &  3.37  &  14.36 & 2.10 &  3.45 \\  
NGC 0300 & SAd     &  114  &  46  &  2.29  &  5.56  &  25.30 & 2.18 &  1.93 \\  
NGC 0925 & SABd    &  104  &  57  &  2.03  &  7.24  &  20.73 & 1.95 &  9.16 \\  
NGC 1365 & SBb     &   20  &  54  &  2.08  &  5.91  &  27.71 & 1.85 & 17.94 \\  
NGC 2366 & IBm     &   32  &  74  &  1.64  &  9.76  &  21.20 & 2.07 &  3.31 \\
NGC 2403 & SABc    &  126  &  60  &  2.30  &  5.52  &  19.51 & 2.47 &  3.19 \\  
NGC 2541 & SAc     &  169  &  63  &  1.48  &  9.16  &  20.22 & 1.60 & 11.22 \\  
NGC 2841 & SAb     &  148  &  66  &  1.84  &  5.70  &  16.90 & 1.79 & 14.01 \\ 
NGC 2976 & SAc     &  143  &  60  &  1.86  &  6.95  &  19.15 & 1.78 &  3.56 \\
NGC 3031 & SAb     &  157  &  59  &  2.33  &  3.40  &  18.68 & 2.15 &  3.59 \\  
NGC 3109 & SBm     &   93  &  90  &  2.20  &  7.76  &  18.25 & 2.08 &  1.30 \\  
NGC 3198 & SBc     &   32  &  70  &  1.81  &  7.44  &  17.37 & 1.61 & 13.80 \\
IC 2574  & SABm    &   50  &  69  &  2.11  &  8.60  &  22.18 & 2.04 &  3.81 \\  
NGC 3319 & SBc     &   34  &  59  &  1.56  &  9.05  &  22.80 & 1.74 & 13.30 \\  
NGC 3351 & SBb     &   10  &  47  &  1.86  &  6.45  &  24.77 & 1.68 & 10.91 \\  
NGC 3370 & SAc     &  143  &  58  &  1.38  &  8.70  &  20.43 & 1.23 &  0.23 \\  
NGC 3621 & SAd     &  161  &  66  &  1.99  &  6.37  &  17.57 & 1.71 &  7.01 \\  
NGC 3627 & SABb    &  173  &  60  &  2.01  &  5.39  &  19.01 & 1.78 & 10.04 \\  
NGC 4244 & SAc     &   42  &  90  &  2.21  &  7.39  &  18.34 & 2.47 &  4.20 \\  
NGC 4258 & SABb    &  150  &  69  &  2.26  &  4.95  &  18.89 & 2.24 &  7.31 \\  
NGC 4414 & SAc     &  166  &  55  &  1.29  &  6.56  &  23.03 & 1.69 & 17.70 \\  
NGC 4535 & SABc    &  180  &  45  &  1.91  &  6.80  &  25.28 & 1.62 & 15.77 \\  
NGC 4536 & SABc    &  118  &  71  &  1.85  &  7.17  &  18.03 & 1.70 & 15.06 \\  
NGC 4605 & SBc     &  125  &  69  &  1.77  &  7.22  &  18.29 & 1.66 &  5.32 \\  
NGC 4639 & SABb    &  130  &  55  &  1.46  &  8.28  &  26.64 & 1.27 & 21.87\\   
NGC 4725 & SABa    &   36  &  58  &  1.99  &  5.93  &  21.91 & 1.79 & 12.76 \\  
NGC 5584 & SAB     &  157  &  44  &  1.50  &  8.87  &  25.27 & 1.39 & 22.69 \\  
NGC 7331 & SAb     &  169  &  66  &  1.96  &  5.45  &  19.12 & 1.79 & 14.72 \\  
NGC 7793 & SAd     &   83  &  53  &  2.02  &  6.43  &  22.30 & 1.85 &  3.94 \\ 
\hline

%
\end{tabular}
\caption{The Tully-Fisher Calibrator Sample.
Column (1): galaxy name (as shown in \href{https://ned.ipac.caltech.edu/forms/byname.html}{NED});
Column (2): Hubble type (as shown in \href{https://ned.ipac.caltech.edu/forms/byname.html}{NED});
Column (3): optical position angle of the disk from \citetalias{TC12};
Column (4): optical inclination of the disk from \citetalias{TC12};
Column (5): optical diameter (as shown in \href{https://ned.ipac.caltech.edu/forms/byname.html}{NED});
Column (6): total apparent magnitude, derived using 3.6 $\mu$m Spitzer band (corrected for Galactic and internal reddening);
Column (7): disk central surface brightness, derived by fitting the exponential disk to the surface brightness profile in 3.6 $\mu$m Spitzer band (corrected for inclination);
Column (8): disk scale length, calculated from the 3.6 $\mu$m Spitzer band;
Column (9): Distance in Mpc, measured using TRGB or Cepheid distance estimation method. An averaged value is taken, when distance is measured by both methods (provided by The Extragalactic Distance Database (EDD) \href{http://edd.ifa.hawaii.edu/}{http://edd.ifa.hawaii.edu/}).
}
\label{tbl_samp}
\end{table*}

\begin{table*}
\begin{tabular}{lllllr@{.}lr@{.}lr@{$\times$}ll}

\hline
Name & Data source & Array/ & Obs. date & FREQ & \multicolumn{2}{l}{B-width} & \multicolumn{2}{l}{Ch-width} & \multicolumn{2}{l}{Beam size} & RMS noise\\
& & configuration & dd/mm/yy & MHz & \multicolumn{2}{l}{MHz} & \multicolumn{2}{l}{km$s^{-1}$} & \multicolumn{2}{l}{arcsec$^{2}$}& mJy/beam\\

\hline
NGC 0224 &R. Braun$^{7}$& WSRT           & 08/10/01 & 1421.82 &  5 & 00 & 2& 06 & 60&60	&2.70\\
NGC 0247 & ANGST$^{2}$  & VLA(BnA/CnB)   & 06/12/08 & 1419.05 &  1 & 56 & 2& 60 &  9&6.2	&0.73\\
NGC 0253 & LVHIS        & ATCA           & 08/03/94 & 1419.25 &  8 & 00 & 3& 30 & 14&5	&1.47\\
NGC 0300 & LVHIS        & ATCA(EW352,367)& 03/02/08 & 1419.31 &  8 & 00 & 8& 00 &180&87	&7.18\\
NGC 0925 & THINGS$^{3}$ & VLA (BCD/2AD)  & 08/01/04 & 1417.79 &  1 & 56 & 2& 60 &  5.9&5.7	&0.58\\
NGC 1365 & S. J\"{o}rs\"{a}ter$^{8}$ &VLA& 26/06/86 & 1412.70 &  1 & 56 &20& 80 & 11.5&6.3	&0.06\\
NGC 2366 & THINGS       & VLA (BCD/2AD)  & 03/12/03 & 1420.02 &  1 & 56 & 2& 60 & 13.1&11.8&0.56\\
NGC 2403 & THINGS       & VLA (BCD/4)    & 10/12/03 & 1419.83 &  1 & 56 & 5& 20 &  8.7&7.6	&0.39\\
NGC 2541 & WHISP$^{4}$  & WSRT           & 16/03/98 & 1417.86 &  2 & 48 & 4& 14 & 13&10&0.59\\
NGC 2841 & THINGS       & VLA (BCD/4)    & 30/12/03 & 1417.38 &  1 & 56 & 5& 20 & 11&9.3	&0.35\\
NGC 2976 & THINGS       & VLA (BCD/4AC)  & 23/08/03 & 1420.39 &  1 & 56 & 5& 20 &  7.4&16.4	&0.36\\
NGC 3031 & THINGS       & VLA (BCD/2AD)  & 23/08/03 & 1420.56 &  1 & 56 & 2& 60 & 12.9&12.4	&0.99\\
NGC 3109 & ANGST        & VLA(2AD/2AC)   & 07/12/08 & 1418.52 &  1 & 56 & 1& 30 & 10.3&8.8	&1.31\\
NGC 3198 & THINGS       & VLA (BCD/4)    & 26/04/05 & 1417.28 &  1 & 56 & 5& 20 & 13&11.5	&0.34\\
IC 2574	 & THINGS       & VLA(BCD)       & 18/01/92 & 1420.13 &  1 & 56 & 2& 60 & 12.8&11.9	&1.28\\
NGC 3319 & WHISP        & WSRT	         & 20/11/96 & 1416.87 &  2 & 48 & 4& 14 & 18.4&11.9	&0.91\\
NGC 3351 & THINGS       & VLA (BCD/4)    & 06/01/04 & 1416.72 &  1 & 56 & 5& 20 &  9.9&7.1	&0.35\\
NGC 3370 & this work    & GMRT           & 06/03/14 & 1414.32 &  4 & 16 &16& 00 & 30&30	&2.39	\\
NGC 3621 & THINGS       & VLA (BnAC/4)   & 03/10/03 & 1416.95 &  1 & 56 & 5& 20 & 15.9&10.2&0.70\\
NGC 3627 & THINGS       & VLA (BCD/2AD)  & 28/05/05 & 1416.96 &  1 & 56 & 5& 20 & 10.6&8.8&0.41\\
NGC 4244 & HALOGAS$^{5}$& WSRT           & 23/07/06 & 1419.24 & 10 & 00 & 2& 60 & 21&13.5	&0.18\\
NGC 4258 & HALOGAS      & WSRT	         & 29/03/10 & 1418.28 & 10 & 00 & 2& 60 & 30&30	&0.25\\
NGC 4414 & HALOGAS      & WSRT           & 25/03/10 & 1417.02 & 10 & 00 & 2& 60 & 30&30	&0.19\\
NGC 4535 & VIVA$^{6}$   & VLA (CD)       & 20/01/91 & 1411.16 &  3 & 12 &10& 00 & 25&24	&0.61\\
NGC 4536 & VIVA	        & VLA (CS)       & 22/03/04 & 1411.89 &  3 & 12 &10& 00 & 18&16	&0.34\\
NGC 4605 & WHISP        & WSRT	         & 11/11/98 & 1419.47 &  2 & 48 & 4& 14 & 13&8.6	&0.93\\
NGC 4639 & this work    & GMRT           & 06/03/14 & 1415.62 &  4 & 16 &16& 00 & 30&30	&2.00\\
NGC 4725 & WHISP        & WSRT	         & 18/08/98 & 1414.49 &  4 & 92 &16& 50 & 30&30	&0.60\\
NGC 5584 & this work    & GMRT           & 10/03/14 & 1412.78 &  4 & 16 &16& 00 & 30&30	&2.83\\
NGC 7331 & THINGS       & VLA (BCD/4)    & 20/10/03 & 1416.55 &  1 & 56 & 5& 20 &  6.1&5.6	&0.44\\
NGC 7793 & THINGS       & VLA (BnAC/2AD) & 23/09/03 & 1419.31 &  1 & 56 & 2& 60 & 15.6&10.8	&0.91\\
\hline
\end{tabular}
\caption{Observational parameters of the sample.
Column (1): galaxy name;
Column (2): name of the survey or of individual author, where from the data were taken (see the reference below); 
Column (3): instrument and the set up;
Column (4): date of observations dd/mm/yy;
Column (5): observed frequency, MHz;
Column (6): band--width, MHz;
Column (7): channel-- width, MHz;
Column (8): synthesised beam, arcsec;
Column (9): RMS noise, mJy/beam.
\textbf{References}: 1. The Local Volume H{\sc i} Survey \citep{lvhis} 2. ACS Nearby Galaxy Survey Treasury \citep{angst} 3. The H{\sc i} Nearby Galaxy Survey \citep{things} 4. Westerbork observations of neutral Hydrogen in Irregular and SPiral galaxies \citep{whisp} 5. Hydrogen Accretion in LOcal GAlaxieS survey \citep{halogas} 6. VLA Imaging of Virgo in Atomic gas \citep{viva} 7. \citet{andromeda} 8.\citet{1365}
}
\label{tbl_obs}
\end{table*}

\begin{table}
\begin{tabular}{lcccl}
\hline
Name & Obs. Date&& Calibrators& $T_{obs}$\\
 &dd/mm/yy& & &hrs\\
\hline
NGC 3370& 06/03/14&& 3C147; 0842+185&4\\
                & 07/03/14&&3C147; 0842+185&4.5\\
                &11/03/14&&3C147; 0842+185&2.5\\
NGC 4639&06/03/14&&3C286; 1254+116&1.1 \\            
                &07/03/14&&3C286; 1254+116&2\\
                &10/03/14&&3C147;3C286;&4\\
               &                &&1254+116      & \\
               &11/03/14&&3C286; 1254+116&3.9\\
NGC 5584&10/03/14&&3C286; 3C468.1;& 6.75\\              
               &                && 1445+099    & \\
               &11/03/14&&3C286; 3C468.1;& 4.25\\
               &                && 1445+099    & \\
\hline         
\end{tabular}
\caption{The GMRT observations.
 Column (1): galaxy name;
 Column (2):  observational dates;
 Column (3): flux and phase calibrators;
 Column (4): integration time.}
\label{tbl_calib}
\end{table}

\section{The Sample}

 A highly accurate calibration of the observed Tully--Fisher
relation involves a detailed analysis of its main statistical properties
such as its slope, zero point, scatter and possible curvature.  As was
mentioned in the introduction, all these properties strongly depend on
the photometry, accurate and independent distances, inclinations, and
the velocity measure, including their observational errors.  For our
calibrator sample we want to take advantage of 21--cm line synthesis
imaging instead of the global H{\sc i} spectral line, and analyse the
spatially resolved kinematics of the gas disks in order to assess the
presence of non--circular motions and to derive accurate rotation
curves.  Eventhough both single--dish and interferometric observations
present the same width of the global H{\sc i} profiles (Figure \ref
{fig_wsd}, upper panel), the circular velocities measured from the
rotation curves may differ from the ones estimated from the global H{\sc
i} profiles.  Therefore, for the calibration of the TFr, it is necessary
to compare the statistical properties of the relation based on different
velocity measures.  However, this approach puts some observational
constraints on the selection of calibrator galaxies because the detailed
analysis of spatially resolved H{\sc i} kinematics cannot be achieved
for a large number of galaxies.
 
Consequently, we adopt a sample of 31 out of 38 spiral galaxies that
constitute the zero-point calibrator sample of \citetalias{TC12} which
they used to study the I-band TFr.  This choice was motivated by the
availability of synthesis imaging data for the galaxies in this sample. 
NGC 3627 was added to the sample as it is satisfies the selection
criteria, although it was not included in the parent sample of
\citetalias{TC12}.  The main advantage of this calibrator sample is that
all galaxies are relatively nearby and have independent distance
measures based on the Cephe\"id Period--Luminosity relation
\citep{freedman01} and/or based on the Tip of the Red Giant Branch
\citep{rizzi07}.  To verify that their selection criteria do not
introduce an artificial scatter, \citetalias{TC12} performed a series of
test simulations and concluded that their criteria do not lead to a
systematic bias.  The adopted criteria of \citetalias{TC12} for their
TFr zero--point calibrator sample are as follows:

(1) Only galaxies more inclined than $45^{\circ}$ are included.  (2)
Galaxies earlier in type than Sa are excluded.  The sample is magnitude
limited ($BTcorr = 15.5$), but covers a wide range of morphological
types as illustrated in Figure \ref{fig_ttype}.  (3) Only galaxies with
regular global H{\sc i} profiles were included to avoid confusion or
contribution by companion galaxies to the flux. 

We emphasise that these criteria were applied to the zero--point
calibrator sample by \citetalias{TC12}, who did not consider the merits
of detailed H{\sc i} kinematics for their sample galaxies.  Using radio
synthesis imaging data, however, we identify various features of their
H{\sc i} kinematics and morphologies such as warps, kinematic
lopsidedness and streaming motions due to the presence of a bar and
spiral arms.  All this is addressed in the following sections.  A brief
overview of the sample is presented in Table \ref{tbl_samp}.

\section{Data collection and analysis}

The collection of H{\sc i} imaging data for 32 galaxies is not trivial. 
We took advantage of the existence of high--quality data obtained with
different observational facilities for 29 galaxies, and observed 3
additional galaxies ourselves with the Giant Metrewave Radio Telescope
(GMRT) in March 2014 (see Section 3.1).  Consequently, the H{\sc i}
imaging data used in this paper were obtained between 1986 and 2014 with
different aperture synthesis imaging arrays, mostly as part of larger
H{\sc i} surveys (THINGS, WHISP, HALOGAS, etc).  Of the galaxies for
which H{\sc i} data is used in this study, 19 were observed with the
VLA, 8 with the WSRT, 2 with the ATCA and 3 with the GMRT.  Data cubes
were kindly made available by authors of previous studies or retrieved
from the appropriate project website.  The collected data cubes were
already continuum subtracted and cleaned.  However, the quality of
these 29 data cubes is very inhomogeneous, considering the long time
period between the individual observations and the relative performance
of each of the interferometers at the time of data taking.  A summary of
observational parameters, together with the date of observations and the
original data source, is presented in Table \ref{tbl_obs}. 

\subsection{The GMRT data} 

The GMRT observations were carried out in March 2014.  In total 36 hours
of observing time were allocated, on account of 12 hours per galaxy. 
Galaxies were observed during 4 tracks: 2 times 12 hours and 2 times 6
hours.  All three galaxies were observed during multiple tracks and
during each track, galaxies were observed in several scans.  Each scan
was 45 minutes long, bracketed by short (5 min) observations of a nearby
phase calibrator.  The correlator settings were kept the same during the
whole observing run, but the frequencies were tuned to match the
recession velocity of each galaxy.  A flux calibrator was observed
separately at the beginning and at the end of each track.  The total
bandwidth of 4.16 MHz was split into 256 channels.  The visibilities
were recorded after an integration time of 16 seconds.  The flux and
phase calibrators used for each galaxy are listed in Table
\ref{tbl_calib}. 

Visibility data were flagged, calibrated and Fourier transformed using
the AIPS \href{http://www.aips.nrao.edu/index.shtml}{ (Astronomical
Image Processing System)} software \citep{aips}, developed by
\href{http://www.nrao.edu/}{NRAO}.  Natural weighting and tapering
resulted in a synthesised beam of 15''$\times$15'' and a velocity
resolution of 3.43 kms$^{-1}$.  The data cubes were cleaned using the
Groningen Imaging Processing SYstem (GIPSY) \citep{gipsy}.  The clean
components were restored with a Gaussian beam after which the data cubes
were smoothed to an angular resolution of 30'' and a velocity resolution
of 25 km/s.  This allowed for the detection of extended low
column--density H{\sc i} emission in the outer regions of the galaxies. 

\subsection{Homogenisation of the sample } 

Preparing the inhomogeneous set of data for an uniform analysis required
some special treatment.  First, all the cleaned data cubes were smoothed
to a circular beam of 60'' to improve the sensitivity for extended, low
column--density gas.  Only the data cube of NGC 300 with its original
angular resolution of 180''$\times$90'' was smoothed to a circular beam
of 180''.  Subsequently the cubes were smoothed to a FWHM velocity
resolution of 25 km$s^{-1}$ with a Gaussian smoothing kernel. 
Subsequently, frequency--dependent masks outlining the regions of H{\sc
i} emission in each channel map were defined by clipping the smoothed
cubes at the $2\sigma$ level.  Noise peaks above this level were removed
interactively.  Defining the emission in the smoothed data cubes allowed
us to identify regions with emission in the very outer, low
column--density parts of the H{\sc i} disks.  The resulting masks were
applied with a conditional transfer to the original data cubes at higher
angular and velocity resolution.  As a result, we have obtained
high--resolution data cubes including low H{\sc i} column--density gas
in the outer regions, so as to reach the flat part of the rotation curve
for almost all of the galaxies.

\begin{figure}
\begin{center}
\includegraphics[scale=0.50]{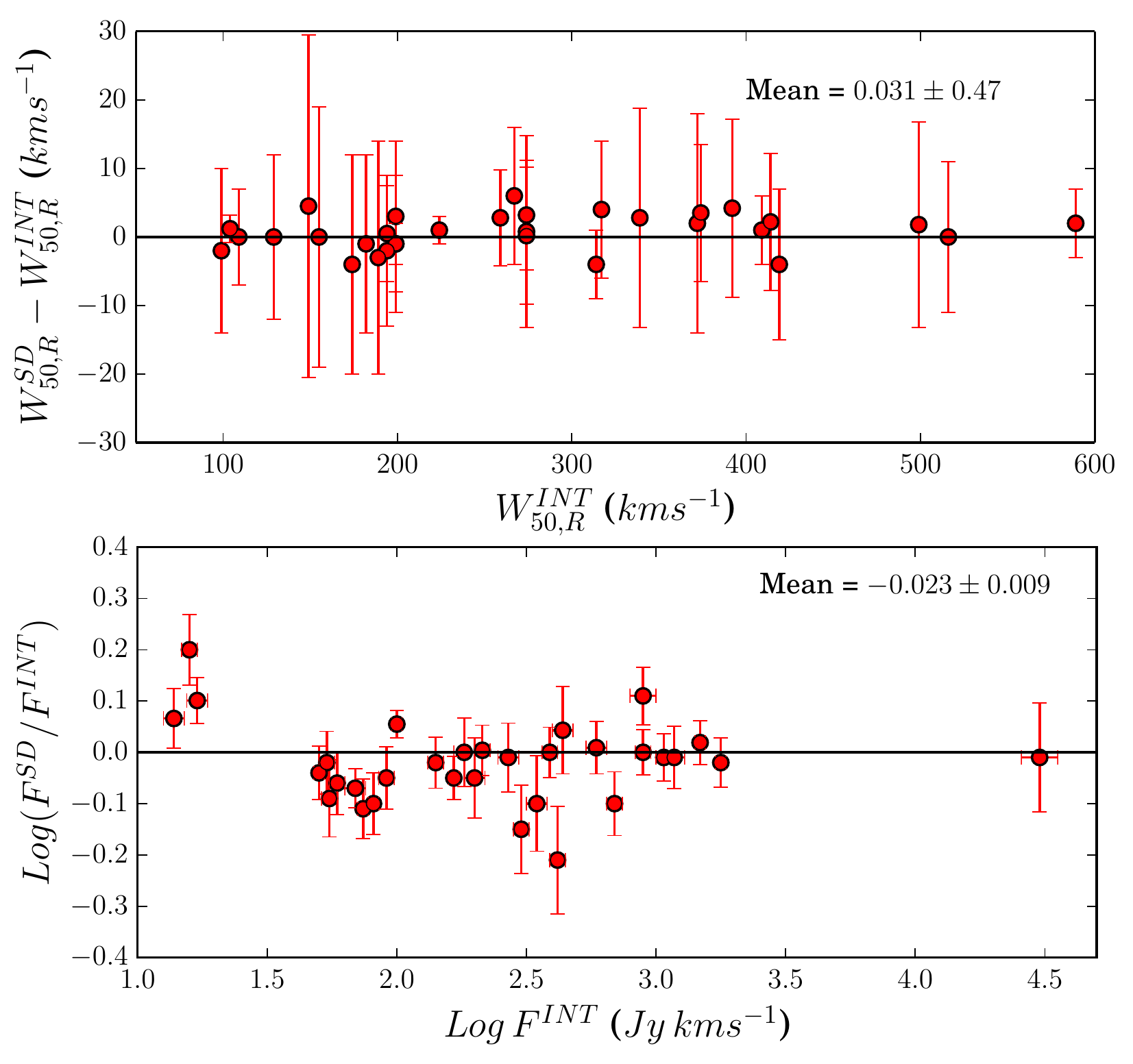}
\caption{Upper panel: A comparison between $W_{50,R}$ derived from the single dish data $W_{50,R}^{SD}$ (EDD) and from the interferometric data $W_{50,R}^{INT}$, corrected for instrumental resolution. 
Lower panel: A comparison between the integrated flux derived from interferometric data and flux from the single--dish observations.
\label{fig_wsd}}
\end{center}
\end{figure}

\subsection{Global H{\sc i} profiles }

The global H{\sc i} profile is one of the main data products of the
H{\sc i} observations.  It contains information about the gas content of
a galaxy, while a galaxy's rotation velocity can be estimated from the
width of the profile.  The H{\sc i} line profiles were obtained by
measuring the primary beam corrected flux density within the masked area
in each velocity channel.  The errors on the flux densities were
determined empirically by calculating the rms noise in the emission free
regions of each channel map. Because features such as warps of the
H{\sc i} disks or non--circular motions do not manifest themselves
clearly and unambiguously in the global H{\sc i} line profile, we do
not obtain a large variety of profile morphologies.  Most of the global
profiles of our sample galaxies are double--peaked, with the exception
of three dwarf galaxies, which show boxy profile shapes: NGC 2366,
IC 2574 and NGC 4605, as can be inferred from Table \ref{tbl_profile}. 
However, some asymmetries in the profiles are still present and cause a
difference in flux between the receding and approaching sides of a
galaxy.  The flux ratios between both sides of the global profiles can
be found in Table \ref{tbl_profile}.  Based on these flux ratios, the
different shapes of the global profiles were classified into 4
categories: symmetric, slightly asymmetric, asymmetric and boxy. 

\subsubsection{Integrated H{\sc i} fluxes}

The integrated H{\sc i} flux was determined as the sum of the primary
beam corrected flux densities in the channel maps, multiplied by the
channel width: $\int S_{v}dv$ \, $[Jy \, kms^{-1}]$.  The integrated
fluxes for our sample galaxies are listed in Table \ref{tbl_param}. 
The error on the integrated flux was calculated as the square--root
of the quadrature--sum of the errors on the flux densities in each
velocity channel (see above).

In the bottom panel of Figure \ref{fig_wsd} we present a comparison
of the integrated fluxes derived from the interferometric data and those
from single--dish observations.  The values of the fluxes from the
single-dish measurements were taken from the ``All Digital H{\sc i}
catalog" \citep{ctf09} as compiled in EDD \citep{EDD}.  The details of
the single--dish measurements can be found in \citet{ksk04, shg05,
ctf09, ctf10, hgm11}.  The weighted mean difference of $-0.023 \pm
0.009$ shows that there is no significant systematic difference between
the single--dish and interferometric flux measurements.  However, some
individual galaxies do show a significant difference which can be
explained in several ways.  First, single--dish observations might be
contaminated due to a contribution to the flux by nearby companion
galaxies within the same beam and at the same recession velocity. 
Second, an interferometric measurement might miss flux in case the
shortest baselines cannot properly sample the largest spatial structures
of an H{\sc i} disk.  Third, a single--dish measurement can miss flux
due to the angular size of a galaxy on the sky.  In particular, many of
our nearby calibrator galaxies are much larger than the primary beam of
the single-dish telescopes. Despite these complications in the flux
measurements, the unweighted average difference between the single--dish
fluxes from the literature and our interferometric fluxes is only
marginal with slightly more flux recovered by the interferomentric
observations.  Consequently we conclude that our results are in good
agreement with previous single--dish measurements.  Note, however, that
accurate flux measurements are not crucial for our purposes. 

\subsubsection{Line widths and rotational velocities}

 Observational studies of galaxies with single--dish telescopes
allow the global H{\sc i} profiles to be collected for thousands of
galaxies.  These profiles hold information not only about the H{\sc i}
gas content of galaxies, but also about their rotational velocities.
The corrected width of the global H{\sc i} profile is a good estimator
of the overall circular velocity of a galaxy, which is usually defined
as 2V$_{\rm circ} = W/\sin(i)$, where V$_{\rm circ}$ is usually
associated with the maximum rotational velocity. The kinematic
information provided by the global H{\sc i} profile, however, is not
very accurate because it provides only a single value for the entire
galaxy.  The rotation curve of a galaxy, on the other hand, represents
the circular velocity as a function of distance from a galaxy's
dynamical centre.  As was already mentioned above, rotation curves of
galaxies are not always flat and featureless, the circular velocity
varies with distance from the dynamical centre of a galaxy.  For
instance, some galaxies show declining rotation curves where V$_{\rm
max}$, the maximal circular velocity of a rotation curve, will be larger
than V$_{\rm flat}$, the circular velocity of the outermost part of the
H{\sc i} disk, which better probes the gravitational potential of the
dark matter halo.  The angular size of a spiral galaxy is usually not
large enough to map its H{\sc i} velocity field point--by--point with a
single--dish telescope, except for the very nearby, large galaxies in
the Local Group (the LMC, SMC and M31).  For most other galaxies, the
ratio of the H{\sc i} diameter to the single--dish beam size is too
small to obtain detailed rotation curves (e.g.  \citet{bosma78}, chapter
3).  However, the global H{\sc i} profile can give a rough estimate for
the kinematics of a galaxy.

The shape of the global H{\sc i} profile can provide some indication for
the shape of the rotation curve of a galaxy.  For example, a
double--peaked profile indicates that the gas disk of a galaxy is
sufficiently extended to sample the flat part of a galaxy's rotation
curve, while a Gaussian or boxy shaped global profile is often observed
for dwarf or low surface--brightness galaxies which usually have a
slowly rising rotation curve with V$_{\rm max}$$<$ V$_{\rm flat}$.  The
wings or steepness of the edges of the profile can be an indication for
a declining rotation curve with V$_{\rm flat}$$<$ V$_{\rm max}$, or for
the presence of a warp \citep{v01}. Our goal is to investigate for
each sample galaxy how the circular velocity derived from the global
H{\sc i} profile compares to the V$_{\rm max}$ and V$_{\rm flat}$ as
measured from the rotation curve.

The width of the global profile is usually defined as a width in $km
s^{-1}$ between the two edges of the profile at the 50\% level of the
peak flux density.  However, different methods for defining the width of
the profile exist in the literature.  For instance the profile width can
be measured at the 20\% level, or by considering the mean peak flux
density.  In our case, for double--peaked profiles, the maximum flux
densities of the two peaks were used separately to determine the full
width of the profile at the 50\% and 20\% levels.  In such a profile the
high velocity peak represents the receding side of the galaxy and the
low velocity peak the approaching side.  Thus, the widths were
calculated using the difference in velocities, corresponding to the 50\%
and 20\% level of each peak of the profile separately:

\begin{equation}
W_{20,50\%} = V_{20,50\%}^{rec} - V_{20,50\%}^{app}. 
\end{equation}

In the other two cases where the profiles are not double--peaked, the
overall peak flux was used to determine the 50\% and 20 \% levels.  The
systemic velocity was determined from the global profile as

\begin{equation}
V_{sys}^{GP} = 0.25\,(V_{20\%}^{app}+V_{50\%}^{app}+V_{20\%}^{rec}+V_{50\%}^{rec}), 
\end{equation}

The adopted error on $V_{sys}^{GP}$ was calculated as half the
difference between the values of $V_{sys}^{GP}$ based on the 20\% and
50\% levels separately. 

All measured widths were corrected for instrumental broadening due to
the finite instrumental velocity resolution. This correction
depends on the instrumental velocity resolution and on the steepness of
the profiles for which the Gaussian shape was assumed.  For a detailed
description of this method see \citet{versanc01}.

Our results can be compared to the corrected widths of the profiles from
single--dish telescopes taken from the literature sources mentioned
above (see section 3.3.1).  The upper panel of Figure \ref{fig_wsd}
demonstrates the comparison between the widths at the 50\% level derived
from the interferometric measure and those from the literature
\citep{ksk04, shg05, ctf09, ctf10, hgm11}.  There is no systematic
offset found and the data are in excellent agreement with the unweighted
average difference of $0.031 \pm 0.47$ $kms^{-1}$ with a 5 $kms^{-1}$
rms scatter.  The measurements obtained from the global H{\sc i}
profiles are presented in Table \ref{tbl_profile}.

\begin{table}
\begin{tabular}{lr@{$\pm$}lllcl}
\hline
Name & \multicolumn{2}{l}{$V_{sys}^{GP}$} & $W_{50}$ & $W_{20}$ & F$_{rec}$/F$_{app}$ & Shape\\
     & \multicolumn{2}{l}{$kms^{-1}$} &   $kms^{-1}$ &$kms^{-1}$ & & \\
\hline
NGC 0055&130&16  &185$\pm$4  & 201$\pm$6  &2.74&A\\
NGC 0224&-297&3  &517$\pm$5  & 533$\pm$7  &1.08&S\\
NGC 0247&161&11  &200$\pm$3  & 227$\pm$3  &0.86&SA\\
NGC 0253&264&20  &410$\pm$3  & 419$\pm$5  &0.93&S\\
NGC 0300&131&10  &160$\pm$5  & 167$\pm$1  &0.82&S\\
NGC 0925&544&6   &200$\pm$8  & 228$\pm$9  &1.08&SA\\
NGC 1365&1630&13 &380$\pm$10 & 405$\pm$7  &1.87&A\\
NGC 2366&92&15   &100$\pm$10 & 120$\pm$11 &1.38&B\\
NGC 2403&120&15  &225$\pm$1  & 257$\pm$4  &0.63&A\\
NGC 2541&568&8   &200$\pm$6  & 211$\pm$6  &0.76&SA\\
NGC 2841&641&14  &590$\pm$3  & 615$\pm$4  &1.27&SA\\
NGC 2976&-7&10   &130$\pm$7  & 164$\pm$8  &1.45&B\\
NGC 3031&-43&13  &415$\pm$6  & 425$\pm$7  &1.06&SA\\
NGC 3109&410&6   &110$\pm$1  & 130$\pm$1  &1.07&SA\\
NGC 3198&644&12  &315$\pm$4  & 322$\pm$1  &1.21&S\\
IC 2574 &40&11   &105$\pm$2  & 134$\pm$3  &1.22&B\\
NGC 3319&748&18  &195$\pm$6  & 221$\pm$3  &0.79&A\\
NGC 3351&761&19  &260$\pm$6  & 284$\pm$3  &1.17&S\\
NGC 3370&1288&8  &275$\pm$4  & 301$\pm$6  &0.91&SA\\
NGC 3621&714&16  &275$\pm$7  & 293$\pm$7  &1.29&A\\
NGC 3627&706&9   &340$\pm$7  & 370$\pm$10 &1.38&A\\
NGC 4244&258&13  &195$\pm$6  & 218$\pm$4  &0.85&SA\\
NGC 4258&455&10  &420$\pm$6  & 434$\pm$7  &0.91&SA\\
NGC 4414&730&5   &375$\pm$6  & 395$\pm$6  &0.94&SA\\
NGC 4535&1945&20 &270$\pm$6  & 278$\pm$5  &1.30&SA\\
NGC 4536&1781&19 &320$\pm$6  & 333$\pm$5  &1.23&SA\\
NGC 4605&155&5   &150$\pm$15 & 186$\pm$5  &0.58&A/B\\
NGC 4639&989&11  &275$\pm$6  & 290$\pm$11 &0.94&S\\
NGC 4725&1230&10 &400$\pm$4  & 405$\pm$2  &1.78&A\\
NGC 5584&1648&8  &190$\pm$10 & 197$\pm$2  &0.74&A\\
NGC 7331&807&8   &500$\pm$10 & 527$\pm$1  &1.12&SA\\
NGC 7793&216&13  &174$\pm$10 & 195$\pm$3  &1.10&SA\\
\hline
\end{tabular}
\caption{Measurements from the global H{\sc i} profiles. 
 Column (1): galaxy name;
 Column (2):  heliocentric systemic velocity;
 Column (3): global profile width at 50\% level;
 Column (4): global profile width at 20\% level;
 Column (5): ratio between integrated flux of receding and approaching side of a galaxy;
 Column (5):Shape of the profile: S--symmetric, SA--slightly asymmetric; A--asymmetric; B--boxy.
 }         
\label{tbl_profile}
\end{table}

\subsection{Velocity fields }

\begin{figure*}
\begin{center}
\includegraphics[scale=1.3]{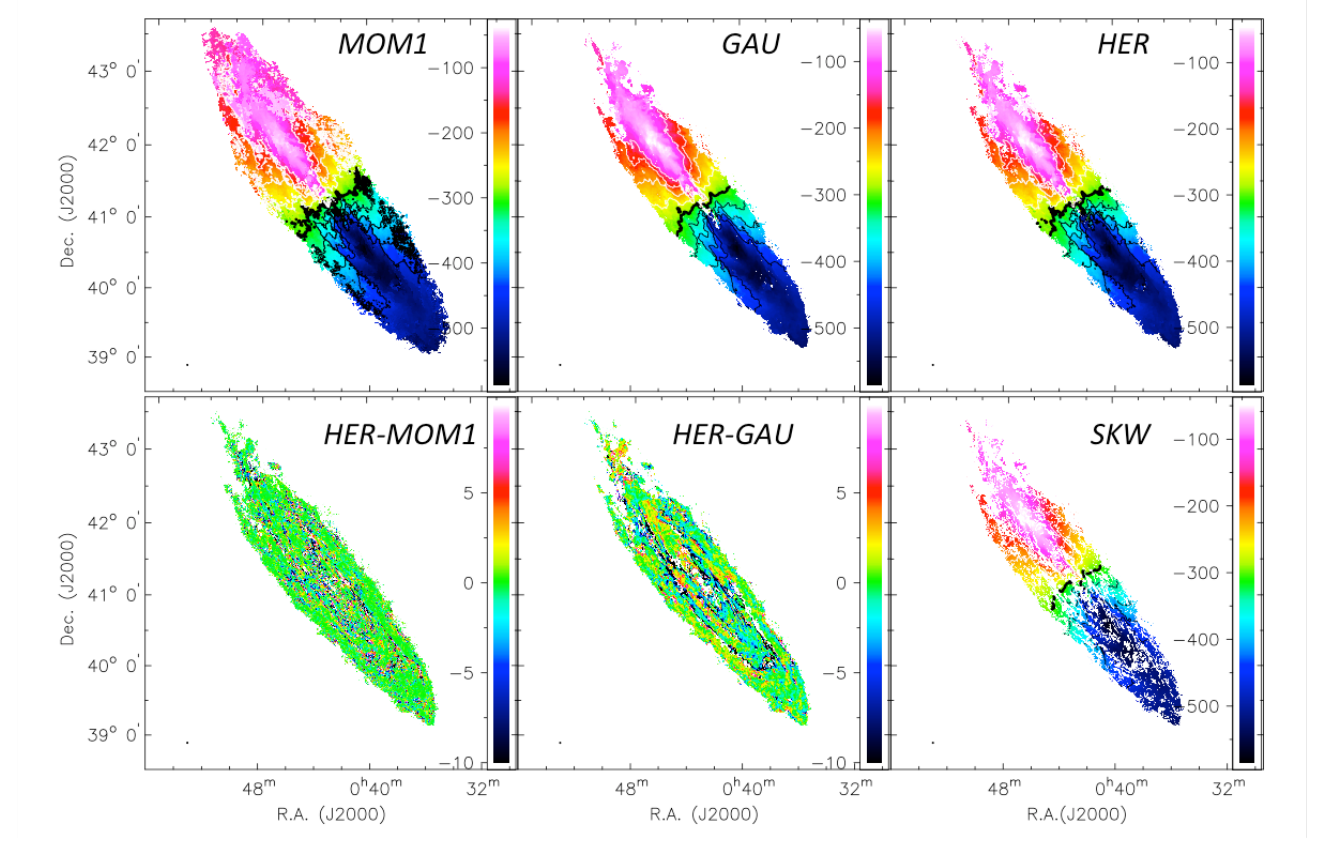}
\caption{Upper panels from left to right: three types of velocity fields -- Moment 1, Gaussian, and Gauss-Hermite velocity fields; 
Bottom panels from left to right: residual maps between Gauss-Hermite and Moment 1 velocity fields, Gauss-Hermite and Gaussian velocity fields, and the Gaussian velocity field with blanked skewed regions of the velocity profile. Data is shown for NGC 224. 
 \label{fig_vifi}}
\end{center}
\end{figure*}

After we estimated the rotational velocity of our sample galaxies
using the width of global H{\sc i} profiles at the 50\% and 20\% levels,
we need to compare these values with the velocities V$_{\rm max}$ and
V$_{\rm flat}$ derived from the rotation curves of the galaxies. 
However, first the velocity field should be constructed for each galaxy,
from which the rotation curve will be derived.

Velocity fields are playing a very important role in analysing the
dynamics of the cold gas in a galaxy.  Usually, the strongest signature
of a velocity field reflects the rotation of the gas disk in the
gravitational potential of a galaxy.  In practice, however, velocity
profiles are often affected by various systematic effects caused either
by limited instrumental resolution or by physical processes within a
galaxy. 

The most common instrumental effect is known as ``beam-smearing'' which
becomes manifest in observations with relatively low angular resolution
(\citet{bosma78}, chapter 3).  This usually happens when a synthesised
beam is relatively large in comparison with the size of the galaxy,
leading to asymmetries in the individual velocity profiles at different
locations in the galaxy, in particular near its dynamical centre.  The
velocity profiles become skewed towards the systemic velocity of the
galaxy and, thereby, the rotational velocity of the galaxy can be
systematically underestimated.  Therefore, one needs to correct for this
effect by identifying the velocity close to the maximum velocity found
in each velocity profile \citep{sancisiallen}.  Otherwise, the velocity
obtained at a particular position will always be closer to the systemic
velocity of the galaxy.  Even though our sample consists of very large,
nearby galaxies, three of the galaxies observed with the GMRT (NGC
3370, NGC 4639, NGC 5584) show ``beam-smearing'' features in their
velocity profiles (see Appendix A for more details).  For those galaxies
we use the envelope--tracing correction method described in
\citet{versanc01}. 

Furthermore, physical processes within a galaxy can have important
effects on the observed velocity field and may lead to
non--axisymmetries and non--circular motions of the gas.  Usually these
processes are associated with gas flows induced by the gravitational
potential of spiral arms and/or bars of galaxies, and result in
so--called streaming motions \citep{visser80, shetty07}.  These appear
as coherent deviations from the circular velocity of the gas in a
galaxy.  Since we are interested in measuring the circular velocity of
the gas to probe the gravitational potential of the dark matter halo,
the flat part of the outer rotation curve V$_{\rm flat}$ should be
measured accurately.  Therefore we identified and removed the signature
of streaming motions from the velocity fields.  There are many ways to
construct velocity fields, and below we describe the most common of
them, as well as our method of identifying streaming motions. 

Calculating the {\it first moment } of the velocity profiles is one of
the most well--known and computationally straight forward methods to
construct a velocity field.  Here the velocity at each position in the
velocity field is an intensity weighted mean of the pixel values along
the operation axis in the data cube.  Despite it being the most commonly
used approach, this method has many disadvantages and its sensitivity to
noise peaks in a H{\sc i} spectrum is one of them.  To obtain a reliable
velocity field one needs to be very careful in identifying the emission
regions and use only those areas to prevent the noise from significantly
influencing the mean velocity.  Another disadvantage is the fact that a
first moment can be sensitive to the effects of severe beam--smearing
which may result in skewed profiles.  Thus, first--moment velocity
fields can be used only as a first order approximation or as initial
estimates for the other methods.  The upper left panel of Figure
\ref{fig_vifi} illustrates a moment--1 velocity field of NGC 224. 

Fitting a {\it Gaussian } function to the velocity profiles is another
way to derive the velocities that are representative of the circular
motion of the gas.  The central velocity of the Gaussian component
defines the velocity at each position in the velocity field.  Where the
first--moment method always yields a velocity regardless of the
signal--to--noise ratio in the profile, fitting a Gaussian to the
profile requires a minimum signal--to--noise ratio.  Therefore, a
velocity field constructed by Gaussian fits usually constitutes of fewer
pixels that have a higher reliability compared to a first--moment map. 
In spite of the fact that a Gaussian--based velocity field is usually
more sparse, this method is most common to construct reliable H{\sc i}
velocity fields for detailed analyses.  The upper--middle panel of
Figure \ref{fig_vifi} shows a Gaussian--based velocity field of NGC 224,
illustrating that velocity profiles at lower signal--to--noise ratio in
the outer regions of galaxies do not allow for acceptable Gaussian fits. 

Using a {\it Gauss--Hermite polynomial} function to fit the velocity
profiles allows the identification of asymmetric velocity profiles that
may be affected by beam--smearing and/or streaming motions and that
cannot be described properly by a single Gaussian.  Gauss--Hermite
polynomials allow to quantify deviations from a Gaussian shape with two
additional parameters describing the skewness ($h3$ term) and kurtosis
($h4$ term) of the profile \citep{hermite}.  In our case we do not take
into account the $h4$ term because instrumental and astrophysical
effects typically result in asymmetric velocity profiles that can be
identified by their $h3$ values. 

For our studies we used both Gaussian and Gauss--Hermite polynomial
functions to fit the velocity profiles and construct the corresponding
velocity fields as indicated by {\it GAU} and {\it HER} in Figure
\ref{fig_vifi}.  We identify asymmetric velocity profiles by considering
the difference between the {\it HER} and {\it GAU} velocity fields as
illustrated in the bottom--middle panel of Figure \ref{fig_vifi}. 
Pixels with values in excess of $\pm3$ km/s in the {\it HER$-$GAU}
difference map were blanked in the {\it GAU} velocity field.  Thereby we
have obtained the third type of velocity field {\it SKW} as illustrated
in the bottom--right panel of Figure \ref{fig_vifi}.  These {\it SKW}
velocity fields will be used to derive the rotation curves from.

\begin{figure}
\begin{center}
\includegraphics[scale=0.60]{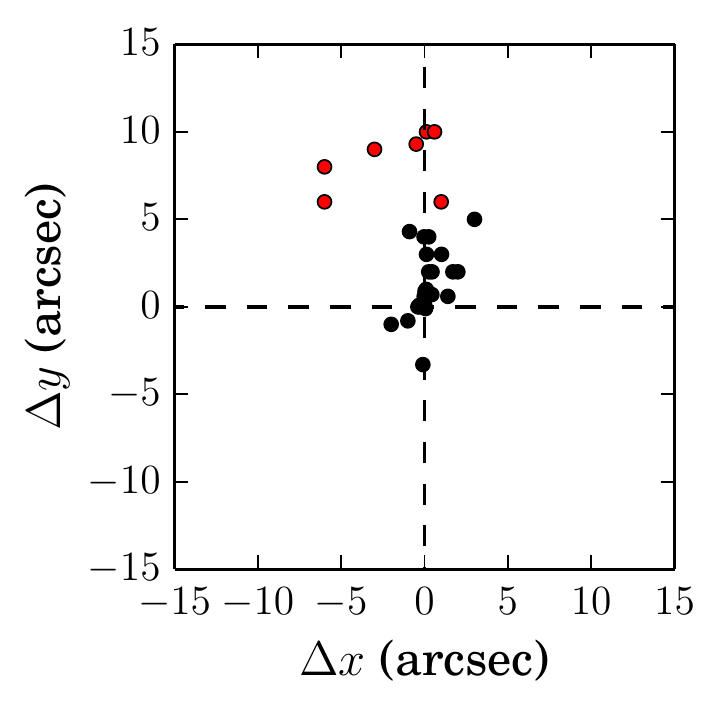}
\caption{Differences between optical and dynamical centres of our sample galaxies. Red symbols indicated the outliers for which optical centres were adopted during the tilted--ring modelling. 
\label{fig_cent}}
\end{center}
\end{figure}

\begin{figure}
\begin{center}
\includegraphics[scale=0.65]{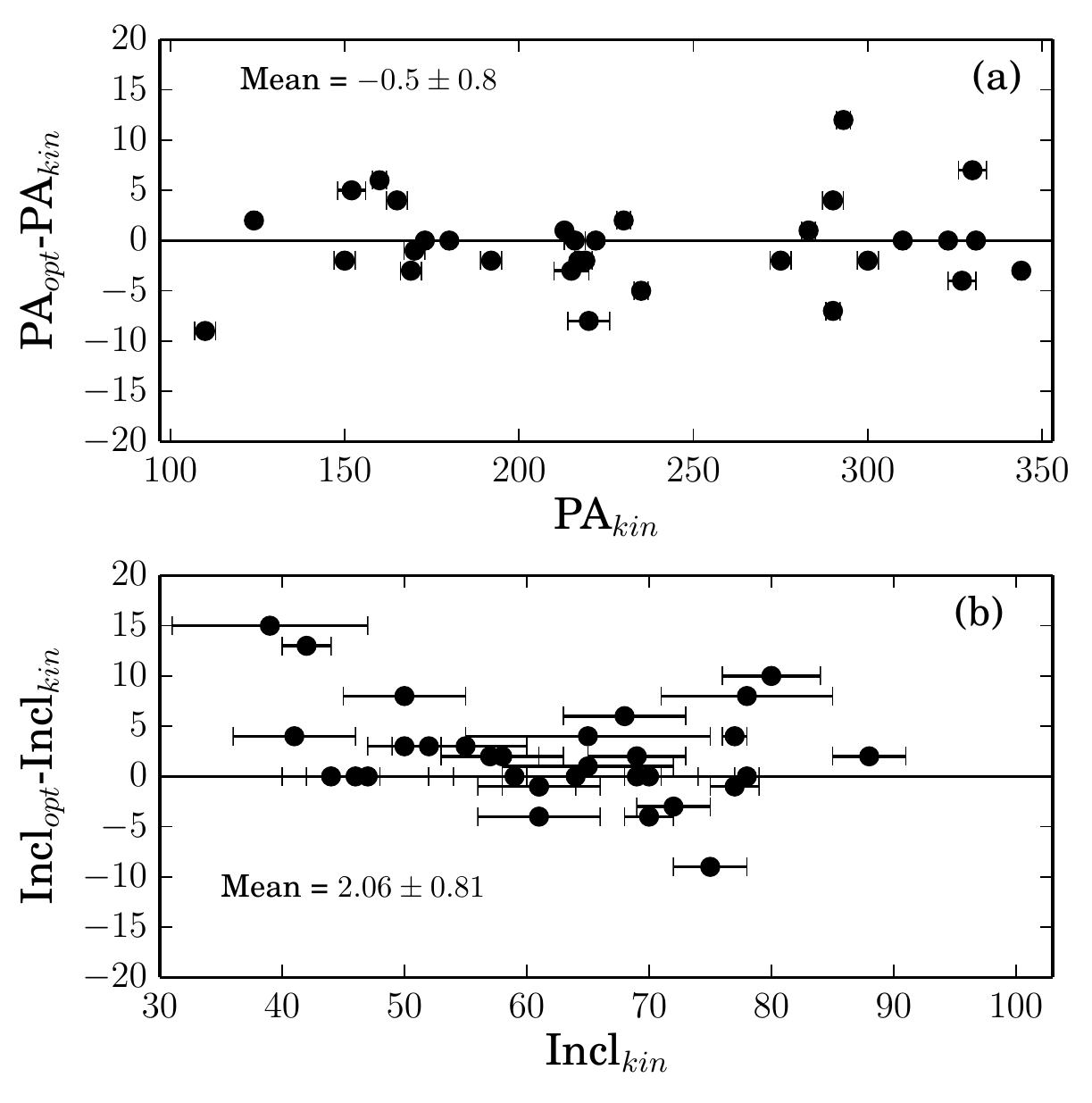}
\caption{Upper panel: the differences between optical position angles from \citetalias{TC12} and the kinematical position angles derived from the tilted--ring modelling. 
Bottom panel: the differences between the optical inclination angles from \citetalias{TC12} and the kinematical inclination angles derived from the tilted--ring modelling. 
\label{fig_incl}}
\end{center}
\end{figure}

\begin{figure*}
\begin{center}
\includegraphics[scale=0.8]{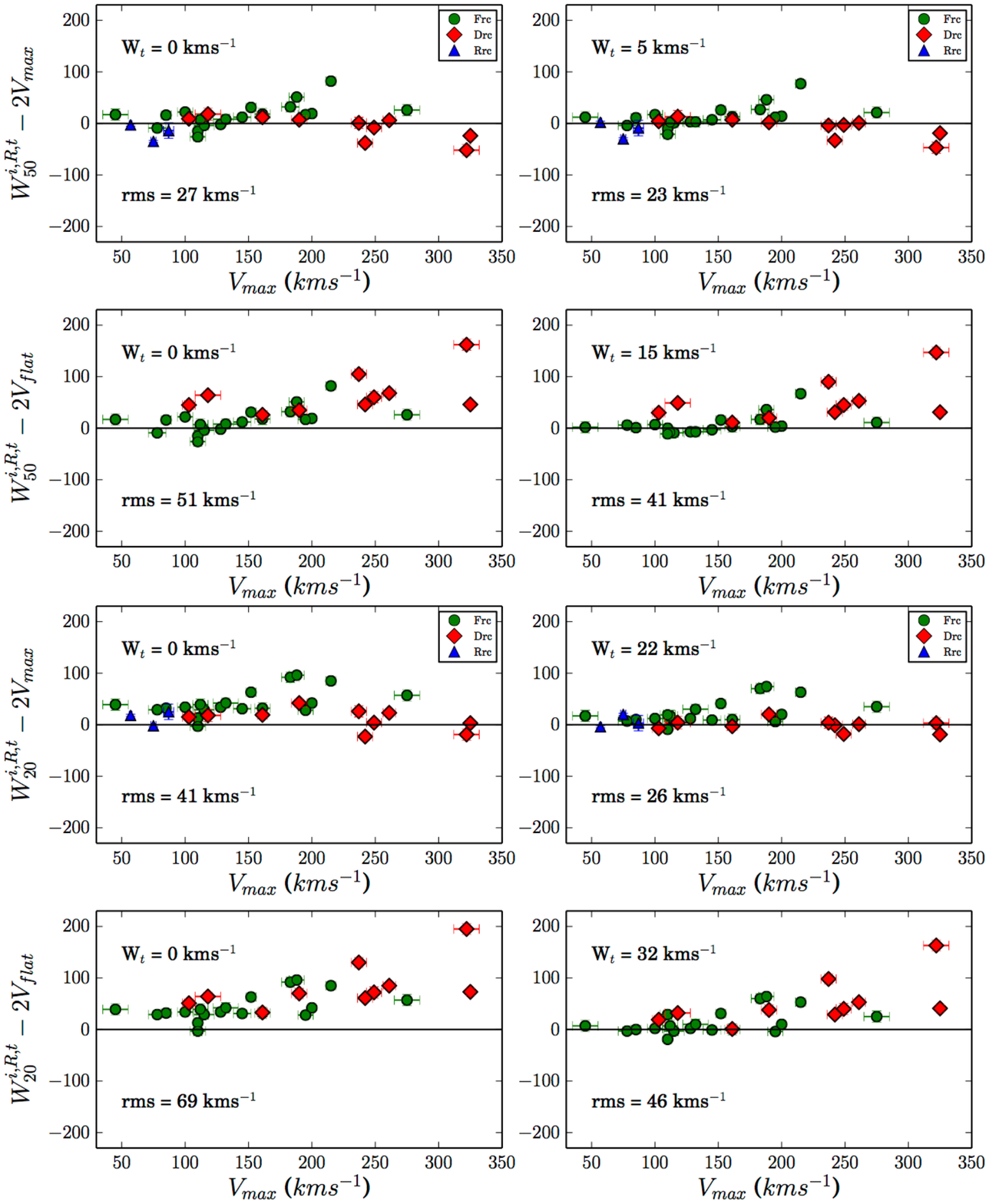}
\caption{Comparison of the global profile widths at the 50\% level (upper four panels) and the 20\% level (lower four panels), corrected only instrumental resolution and inclination $W_{50,20}^{R, i}$ with 2V$_{\rm max}$ and 2$V_{\rm flat}$ (left panels), and corrected also for random motions $W_{50,20}^{R,t,i}$ (right panels). Blue symbols indicate galaxies with rising rotation curves (V$_{\rm max}$$<$V$_{\rm flat}$) and red symbols indicate galaxies with declining rotation curves (V$_{\rm max}$$>$V$_{\rm flat}$). Green symbols show flat rotation curves without declining part (V$_{\rm max}$$=$V$_{\rm flat}$).
\label{fig_vres}}
\end{center}
\end{figure*}

\begin{table}
\begin{tabular}{lr@{$\pm$}lr@{$\pm$}lr@{$\pm$}lr@{$\pm$}ll}
\hline
Name & \multicolumn{2}{l}{$V_{sys}$}&\multicolumn{2}{l}{$P.A.$}&\multicolumn{2}{l}{$Incl.$}&\multicolumn{2}{l}{$V_{max}$}& $V_{flat}$ \\
&\multicolumn{2}{l}{$km s^{-1}$}&\multicolumn{2}{l}{$deg.$}&\multicolumn{2}{l}{$deg.$ }&\multicolumn{2}{l}{$km s^{-1}$}&$km s^{-1}$\\
\hline
NGC 0055	&130&5 	&110&3	 &78&7	     &85 &1	      &85$\pm$2   \\
NGC 0224	&-300&3	&37&1	 &78&1	     &261&2    &230$\pm$7	\\
NGC 0247	&160&10	&169&3	 &77&2      &110&5     &110$\pm$5	\\
NGC 0253	&240&5	&230&2	 &77&1	     &200&4    &200$\pm$4	\\
NGC 0300	&135&10	&290&3	 &46&6    &103&3	      &85$\pm$7   \\
NGC 0925	&550&5	&283&2	 &61&5	     &115&4	    &115$\pm$4	\\
NGC 1365	&1640&3	&218&2	 &39&8	     &322&6    &215$\pm$4	\\
NGC 2366	&107&10	&42&6	 &68&5	       &45&5      &45$\pm$5   \\
NGC 2403	&135&1	&124&1	 &61&3	     &128&1    &128$\pm$1	\\
NGC 2541	&560&5	&170&3	 &64&4	     &100&4    &100$\pm$4	\\
NGC 2841	&640&20	&150&3	 &70&2	     &325&2    &290$\pm$6	\\
NGC 2976	&5&5	   &323&1 &61&5	       &78&4      &78$\pm$4   \\
NGC 3031	&-40&10	&330&4	 &59&5	     &249&3    &215$\pm$9	\\
NGC 3109	&404&5	&92&3	 &80&4	     &57&2          &  --	\\
NGC 3198	&660&10	&215&5	 &70&1	     &161&2	&154$\pm$4	\\
IC 2574  	        &51&3  	&55&5	      &65&10	&75&5       &  --	\\
NGC 3319	&730&4	&33&2      &57&4    &112&10	&112$\pm$10	\\
NGC 3351	&780&5        &192&1	 &47&5	     &190&5	&176$\pm$8	\\
NGC 3370	&1280&15     &327&3	 &55&5    &152&4	     &152$\pm$4	\\
NGC 3621	&730&13	&344&4	 &65&7    &145&5	      &145$\pm$5	\\
NGC 3627	&715&10	&172&1	 &58&5	     &183&7	&183$\pm$7	\\
NGC 4244	&245&3	&222&1	 &88&3      &110&6	      &110$\pm$6	\\
NGC 4258	&445&15	&331&1	 &72&3	     &242&5	&200$\pm$5	\\
NGC 4414	&715&7	&160&2	 &52&4	     &237&10	&185$\pm$10	\\
NGC 4535	&1965&5	&180&1	 &41&5      &195&4	      &195$\pm$4	\\
NGC 4536	&1800&6	&300&3	 &69&4	     &161&10	&161$\pm$10	\\
NGC 4605	&160&15	&293&2	 &69&5	      &87&4       &  --  \\
NGC 4639	&978&20	&311&1	 &42&2	     &188&1	&188$\pm$1	\\
NGC 4725	&1220&14      &30&3	 &50&5    &215&5	      &215$\pm$5	\\
NGC 5584	&1640&6	&152&4	 &44&4    &132&2	      &132$\pm$2	\\
NGC 7331	&815&5	&169&3	 &75&3	     &275&5	&275$\pm$5	\\
NGC 7793	&228&7	&290&2	 &50&3	     &118&8	      &95$\pm$8   \\
\hline
\end{tabular}
\caption{Tilted--ring modeling results. 
Column (1): galaxy name;
Column (2): systemic velocity;
Column (3): kinematic position angle, calculated as an angle between the North direction and receding side of a galaxy;
Column (4): kinematic inclination from face--on;
Column (5): maximal rotational velocity;
Column (6): rotational velocity of the flat part of rotation curve.
}
\label{tbl_rot}
\end{table}

\subsection{Rotation curves and velocity field models }

 From the previous section it is clear that velocity fields tend to
have pixels with skewed velocity profiles mostly due to beam--smearing
and non--circular motions.  Thus it is important to stress that we
derive rotation curves for our galaxies from the {\it SKW} velocity
fields which were censored for such effects (see Section 3.4). For this
purpose, we fitted a tilted--ring model to the {\it SKW} velocity field
to derive the rotation curve of each galaxy. Details on the
tilted--ring modelling method, its parameter fitting and error
calculations, are described by \citep{begeman89}. The derivation of a
rotation curve was done in four steps.  The widths of the tilted--rings
were adjusted separately for each galaxy, taking into account the H{\sc
i} morphology and the size of the synthesised beam. 

As a first step, for each ring we fitted to the {\it SKW} velocity
field, only the position of its dynamical centre $x_{0}$ and $y_{0}$,
and the systemic velocity of a galaxy $V_{sys}$ were fitted.  All other
parameters, such as the position angle ($pa$), inclination angle ($i$)
and the rotation velocity V$_{\rm rot}$ were fixed and remained the same
for all tilted rings. In this first step, the position ($pa$) and
inclination ($i$) angles were adopted from the optical measurements
  done by \citetalias{TC12} and listed in Table \ref{tbl_samp} while
  $V_{rot}$ was estimated from the measured width of the global H{\sc
    i} profile. All the data points within a tilted ring were weighted
uniformly at this step.  After the fitting, the position of the
dynamical centre ($x_{0},y_{0}$) and the systemic velocity $V_{sys}$
were calculated as the weighted means of the solutions for all fitted
rings.  For cases where the fitted position of the dynamical centre had
large deviations from ring to ring, the position of the optical centre
was adopted.  The offsets between optical and dynamical centres are
shown in Figure \ref{fig_cent} and are typically smaller than the
angular size of the interferometric synthesised beam. 

At the second step, we fitted only the position angle ($pa$) of a tilted
ring, indicating the angle between north and the kinematically receding
side of a galaxy, measured eastward.  All the other parameters were kept
fixed whereby the position of the dynamical centre and the systemic
velocity $V_{sys}$ were adopted from the previous step.  Again, all the
data points within each tilted--ring were weighted uniformly.  While the
position angle can be determined very accurately and is mostly invariant
from ring to ring, we paid attention to the possible existence of a
trend with radius so as to include a geometric warp in the model when
warranted.  The weighted average value of the position angle within the
optical radius of a galaxy, as well as a possible trend with radius were
fixed at the next step.  A comparison between the optical and the
kinematical position angles is presented in Figure \ref{fig_incl}{\it
a}.  Overall, the optical and kinematic position angles are in a good
agreement with a weighted mean difference of $-0.5\pm0.8$ {\it degrees}. 

In the third step, adopting and fixing the results from the previous
steps, we fitted both the inclination angle ($i$) and the rotational
velocity (V$_{\rm rot}$) of each ring, weighing all the data points in a
ring according to $cos(\theta)$, where $\theta$ is the angle in the
plane of the galaxy measured from its receding side.  The values of the
inclination angle and the rotation velocity of a tilted--ring are highly
covariant in the fitting algorithm and rather sensitive to pixel values
in the {\it SKW} velocity field that are still affected by non--circular
motions.  Hence, the fitted inclination may vary significantly from ring
to ring within a galaxy.  We assume, however, that the cold gas within
the optical radius of a galaxy is largely co--planar and we calculated
the weighted mean inclination of the rings within the optical radius of
the galaxy. We refer to this as the kinematic inclination of a
galaxy. If the more accurately determined position angle indicates a
warp in the outer regions, we also allow for an inclination warp if the
fitted values hint at this. Adopting the inclination angles derived from the optical
images as reported by \citetalias{TC12}, we compare the optical and
kinematic inclination angles in the bottom panel of Figure \ref{fig_incl}. 
We find that the kinematical inclination angles are
systematically lower than the optical ones, with an average weighted
difference of $2.06\pm0.81$ {\it degrees}.  It is important to point out
that the derived optical inclination angles depend on the assumed
thickness of the stellar disk when converting isophotal ellipticities to
inclinations, and therefore it can cause the systematic offset between
optical and kinematical inclinations.  Such a systematic bias is not
expected in deriving kinematic inclinations. We note that the
systematic difference we find is very close to an empirical correction
of 3$^{\circ}$ as determined by \citet{aar80}, which was usually applied
when interpreting optical and kinematical inclination angles.

In the fourth and final step, V$_{\rm rot}$ was fitted for each
tilted--ring while all the previously determined parameters ($x_{0},
y_{0}, V_{sys}, pa_{kin}, i_{kin}$) were kept fixed, including the
geometry of a possible warp. Hence, in deriving the rotation curve
of a galaxy we use its global systemic velocity and geometry (except for
the five cases where optical centres were adopted, Figure
\ref{fig_cent}) based on the characteristics of the {\it SKW} velocity
field. The values of V$_{\rm rot}$ as measured during the previous step
were adopted as an initial estimate while all the points within each
tilted--ring were considered and weighted with $cos(\theta)$ when
fitting V$_{\rm rot}$ for each ring, yielding the circular velocity of
the H{\sci} gas as a function of distance from a galaxy's dynamical
centre.  To investigate possible kinematic asymmetries we fitted V$_{\rm
rot}$ not only to the full tilted--ring but also to the receding and
approaching sides of a galaxy separately.  Subsequently, we used these
rotation curves to identify the maximum circular velocity V$_{\rm max}$
of the gas disk, as well as the circular velocity of the outer gas disk
V$_{\rm flat}$. 

To verify our final rotation curves, they were projected onto
position--velocity diagrams extracted from the data cubes for visual
inspection.  To further verify the results from the tilted--ring fits to
the {\it SKW} velocity field, they were also used to construct an
axisymmetric, model velocity field of a regular, rotating gas disk. 
This model was subsequently subtracted from the Gauss--Hermite
polynomial velocity field to construct a map of the residual velocities. 
This residual velocity field highlights the locations of velocity
profiles with significant skewness and could have revealed possible
systematic residuals that should have been accommodated by the
tilted--ring model. 

Table \ref{tbl_rot} summarises the final results obtained from the
tilted--ring fitting process based on the {\it SKW} velocity field. 
Errors on $V_{sys}$, the position angle and the inclination are
based on the variance in the ring--to--ring solutions. Errors on
V$_{\rm max}$ and V$_{\rm flat}$ were measured as the difference between
the velocities of the approaching and receding sides of the galaxy. 
All final data products are presented in the accompanying Atlas that
will be discussed below. 

We conclude this section by recalling that the global geometric
properties of the H{\sc i} gas disks (dynamical centre, position angle
and inclination angle) as derived from the {\it SKW} velocity fields are
in good agreement with the same geometries derived from photometric
images of these galaxies.  Moreover, from the fitting procedure we
derived high--quality rotation curves of the cold gas, presenting the
circular velocity of a galaxy as a function of distance from its
dynamical centre.  These rotation curves will be used in a forthcoming
paper analysing the detailed mass distributions within our sample
galaxies.  For the purpose of studying the statistical properties of the
Tully--Fisher relation in a forthcoming paper, we measure two values of
the rotational velocity of each galaxy: the maximal rotational velocity
V$_{\rm max}$ and the velocity at the flat part of the outer rotation
curve V$_{\rm flat}$.  In the next section we will compare these values
with the rotational velocity as estimated from the corrected width of
the single--dish profiles.

\begin{figure}
\begin{center}
\includegraphics[scale=0.75]{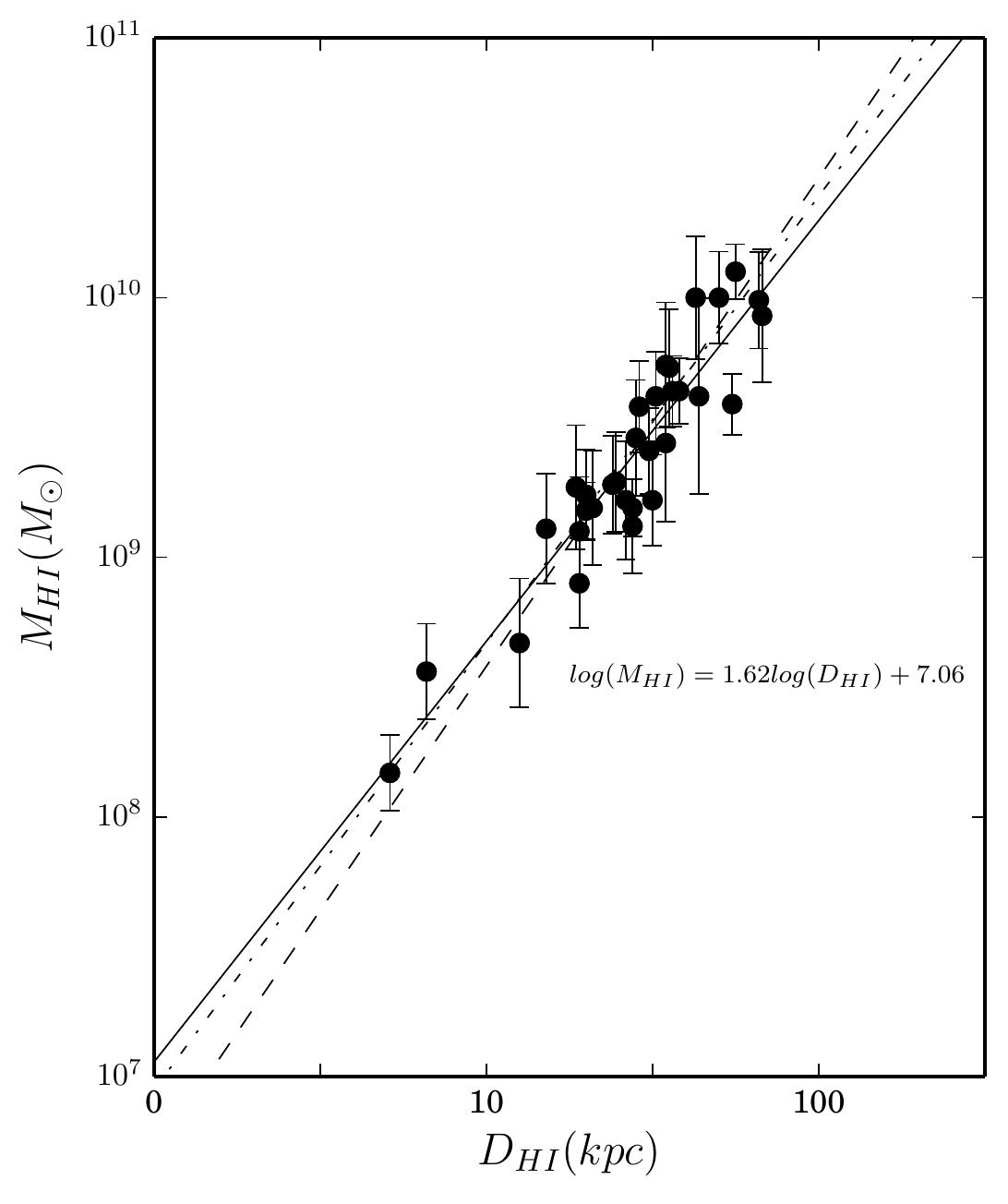}
\caption{Correlation between H{\sc i} mass and isophotal diameter of the H{\sc i} disks. The solid line indicates the linear fit to the relation. Dashed and dashed--dot lines show the results found by \citet{versanc01} and by \citet{martinsson11}, respectively.
\label{fig_DHI}}
\end{center}
\end{figure}

\subsection{Comparison of different velocity measures}

As was already mentioned above, the width of the global H{\sc i} profile
can give a good estimate of the typical circular velocity of the cold
gas in a galaxy.  However the measured width of the global profile
should be corrected not only for finite instrumental spectral
resolution, but also for astrophysical effects such as the turbulent
motions of the gas and the inclination of the rotating gas disk. 

\subsubsection{Turbulent motion correction.} 

The empirical correction of the measured line width for turbulent motion
of the gas was investigated previously by several authors
\citep{bottinelli83, broeils92, rhee96} by matching the corrected global
H{\sc i} line widths to the rotational velocities as measured from
rotation curves. 

In this work we adopt the recipe and parameter values for turbulent
motion correction from \citet{versanc01}.  While adopting $W_{c}=120$
km/s, they showed that, in order to match the amplitudes of the rotation
curves, the empirical values of the turbulence parameter $W_{t}$ as
proposed in previous studies, should be adjusted.  Its values depend on
the level at which the line width was measured (20\% or 50\%) and on the
velocity measure from the rotation curve (V$_{\rm max}$ or V$_{\rm
flat}$).  Therefore, when applying this correction to our galaxies, we
adopt their values: \newline

$W_{t,20}^{max} = 22\,kms^{-1}, \, W_{t,50}^{max} = 5\,kms^{-1}$;
\newline

$W_{t,20}^{flat} = 32\,kms^{-1}, \, W_{t,50}^{flat} = 15\,kms^{-1}$.
\newline

\subsubsection{Inclination correction.} 

The correction of the global profile width for inclination was done
using the kinematic inclinations obtained from the tilted--ring
modelling of the {\it SKW} velocity fields as listed in Table \ref{tbl_rot}:
\newline

$W_{50,20}^{R,t,i} = W_{50,20}^{R,t}/\, sin(i_{kin})$.\\\\ 
The line widths corrected for instrumental resolution, turbulent motion,
and inclination are presented in Table \ref{tbl_param}.

\subsubsection{Comparing $W_{50,20}^{R,t,i}$,  V$_{\rm max}$ and V$_{\rm flat}$}

We start this subsection by stressing, again, that a single--dish
measurement of a global H{\sc i} profile does not inform the observer
whether a galaxy's rotation curve is declining or not. 

Figure \ref{fig_vres} shows the differences between the corrected global
profile widths and the values of V$_{\rm max}$ and V$_{\rm flat}$
derived from the rotation curves.  Green symbols correspond to galaxies
with rotation curves that monotonically rise to a extended flat part
(V$_{\rm max}$ = V$_{\rm flat}$).  Red symbols correspond to galaxies
with rotation curves that rise to a maximum beyond which they decline to
a more or less extended flat part (V$_{\rm max}$$>$V$_{\rm flat}$). 
Blue symbols correspond to galaxies with rotation curves that are still
rising at theit last measured point and do not reach a flat part
(V$_{\rm max}$$<$V$_{\rm flat}$ while V$_{\rm flat}$ is not actually
measured).  The left panels show the differences without any random
motion corrections ($W_{t}=0$) applied, and the right panels show the
differences with the random motion corrections applied for variouss
values of $W_{t}$. 

We see that the more--massive galaxies tend to have declining rotation
curves while the three galaxies with rising rotation curves are all of
low mass.  From the panels in the left column we conclude that the
global profile width systematically overestimates the rotational
velocity if no correction for turbulent motions is applied, with the
exception of $W_{50}^{R,i}$ as an estimate for 2V$_{\rm max}$ (upper
panel).  Furthermore, by comparing the corresponding panels in the left
and right columns, we see that the variance in the differences is
significantly reduced by applying the corrections for turbulent motions:
Both $W_{50}^{R,t,i}$ and $W_{20}^{R,t,i}$ match the maximum velocity
notably better after corrections for random motions, although systematic
offsets between the red and green symbols persist. 

Focusing on the panels in the right column, we see that $W_{t}$ values
of 5 and 22 km/s allow the recovery of 2V$_{\rm max}$ from the 50\% and
20\% line widths respectively, at least in a statistical sense.  In
these cases, however, 2V$_{\rm flat}$ for galaxies with monotonically
rising rotation curves (green symbols) tend to be systematically
overestimated, especially for the 20\% line widths $W_{20}^{R,t,i}$.  We
also see that $W_{t}$ values of 15 and 32 km/s allow a somewhat better
recovery of 2V$_{\rm flat}$ for galaxies with monotonically rising
rotation curves (green symbols) but the corrected global profile line
widths systematically overestimate V$_{\rm flat}$ for galaxies with
declining rotation curves due to the fact that V$_{\rm max}$$>$V$_{\rm
flat}$. 

An important conclusion from this comparison is that, for a sample of
galaxies with flat and declining rotation curves, the corrected width of
the global H{\sc i} line profile cannot be unambiguously corrected to
recover V$_{\rm flat}$ which is probing the potential of the dark matter
halo without introducing a systematic bias that is largely correlated
with the mass of a galaxy.  This has implications for the slope of, and
may possibly introduce a curvature in, the Tully--Fisher relation when
using $W_{20,50}^{R,t,i}$ from global H{\sc i} profiles instead of
V$_{\rm flat}$ from extended H{\sc i} rotation curves.

\begin{table}
\begin{tabular}{lr@{.}lr@{.}lr@{.}lr@{.}l}
\hline
Name &\multicolumn{2}{l}{$\int S_{v}dv$}&\multicolumn{2}{l}{$M_{HI}$}&\multicolumn{2}{l}{$D_{HI}$}&\multicolumn{2}{l}{$\Sigma_{HI}^{max}$} \\
& \multicolumn{2}{l}{Jy $ km s^{-1}$}&\multicolumn{2}{l}{$10^{9} M_{\odot}$}&\multicolumn{2}{l}{$kpc$}&\multicolumn{2}{l}{$M_{\odot}pc^{-2}$} \\
\hline
NGC 0055	&1786&7   &1&54	&20&3	              & 14&7 \\
NGC 0224	&30292&5   &4&18	&44&5	          & 5&4 \\
NGC 0247	&594&4	    &1&67   &26&7	         & 6&1 \\
NGC 0253	&693&7       &1&95	&25&1              & 4&6 \\
NGC 0300	&1779&2	    &1&57	&28&1            & 5&6   \\
NGC 0925	&274&7	    &5&44	&35&5        & 4&4  \\
NGC 1365	&167&3	    &12&71	&57&4          & 7&8  \\
NGC 2366	&307&1	    &0&79	&19&2	         & 12&3  \\
NGC 2403	&1088&6	    &2&61	&30&9	  	     & 9&4  \\
NGC 2541	&143&3	    &4&25	&32&6          & 1&6  \\
NGC 2841	&183&5	    &8&56	&68&1         & 4&3  \\
NGC 2976	&50&2         &0&15	&5&1          & 12&1 \\
NGC 3031	&907&1	    &2&77	&34&8     	& 4&1 \\
NGC 3109	&1181&2	    &0&47	&12&6           & 7&1 \\
NGC 3198	&218&2	    &9&81	&66&9           & 4&9 \\
IC 2574	     &389&4	 &1&34	&27&7            & 8&0\\
NGC 3319	&93&1          &3&88	&29&1         & 1&4 \\
NGC 3351	&59&9	         &1&68	&31&7        	& 6&4 \\
NGC 3370	&17&3	         &2&90	&28&4         & 4&9  \\
NGC 3621	&904&1	   &10&05	&51&0	        & 9&3  \\
NGC 3627	&53&7         &1&27	&19&4         & 11&3  \\
NGC 4244	&423&9	   &1&77	&20&3           & 1&2  \\
NGC 4258	&440&4	   &5&55	&35&4            & 4&8   \\
NGC 4414	&25&2         &1&86	&18&8	  	       & 2&0   \\
NGC 4535	&75&7	        &4&44	&38&2   	        & 5&1    \\
NGC 4536	&82&2	        &4&40	&36&5	        & 6&2       \\
NGC 4605	&55&1	        &0&36	&6&7     	       & 3&1    \\
NGC 4639	&13&8	        &1&56	&21&2	  	       & 4&3      \\
NGC 4725	&102&2	  &3&92	&80&4	        & 2&6          \\
NGC 5584	&15&9	        &1&93	&24&2	  	  & 5&1   \\
NGC 7331	&202&1	  &10&03	&42&8          & 22&1      \\
NGC 7793	&352&5      &1&29	&15&2            & 10&4    \\
\hline
\end{tabular} 
\caption{Measured H{\sc i} properties. 
Column (1): galaxy name;
Column (2): integrated H{\sc i} flux;
Column (3): H{\sc i} mass;
Column (4): H{\sc i} diameter;
Column (5): maximum H{\sc i} column density, projected towards face-on
}
\label{tbl_param}
\end{table}

\subsection{Total H{\sc i} maps and radial H{\sc i} surface--density profiles }

H{\sc i} integrated column--density maps were created by adding the
primary beam corrected, non--zero emission channels of the cleaned data
cubes with applied masks.  The pixel values in the resulting map were
converted from flux density units $[Jy/beam]$ to column--densities
$[atoms$ $cm^{-2}]$, according to the formula:

\begin{equation} 
N_{HI} = 1.823 \times 10^{18} \int T_{b} dv,
\end{equation}

where $T_{b}$ is the brightness temperature and $dv$ is the velocity
width over which the emission was integrated.  $T_{b}$ in Kelvin is
calculated as follows:

\begin{equation} 
T_{b} = \frac{605.7}{\Theta_{x} \Theta_{y}}  \, S_{\nu} \,  \bigg( \cfrac{\nu_{0}}{\nu} \bigg)^{2},
\end {equation}

where $\Theta_{x}$ and $\Theta_{y} $ are the major and minor axes of the
Gaussian beam in arcseconds, $S_{\nu}$ is the flux density in
$mJy/beam$, and $\nu_{0} / \nu $ is the ratio between the rest--frame
and observed frequency of the H{\sc i} line. 

The radial H{\sc i} surface density profiles were derived by azimuthally
averaging pixel values in concentric ellipses projected onto the H{\sc
i} column--density map for both the receding and approaching side of a
galaxy separately.  The position and inclination angles of the ellipses
were taken from the tilted--ring fitting results (see Section 3.5),
taking a warp, if present, into account.  Note that the column--density
profiles are deprojected to represent face--on values.  Subsequently the
conversion from $[atoms$ $cm^{-2}]$ to $[M_{\odot}pc^{-2}]$ was applied:

\begin{equation} 
1 [M_{\odot} \, pc^{-2}] = 1.249 \times 10^{20} [atoms \, cm^{-2}] .
\end{equation}

The H{\sc i} column--density maps and the radial, face--on surface
density profiles are presented in the Atlas while the azimuthally
averaged peak column--density of each galaxy is listed in Table
\ref{tbl_param}.

\section{Global H{\sc i} properties}

In this section we investigate the global properties of the gas
disks of our sample galaxies and compare these to the same properties of
galaxies in the volume--limited Ursa Major sample \citep{versanc01} and
the sample of \citep{martinsson11}.  Although our sample of
Tully--Fisher calibrator galaxies is by no means statistically complete,
it is important to demonstrate that our sample is not biased and
representative of late--type galaxies, at least from an H{\sc i}
perspective.  The main properties of the H{\sc i} gas disks of our
galaxies are listed in Table \ref{tbl_param}.

\begin{figure}
\begin{center}
\includegraphics[scale=0.7]{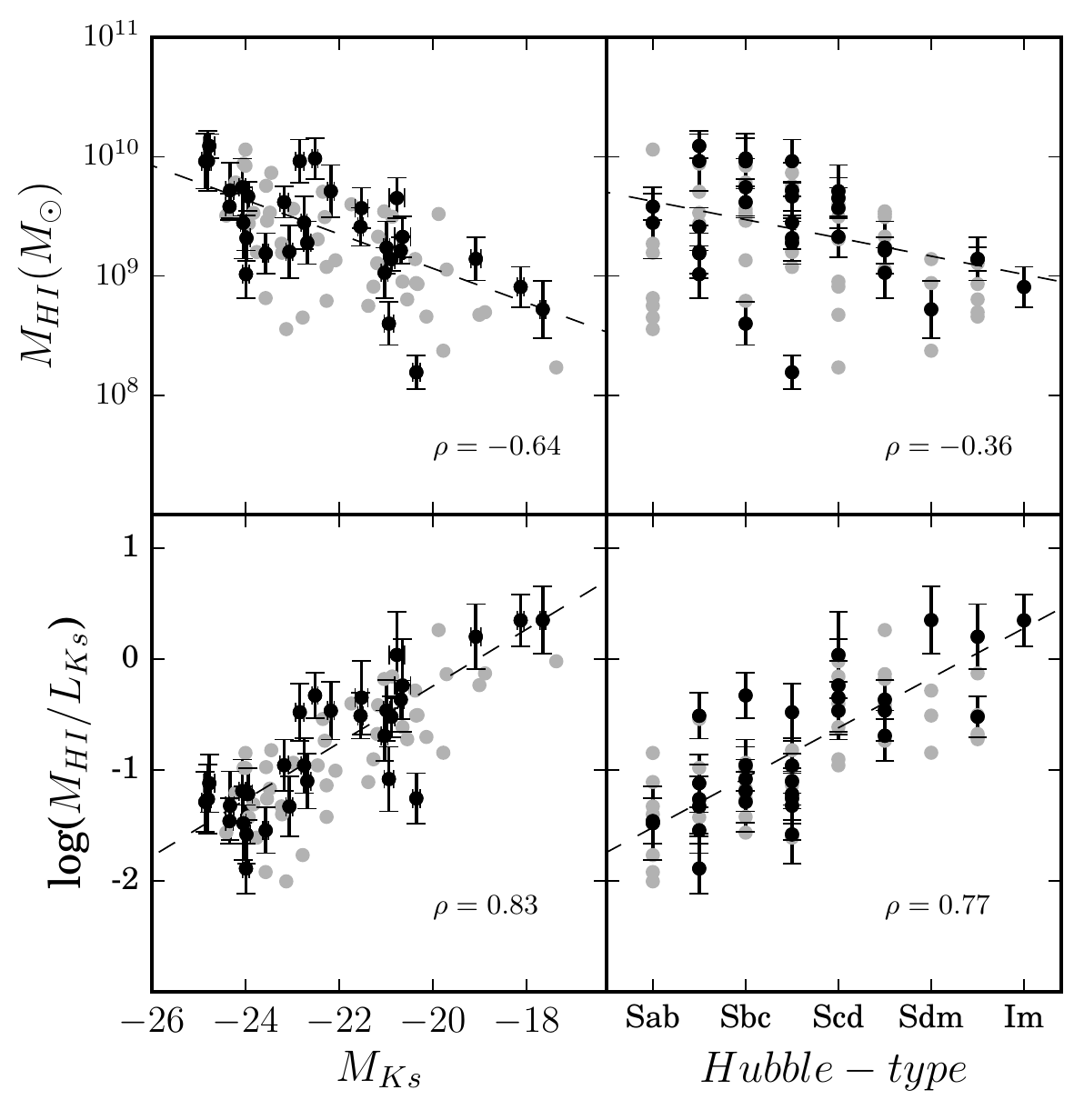}
\caption{Upper panels: correlation between H{\sc i} mass and absolute K$_{s}$ magnitude (left) and morphological type (right). For comparison, we show these values also for galaxies in the Ursa Major volume--limited sample from \citet{versanc01}.
Lower panels: correlation between H{\sc i} mass--to--light ratio and absolute K$_{s}$ magnitude (left) and morphological type (right). Black dashed lines are fits to the relations with $\rho$ indicating Pearson's correlation coefficient. Solid grey line show the fit to the Ursa Major sample.
\label{fig_mlt}}
\end{center}
\end{figure}

\subsection{The masses and sizes of H{\sc i} disks}

The masses and sizes of H{\sc i} disks are among the main global
parameters of spiral galaxies.  A remarkably tight correlation exists
between the mass and the size of a galaxy's H{\sc i} disk as
demonstrated before by various authors \citep{versanc01, swaters02,
noord06}, although the slope and zero point of this correlation vary
slightly from sample to sample.  Here we investigate whether the
galaxies in our sample adhere to this tight correlation.

\begin{figure*}
\begin{center}
\includegraphics[scale=0.75]{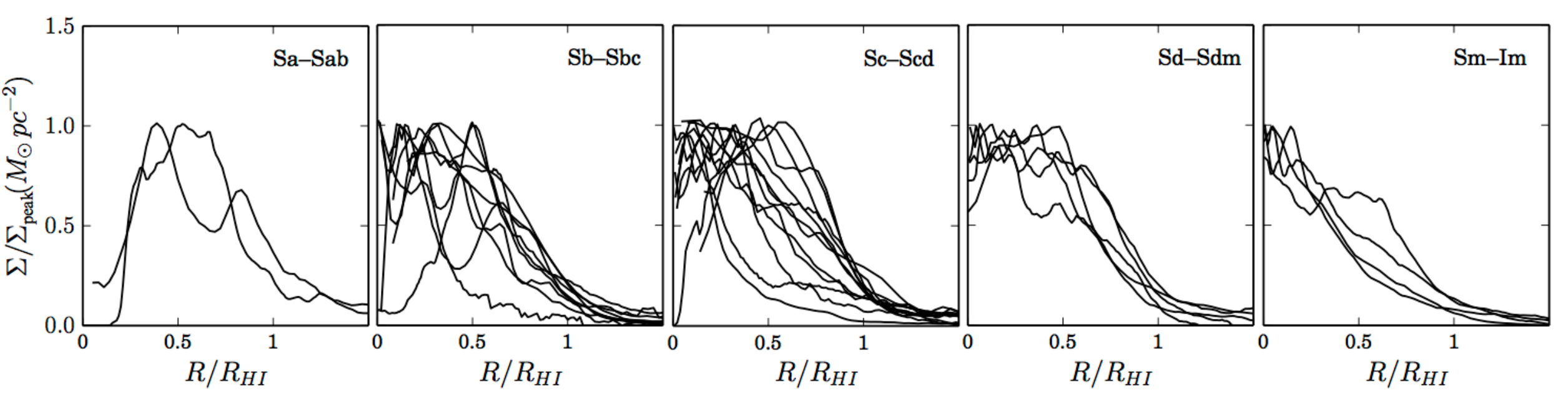}
\caption{ Azimuthally averaged H{\sc i} surface--density profiles, scaled in radius by $R_{HI}$ measured at $1\,M_{\odot}\,pc^{-2}$ and in amplitude by the peak surface--density. The profiles are presented in five bins of morphological types. 
\label{fig_profs}}
\end{center}
\end{figure*}

The total H{\sc i} mass of each galaxy is calculated according to:

\begin{equation}
M_{HI}\,[M_{\odot}]= 2.36 \times 10^{5} \,D^2\, \int S_{\nu} dv,
\end{equation}

where $D$ is the distance to a galaxy in {\it [Mpc]}, as listed in Table
\ref {tbl_samp}, and $ \int S_{\nu} dv\,[Jy\,kms^{-1}]$ is the
integrated flux density, as presented in Table \ref {tbl_param} and
described in Section 3.3.1. 

Next, we measure the diameters of the H{\sc i} disks.  We do not measure
$D_{HI}$ from the H{\sc i} maps, which are often irregular in their
outer regions.  Instead, we measure the H{\sc i} radius as the radius
where the azimuthally--averaged, face--on H{\sc i} surface--density has
dropped to $1\,M_{\odot}\,pc^{-2}$. 

The $M_{HI}$ versus $D_{HI}$ correlation for our sample galaxies is
shown in Figure \ref{fig_DHI}.  The solid line illustrates a linear fit
and can be described as:

\begin{equation}
log(M_{HI})=1.62log(D_{HI})+7.06,
\end{equation}

with a rms scatter of 0.13.  The dashed and dashed--dot lines correspond
to the relations found by \citet{versanc01} and by \citet{martinsson11},
respectively.  We find an insignificantly shallower slope for this
correlation which may be due to a relatively small number of
intrinsically small galaxies in our sample.

\subsection{H{\sc i} mass versus luminosity and Hubble type}

It is well known that H{\sc i} masses and $M_{HI}/L$ ratios of galaxies
correlate well with the luminosity and morphological type of a galaxy. 
Our galaxy sample is not an exception and shows good agreement with
previous studies \citep{roberts94, versanc01, swaters02}.  In our work
we consider the H{\sc i} masses and $M_{HI}/L_{Ks}$ ratios, and compare
the trends observed in our sample with the volume--limited sample of
Ursa Major galaxies \citep{versanc01} for which deep $K_{s}$ photometry
is available and obtained in a similar fashion as for our sample
galaxies, as will be discussed in a fortcoming paper. 

Correlations of the total H{\sc i} mass and the H{\sc i} mass--to--light
ratio with $M_{Ks}$ and the morphological types of the galaxies are
shown in Figure \ref{fig_mlt}.  The light--gray symbols correspond to
galaxies in the Ursa Major reference sample. 

We see that our sample galaxies follow the same known correlations
with a similar scatter as the reference sample.  The upper--left panel
of Figure \ref{fig_mlt} shows that more luminous galaxies tend to
contain more H{\sc i} gas (Pearson's correlation coefficient $\rho=
-0.64$ ), while the lower--left panel shows that the H{\sc i}
mass--to--light ratio decreases with luminosity ($\rho= 0.83$).  The
upper--right panel of Figure \ref{fig_mlt} illustrates that a weaker
trend exists between a galaxy's H{\sc i} mass and its Hubble type, both
for our sample galaxies ($\rho= -0.36$) and the Ursa Major reference
sample ($\rho= -0.38$).  The lower--right panel of Figure \ref{fig_mlt},
however, presents a strong correlation between H{\sc i} mass--to--light
ratio and Hubble type with later--type galaxies containing more gas per
unit luminosity ($\rho= 0.77$). 
  
We conclude that the galaxies in our TFr calibrator sample are
representative of normal field galaxies as found in a volume--limited
sample.

\subsection{Radial surface density profiles}

 As can be seen in Figure \ref{fig_ttype}, our sample covers a broad
range of morphological types from $Sa$ to Irregular.  We divide our
sample into five bins of morphological types $Sa- Sab$, $Sb-Sbc$,
$Sc-Scd$, $Sd-Sdm$ and $Sdm-Im$ and investigate possible difference
between the H{\sc i} surface density profiles of galaxies of different
morphological types.

The azimuthally averaged radial H{\sc i} surface--density profiles,
normalised to their peak values and their H{\sc i} diameters, are
presented in Figure \ref{fig_profs}.  Clearly the shapes and amplitudes
of the profiles are very diverse, and cannot be described with a uniform
profile, in disagreement with the results by \citet{martinsson11} for
the Disk--Mass sample which is dominated by $Sc$--type galaxies. 
Although there is a large variety in profile shapes, we confirm the
overall trend that early--type spiral galaxies tend to have central
holes in their H{\sc i} disks while late--type and irregular galaxies
tend to have central concentrations of H{\sc i} gas.  The gas disks of
spiral galaxies of intermediate morphological types have a more constant
surface--density or may show mild central depressions in their
surface--densities

\section{Summary}

We have presented the analysis of 21-cm spectral--line aperture
synthesis observations of 32 spiral galaxies, which are representing a
calibrator sample for studying the statistical properties of the
Tully--Fisher relation.  The data were collected mostly from the
literature and obtained with various observational facilities (VLA,
ATCA, WSRT).  We observed for the first time three galaxies in our
sample ourselves with the GMRT.  The most important aspect of this work
is that, despite the broad range in data quality, we analysed the entire
sample in the same manner.  Although previously many of these galaxies
were studied individually, we present for the first time a set of
H{\sc i} synthesis imaging data products for all these galaxies
together, analysed in a homogeneous way.  The data products consist of
total H{\sc i} maps, velocity fields in which asymmetric velocity
profiles are identified and removed, H{\sc i} global profiles, radial
H{\sc i} surface--density profiles, position--velocity diagrams and,
most importantly, high--quality extended H{\sc i} rotation curves, all
presented in an accompanying Atlas.  The rotation curves and the derived
measurements of V$_{\rm max}$ and V$_{\rm flat}$ will be used in
forthcoming papers investigating the statistical properties of the
Tully--Fisher relation.  The radial H{\sc i} surface--density profiles
will be used for rotation curves decompositions, aimed at studying the
mass distributions within spiral galaxies. 

Overall, our kinematical study of the gas disks of our sample galaxies
shows excellent agreement with previously reported values for the
position and inclination angles, and systemic velocities.  We find a
good agreement with literature values for both H{\sc i} fluxes and the
widths of the global H{\sc i} profiles at the 50\% level, obtained from
single--dish observations.  However, in some cases we find somewhat
larger fluxes than previously reported in the literature (see Section
3.3.1 for details).  This can be easily explained given the large
angular size of our galaxies, and the fact that single--dish
observations can miss some flux due to the relatively small beam sizes. 

All galaxies in our sample have extended H{\sc i} disks, which serve our
purpose to probe the gravitational potential of their dark matter halos. 
Moreover, our galaxies follow the well--known correlation between their
H{\sc i} mass and the diameter of their H{\sc i} disks, but with a
slightly shallower slope then shown in previous studies.  Studies of the
H{\sc i} mass and its correlation with the absolute $K_{s}$ magnitude
show that while more luminous galaxies tend to have more H{\sc i} gas,
its fraction decreases with luminosity.  We find hints for similar
trends with Hubble--type: late--type spiral galaxies tend to have a
larger fraction of H{\sc i} gas than more early--type spirals, but the
total mass of the H{\sc i} gas decreases.  A qualitative comparison with
the volume--limited sample of Ursa Major galaxies shows that our
calibrator sample is respresenatative for a population of field
galaxies. 

The radial H{\sc i} surface--density profiles were scaled radially with
$R_{HI}$ and divided into five groups according to the morphological
type of the galaxies.  The gas disks of early--type spirals tend to have
central holes while the has disks of late--type spirals and irregulars
tend to be centrally concentrated. 

Velocity fields were constructed by fitting Gaussian and Gauss--Hermite
polynomial functions to the velocity profile at each position in the
data cube.  We identified the pixels with skewed velocity profiles,
using the difference between these two velocity fields, and censored
them in the Gaussian velocity field, thereby creating a third type of
velocity field. 
 
Rotation curves were derived from these censored velocity fields using
tilted--ring modelling.  The procedure was done in four steps, which
allowed to follow the geometry of a possible warp in the outer region. 
Rotation curves were identified into 3 categories: rising, flat and
declining.  The obtained values of V$_{\rm max}$ and V$_{\rm flat}$ were
compared with the velocities derived from the corrected H{\sc i} global
profiles at the 50\% and 20\% level.  The comparison tests show that
while the width of the H{\sc i} profile can be a good representation for
the maximal rotation velocity of the galaxy, it may significantly
overestimate the velocity at the outer, flat part of the rotation curve,
especially for high--mass galaxies which tend to have declining rotation
curves.  For the rising rotation curves, where V$_{\rm max}$ is lower
than V$_{\rm flat}$, the width of the profile tends to underestimate
V$_{\rm flat}$. 

The statistical properties of the Tully-Fisher relation based on
different velocity measures ($W$, V$_{\rm max}$ and V$_{\rm flat}$) will
be studied in forthcoming papers.

\section*{acknowledgements}

We would like to give a special thanks to Robert Braun, George Heald,
Gustav van Moorsel and Tobias Westmeier for kindly making their data
available.  AP is grateful to Erwin de Blok for help with the THINGS
data and for fruitful discussions.  We thank the staff of the GMRT that
made our observations possible.  GMRT is run by the National Centre for
Radio Astrophysics of the Tata Institute of Fundamental Research.  This
research has made use of the NASA/IPAC Extragalactic Database (NED)
which is operated by the Jet Propulsion Laboratory, California Institute
of Technology, under contract with the National Aeronautics and Space
Administration.  We acknowledge financial support to the DAGAL network
from the People Programme (Marie Curie Actions) of the European Union's
Seventh Framework Programme FP7/2007-2013/ under REA grant agreement
number PITNGA-2011-289313.  We acknowledge the Leids Kerkhoven--Bosscha
Fonds (LKBF) for travel support.

\appendix 

\section{Notes on individual galaxies}

\textbf{NGC 55 :} This galaxy is morphologically and kinematically
asymmetric, with the receding side extending farther into the halo than
the approaching side.  It was previously studied by \citet{ngc55} and
\citet{ngc55a}. 

\textbf{NGC 224 :} The central region of this galaxy is lacking H{\sc i}
gas and thus the rotation curve could not be constructed in the inner
regions which is indicated by the vertial, dashed line in the rotation
curve panel at a radius of 1500 {\it arcsec}.  The rotation curve is
declining.  These data were first presented in \citet{andromeda} and an
alternative kinematic analyses can be found in \citet{corbelli10}. 

\textbf{NGC 247 :} This galaxy was previously studied by
\citet{carignan90}. 

\textbf{NGC 253:} This galaxy has a lack of H{\sc i} gas in its centre. 
The inner points of the rotation curve suffer from beam smearing,
therefore the rotation curve is not properly recovered in the centre. 
The region affected by beam--smearing is indicated with a vertical,
dashed line at a radius of 150 {\it arcsec}.  It was previously studied
by \citet{ngc253}. 

\textbf{NGC 300:} This galaxy has a very extended H{\sc i} disk in
comparison with the stellar disk, and is warped in position angle. 
Traces of Galactic emission can be seen at the zero--velocity channels
in the position--velocity diagrams.  The rotation curve is declining. 
It was previously studied by \citet{ngc300} and \citet{ngc300a}. 

\textbf{NGC 925:} This galaxy has an extended H{\sc i} low
surface--density disk.  It was previously studied by \citet{ngc925} and
as part of the THINGS survey \citep{deblok08, things}. 

\textbf{NGC 1365:} A grand--design spiral galaxy with a strong bar.  The
central regions are completely dominated by non--circular motions due to
the bar, and suffer from a lack of H{\sc i} gas for a proper recovery of
the inner rotation curve.  A warp can be identified in both position
angle and inclination.  It was previously studied by \citet{1365} and
re-analysed lately by \citet{1365a}. 

\textbf{NGC 2366:} This flocculent, barred galaxy is kinematically
lopsided.  It was previously studied as a part of the THINGS survey
\citep{deblok08, things, oh08}. 

\textbf{NGC 2403:} This galaxy was previously studied by
\citet{begeman87} and is part of the THINGS survey \citep{deblok08,
things}. 

\textbf{NGC 2541:} The H{\sc i} disk of this galaxy displays a warp in
position angle. 

\textbf{NGC 2841:} This galaxy has an extended, warped gas disk with a
declining rotation curve.  It was previously studied by \citet{bosma81,
begeman87} and is a part of the THINGS survey \citep{deblok08, things}. 

\textbf{NGC 2976:} This galaxy has an extended very low surface--density
H{\sc i} disk.  It has a solid--body rotation curve which does not reach
the flat part.  NGC 2976 was previously studied as part of the THINGS
survey \citep{deblok08, things}.  Our kinematic study provides results
consistent with the results derived from $H\alpha$ and $CO$ observations
\citep{simon03}. 
 
\textbf{NGC 3031:} This is a grand--design spiral galaxy.  There is a
lack of H{\sc i} gas in the center.  The H{\sc i} disk displays a warp
in both position angle and inclination.  The rotation curve of this
galaxy is dominated by non--circular motions in the inner regions
\citep{visser80} .  NGC 3031 was previously studied as part of the
THINGS survey \citep{deblok08, things}. 
 
\textbf{NGC 3109:} This is a highly--inclined spiral galaxy.  It has a
rising rotation curve.  We used available VLA data although it was also
studied with deeper observations using KAT-7 \citep{ngc3109}. 
Eventhough the KAT-7 data derived a more extended rotation curve, it is
still rising at the outermost measured point. 
 
\textbf{NGC 3198:} This is a barred galaxy with a gas disk warped in
position angle.  It has a declining rotation curve.  It was previously
studied by \citet{bosma81, begeman89} and is part of the THINGS survey
\citep{deblok08, things}. 
  
\textbf{IC 2574:} This galaxy has a rising rotation curve at its
outermost measured point.  It was previously studied as part of the
THINGS survey \citep{deblok08, things, oh08}. 
  
\textbf{NGC 3319:} This is a barred galaxy.  In the atlas, a vertical,
dashed line at a radius of 80 {\it arcsec} indicates the region
dominated by non--circular motions due to the bar.  It was previously
studied by \citet{ngc3319} and \citet{ngc3319a}
   
\textbf{NGC 3351:} This is a barred galaxy with a H{\sc i} disk that
displays a warp in position angle.  There is a lack of H{\sc i} gas in
the centre.  The rotation curve of this galaxy is declining.  It was
previously studied as part of the THINGS survey \citep{deblok08,
things}. 
   
\textbf{NGC 3370:} The observation of this galaxy suffers from
beam--smearing.  It has a lack of H{\sc i} gas in the centre. 
    
\textbf{NGC 3621:} This galaxy has an extended H{\sc i} disk, displaying
a warp in position angle.  It was previously studied as part of the
THINGS survey \citep{deblok08, things}. 
   
\textbf{NGC 3627:} This galaxy is kinematically lopsided.  It has a
rather small H{\sc i} disk.  There is a lack of H{\sc i} gas in the
centre.  It was previously studied as part of the THINGS survey
\citep{deblok08, things}. 
   
\textbf{NGC 4244:} This is an edge--on oriented galaxy.  Its rotation
curve was derived using the envelope--tracing method
\citep{sancisiallen}.  It was previously studied as part of the HALOGAS
survey \citep{ngc4244,halogas}. 
   
\textbf{NGC 4258:} This is a barred galaxy.  The gas kinematics in the
central region is dominated by non--circular motions due to the bar.  It
has a declining rotation curve.  NGC 4258 was previously studied by
\citet{ngc4258}.  It was observed as part of the HALOGAS survey
\citep{halogas}. 
   
\textbf{NGC 4414:} This galaxy has an extremely extended low
surface--density H{\sc i} disk.  The rotation curve is declining.  It
was previously studied as part of the HALOGAS survey \citep{ngc4414,
halogas}. 
    
\textbf{NGC 4535:} This barred galaxy has a lack of H{\sc i} gas in its
centre.  It was previously studied as part of the VIVA survey
\citep{viva}. 
    
\textbf{NGC 4536:} This barred galaxy has a lack of H{\sc i} gas in its
centre.  It was previously studied as part of the VIVA survey
\citep{viva}. 
    
\textbf{NGC 4605:} This galaxy is kinematically lopsided.  It has an
extended, H{\sc i} low surface--density gas disk.  Its rotation curve
does not reach the flat part.  Our H{\sc i} kinematics study is
consistent with results derived from $H\alpha$ and $CO$ observations
\citep{simon05}.  It was previously studied by \citet{broeils92}. 
   
\textbf{NGC 4639:} The observation of this galaxy suffered from
beam--smearing.  It has a lack of H{\sc i} gas in the centre.  The gas
disk is warped both in position angle and in inclination. 
   
\textbf{NGC 4725:} This barred galaxy has a lack of the H{\sc i} gas in
its centre.  The gas disk displays a position angle warp.  It has a
tidally interacting companion.  NGC 4725 was previously studied by
\citet{ngc4725}. 
   
\textbf{NGC 5584:} This flocculent spiral galaxy has a lack of H{\sc i}
gas in its centre.  The observation of this galaxy suffered from
beam--smearing effects. 
   
\textbf{NGC 7331:} This barred galaxy has lack of the H{\sc i} gas in
its centre.  It was previously studied as part of the THINGS survey
\citep{deblok08, things}. 
   
\textbf{NGC 7793:} This flocculent spiral has a gas disk warped in
position angle.  It has a declining rotation curve.  NGC 7793 was
previously studied by \citet{ngc7793} and is part of the THINGS survey
\citep{deblok08, things}.

\section{The ATLAS}

An atlas of the final data products is presented below.  Each page
presents the same data for an individual galaxy, where the following
information is given:

\textbf{Optical image} -- to introduce the morphology of each galaxy,
its Hubble type and optical diameter, the blue POSS image of each galaxy
is shown on the same scale as the interferometric images.  Due to the
large angular extent of the H{\sc i} disk of M31, we present its NUV
image from GALEX.  The white ellipse indicates the optical size and
orientation as reported by NED. 

\textbf{H{\sc i} integrated map} -- gives information about the H{\sc i}
column--density distribution, presented at the same scale as the optical
image.  H{\sc i} column--density maps are presented on the same grey
scale stretch.  The lowest H{\sc i} column-density contour for all
galaxies is drawn at $0.5\,M_{\odot}pc^{-2}$ and the highest contour
represents the maximum column--density as listed in Table
\ref{tbl_param}. The synthesised beam is shown in the bottom--left
corner of the H{\sc i} map.  Due to the very high spatial resolution of
some data, the synthesised beam is often too small compared to the size
of the galaxy and can be hardly seen in several cases. 

\textbf{Velocity field} -- the {\it SKW} velocity field is shown (see
Section 3.5).  The receding and approaching sides of each galaxy are
indicated by grey shades.  The thick black line in the middle indicates
the systemic velocity obtained from the tilted--ring fits. 

\textbf{Residual map} -- the residual map between the observed velocity
field and the modelled velocity field based on the derived rotation
curves.  The difference range runs from -30 kms$^{-1}$ to +30 kms$^{-1}$
and is represented by grey shades.  White contours show negative
residuals and black contours positive.  Residual contours are drawn in
steps of 5 $kms^{-1}$. 

\textbf{Global H{\sc i} profile} -- shows the total, spatially
integrated H{\sc i} flux density, where every point corresponds to the
primary beam corrected H{\sc i} flux density for a single channel in the
data cube.  The vertical arrow indicates the systemic velocity of the
galaxy as measured from the global profile (see Section 3.3.2.)

\textbf{Surface--density profile} -- presents the surface--density of
H{\sc i} gas in a galaxy as derived from the total H{\sc i} map, scaled
by the H{\sc i} mass.  The red curve corresponds to the receding side of
the galaxy and the blue curve to the approaching side. 

\textbf{Tilted--ring fitting results} -- three combined panels show the
results from the various steps in the tilted--ring fitting procedure. 
From top to bottom: fitted position angles in step 2, fitted inclination
angles in step 3, and the fitted rotational velocity in step 4.  The red
curve in the velocity panel shows the rotation curve of the receding
side and the blue curve that of the approaching side.  The black curve
corresponds to the rotation curve obtained by fitting titlted rings to
both sides of the galaxy simultaneously.  The size of the optical disk
is indicated with a dashed vertical line.  For some galaxies the first
dashed vertical line shows the region where the gas flow is dominated by
non--circular motions due to the bar.  The black line in the two upper
panels indicates the adopted position and inclination angles.  The
dashed horizontal lines show the optical measure of the position and
inclination angles. 

\textbf{Position--Velocity diagrams } -- the position--velocity diagrams
shown in the two bottom left panels.  They were extracted from the data
cubes along two orthogonal cuts through the dynamical centre of the
galaxy following the inner kinematical minor (upper) and major (bottom)
axes.  Contours correspond to levels of 2, 3, 4.5, 6, 9, 12, 15, ... 
$\sigma$.  The rotation curves of the receding and approaching sides of
a galaxy are projected onto these position--velocity diagrams, taking
the geometry of a possible warp into account.  In the ideal case if pure
circular orbits, the slice along the kinematic minor axis should be
perpendicular to the slice along the kinematic major axis.  However, the
non-circular motions in the centers of strong barred galaxies manifest
themselves by slight deviation from this geometry.  The grey vertical
lines on the major axis cut indicate the size of the optical disk.

\begin{figure*}
\caption{NGC 55 (VLA)}
\includegraphics[scale=0.95]{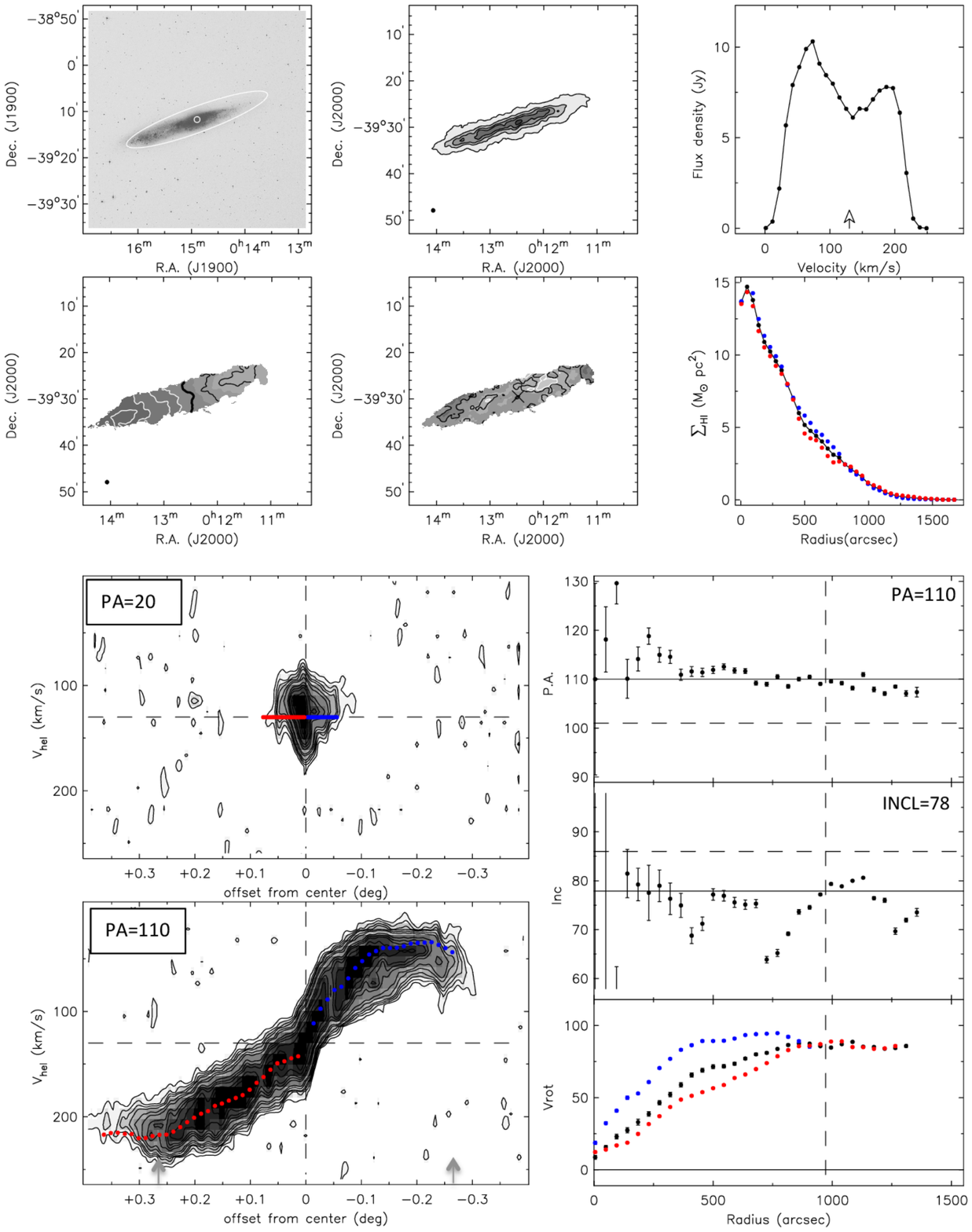}
\end{figure*}
\clearpage

\begin{figure*}
\caption{NGC 224 (WSRT)}
\includegraphics[scale=0.95]{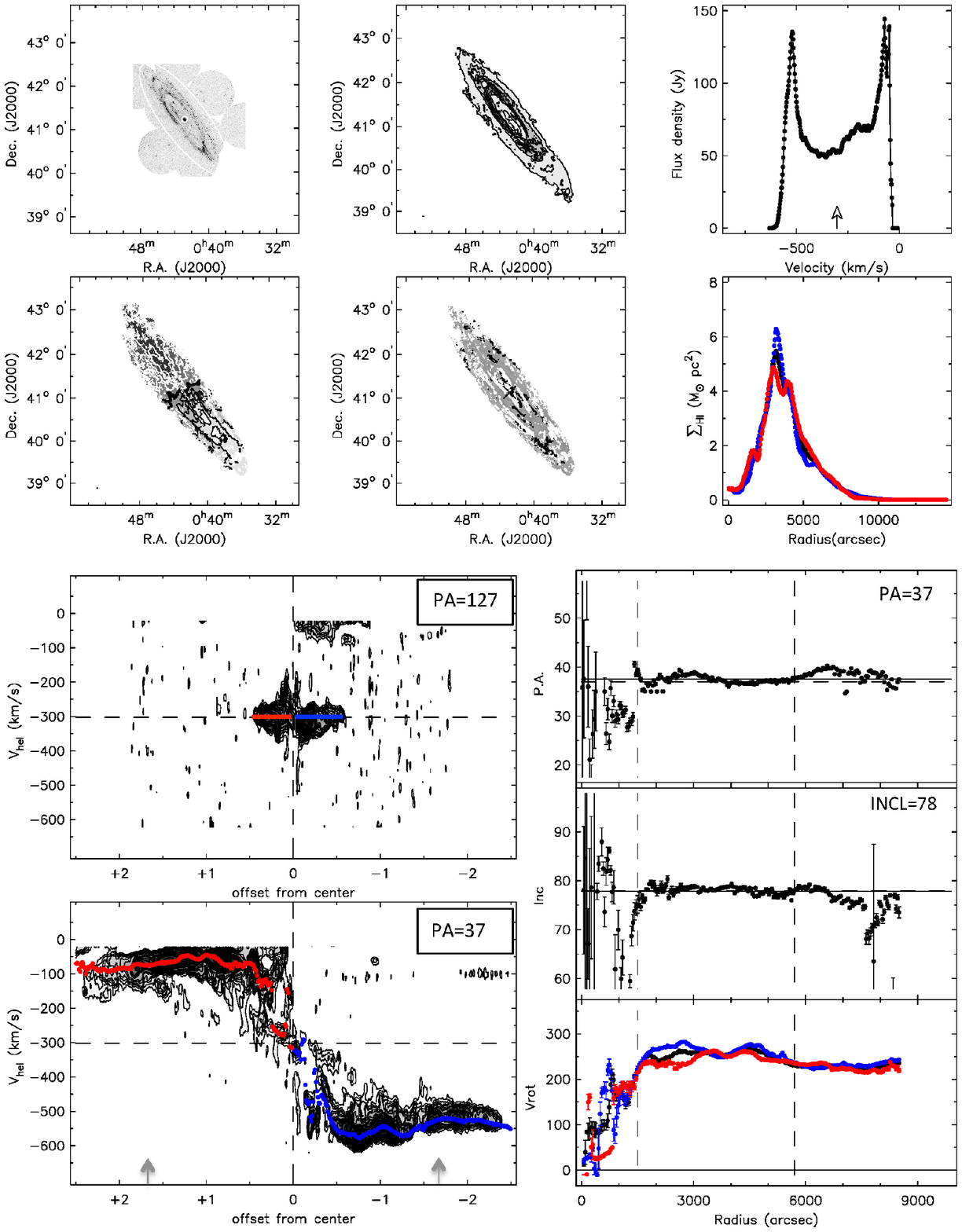}
\end{figure*}
\clearpage

\begin{figure*}
\caption{NGC 247 (VLA)}
\includegraphics[scale=0.95]{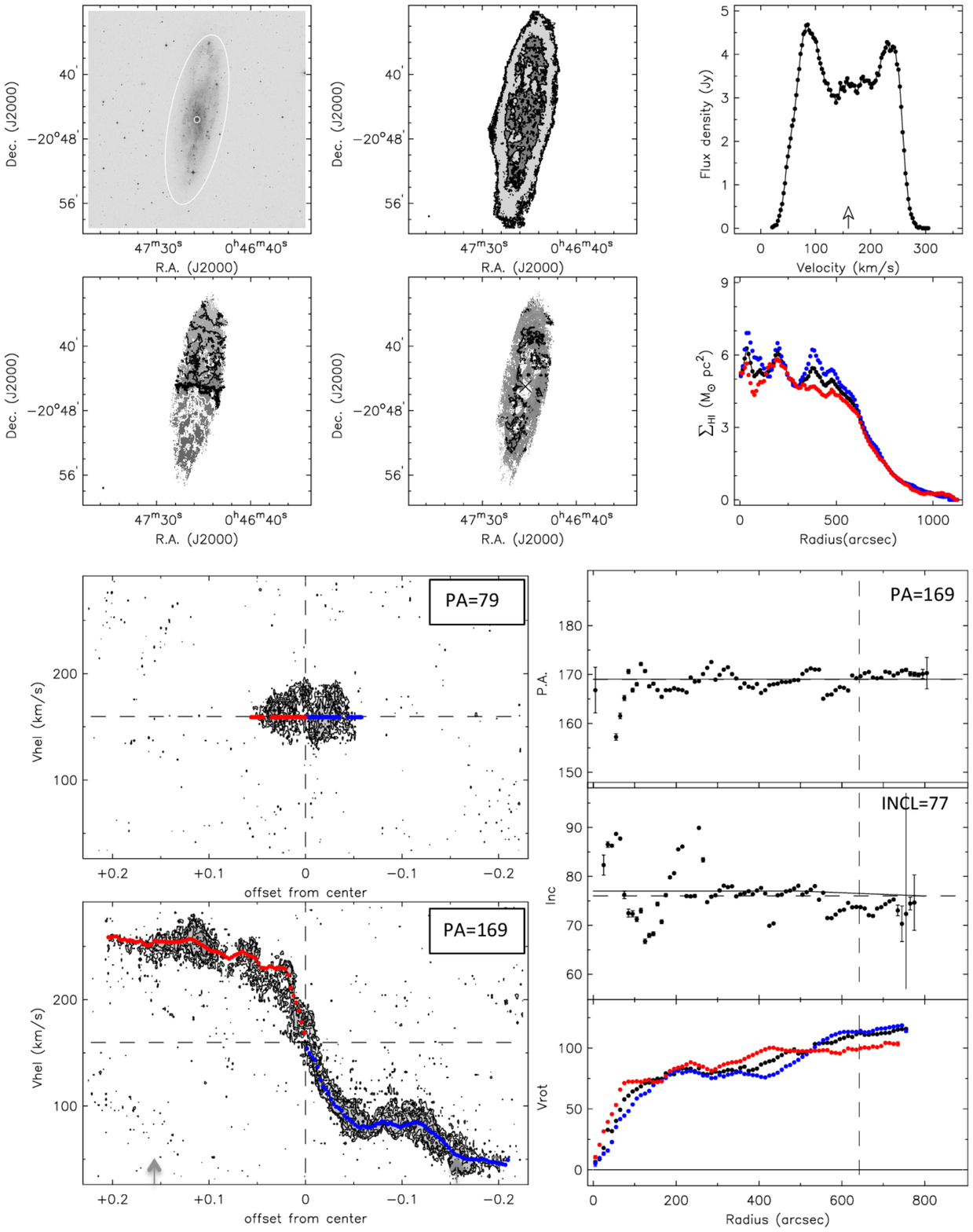}
\end{figure*}
\clearpage

\begin{figure*}
\caption{NGC 253 (ATCA)}
\includegraphics[scale=0.95]{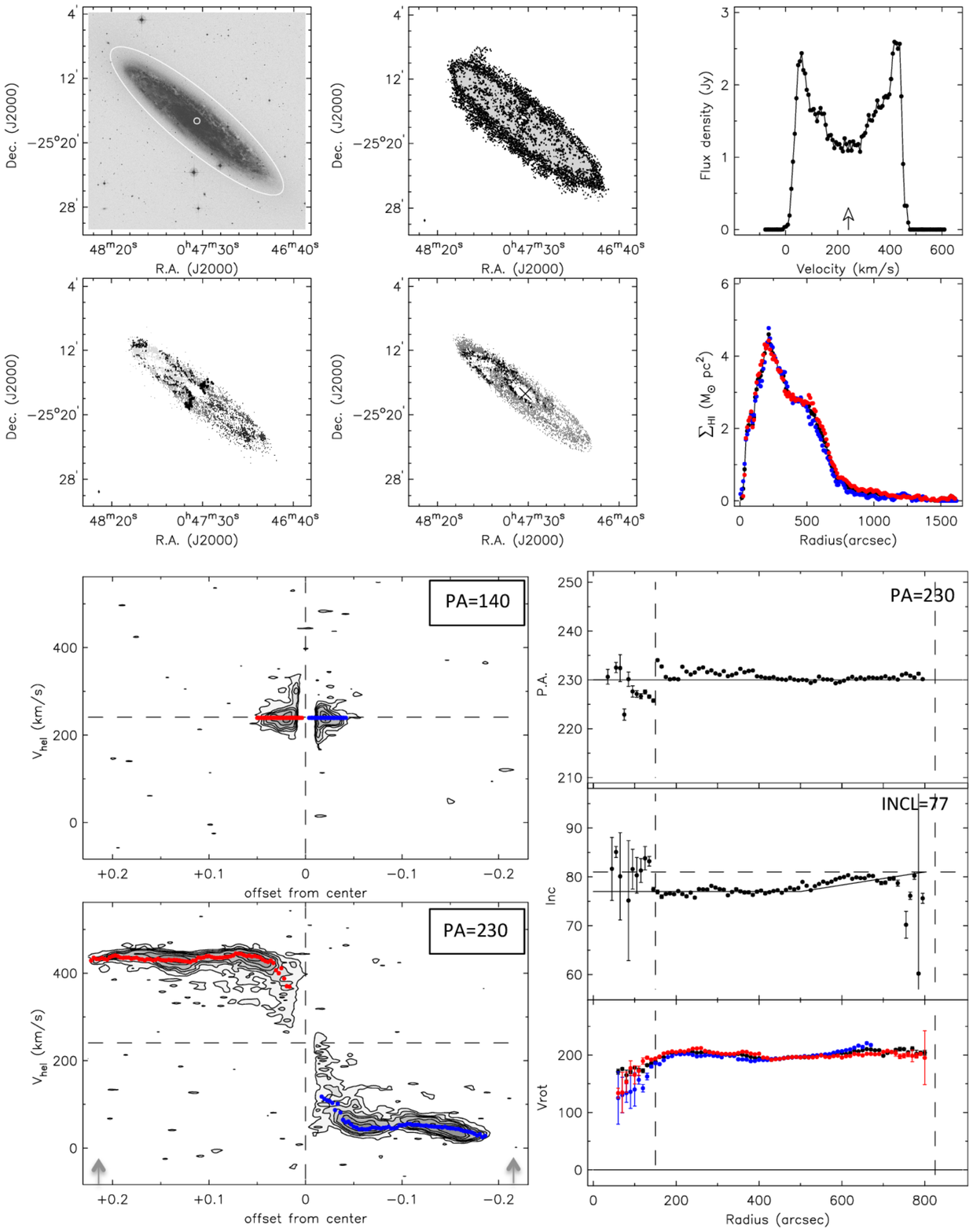}
\end{figure*}
\clearpage

\begin{figure*}
\caption{NGC 300 (ATCA)}
\includegraphics[scale=0.95]{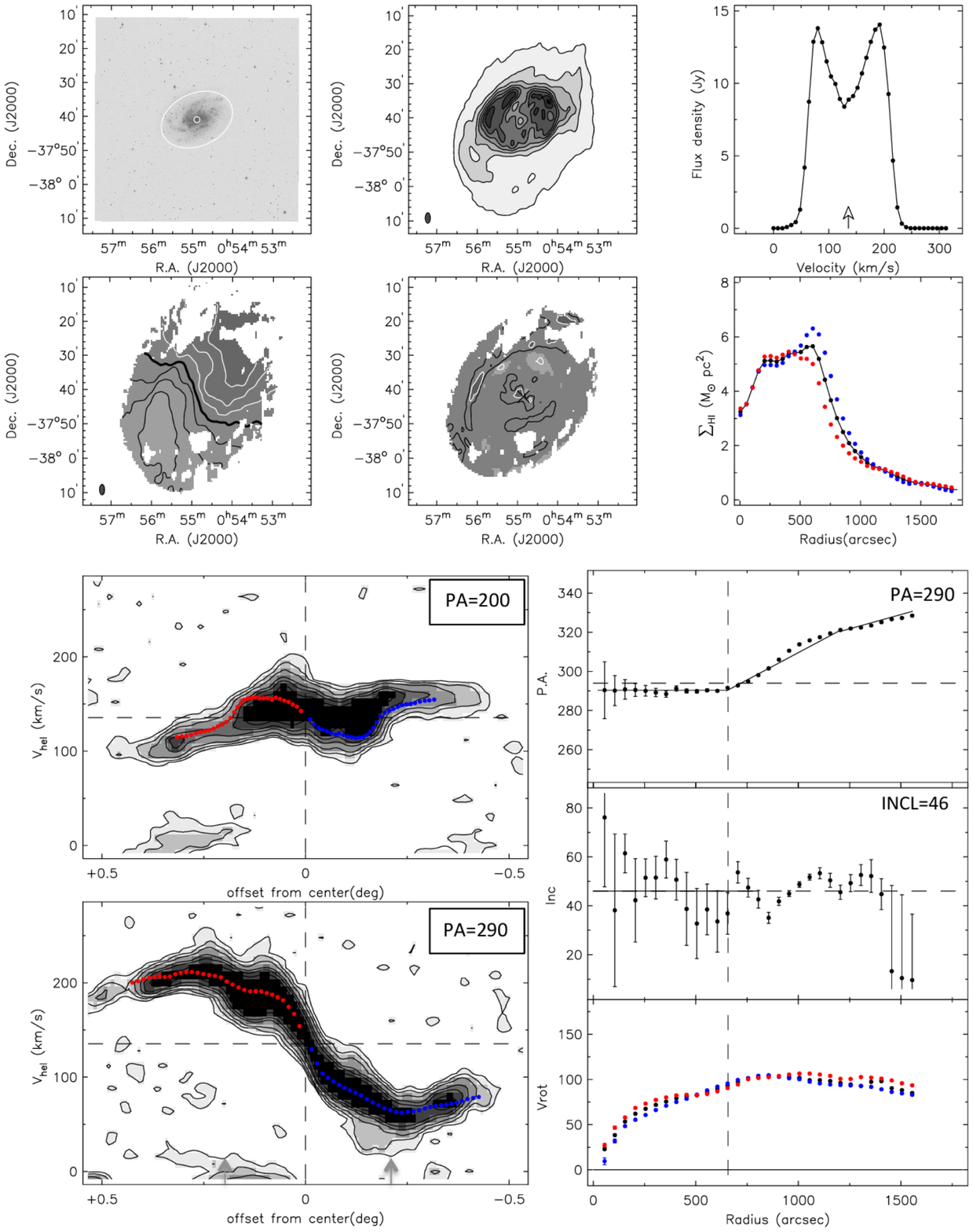}
\end{figure*}
\clearpage

\begin{figure*}
\caption{NGC 925 (VLA)}
\includegraphics[scale=0.95]{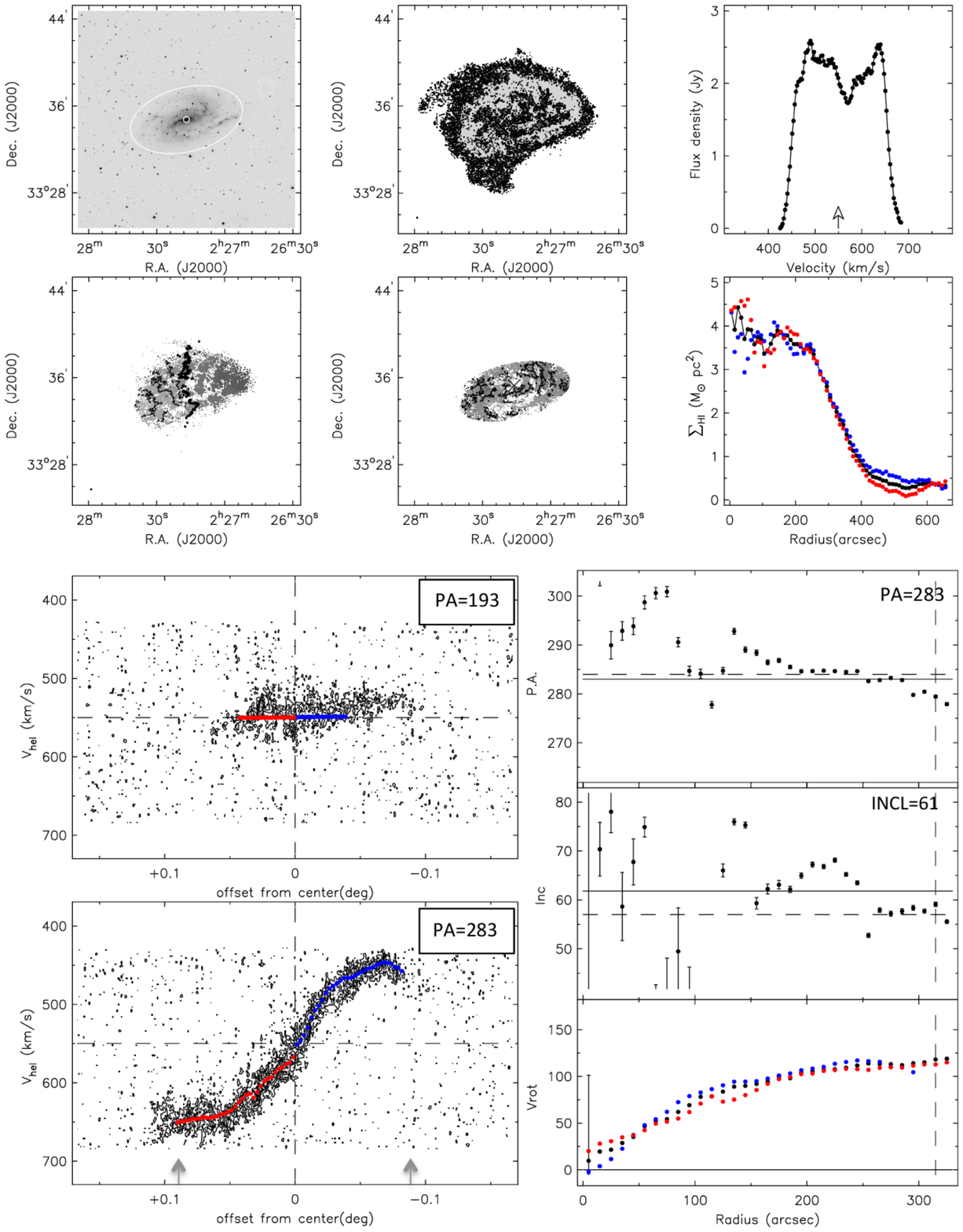}
\end{figure*}
\clearpage

\begin{figure*}
\caption{NGC 1365 (VLA)}
\includegraphics[scale=0.95]{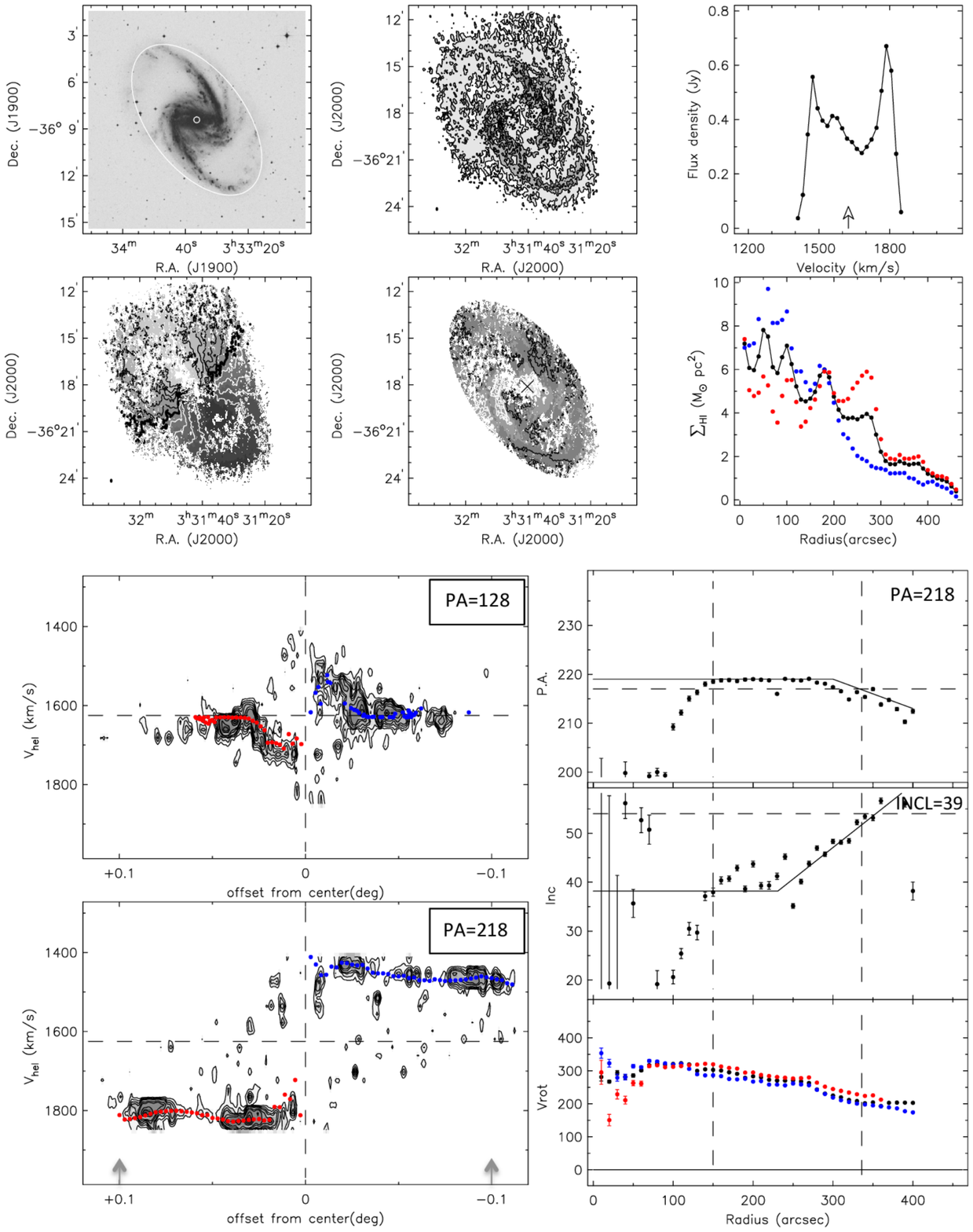}
\end{figure*}
\clearpage

\begin{figure*}
\caption{NGC 2366 (VLA)}
\includegraphics[scale=0.95]{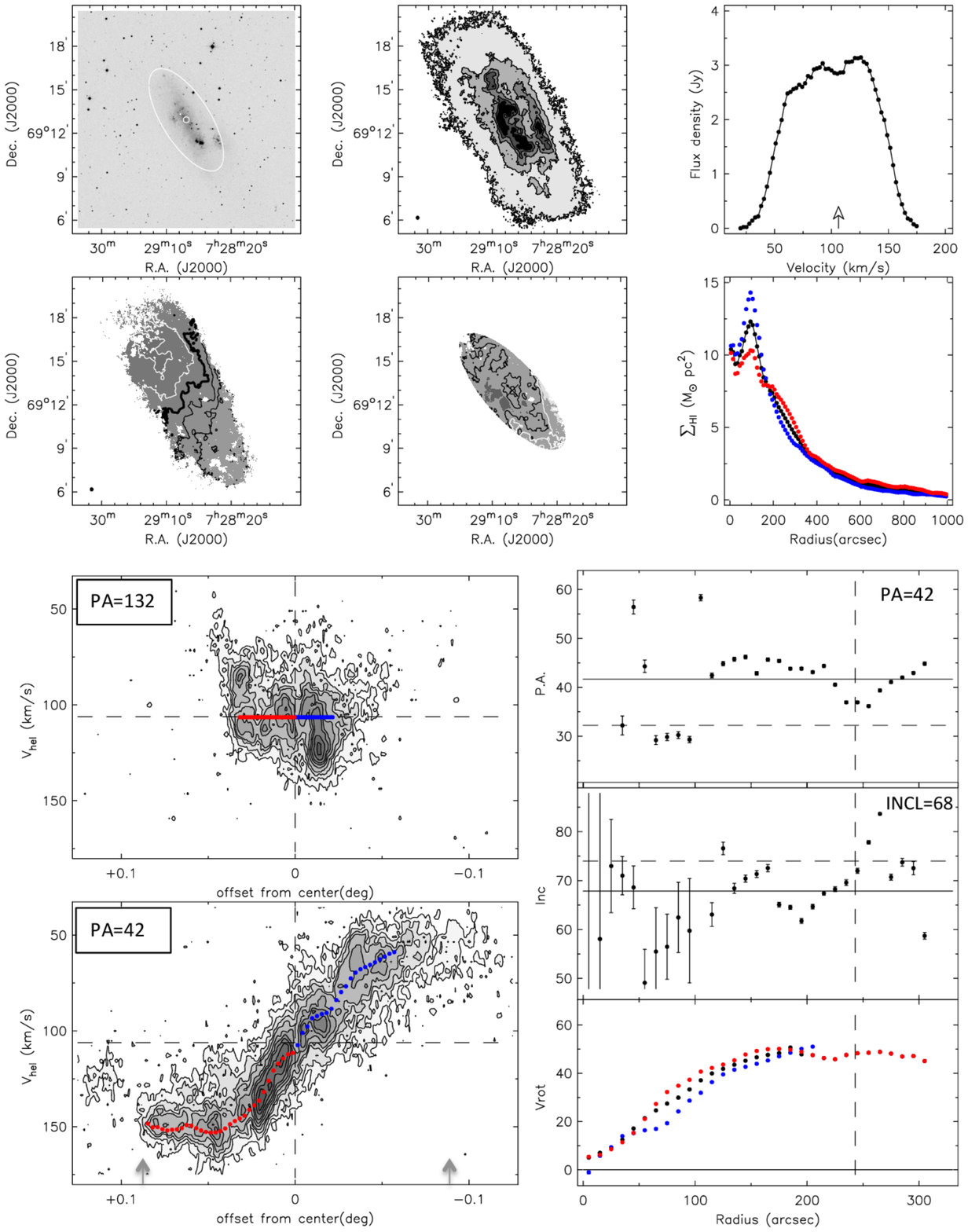}
\end{figure*}
\clearpage

\begin{figure*}
\caption{NGC 2403 (VLA)}
\includegraphics[scale=0.95]{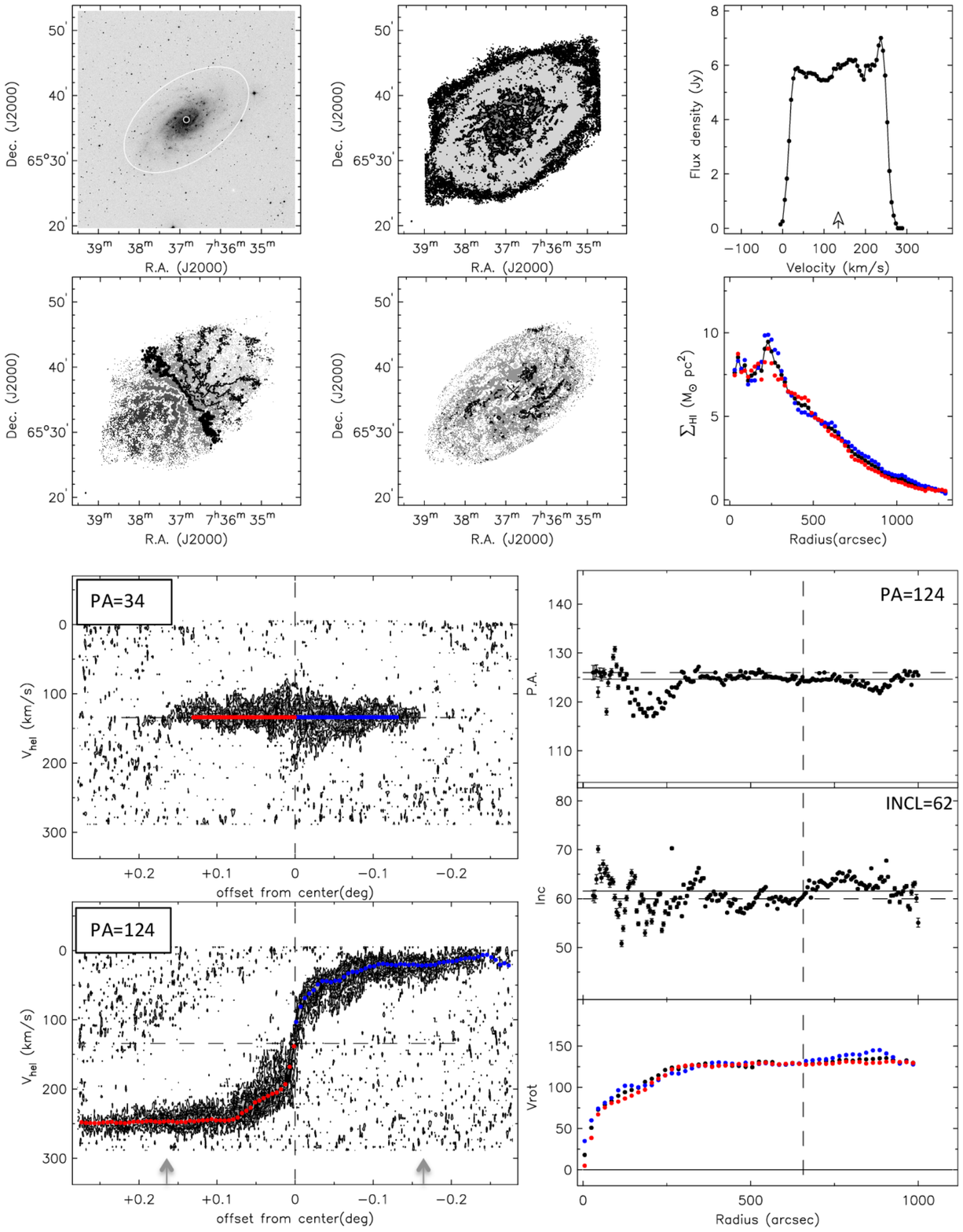}
\end{figure*}
\clearpage

\begin{figure*}
\caption{NGC 2541 (WSRT)}
\includegraphics[scale=0.95]{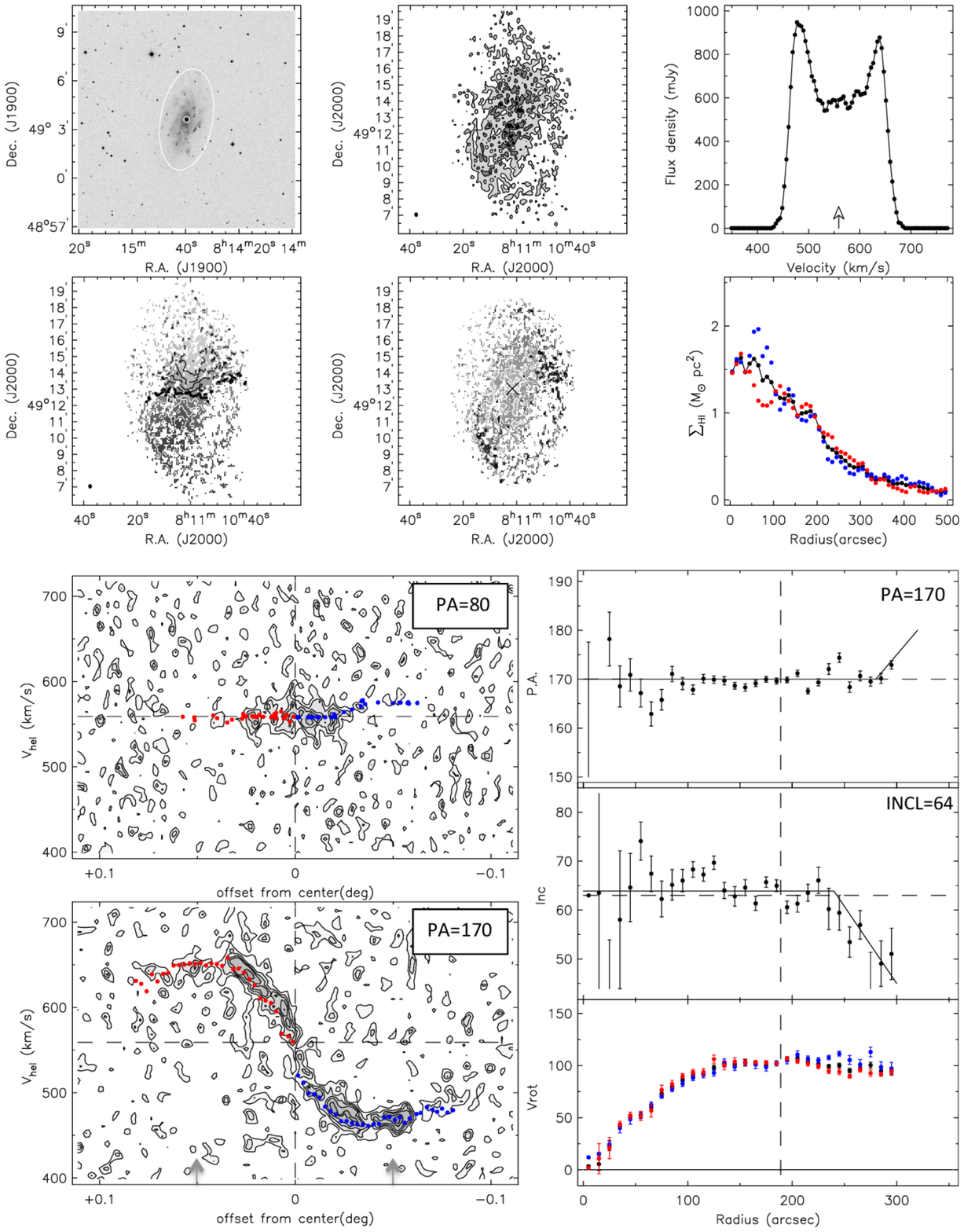}
\end{figure*}
\clearpage

\begin{figure*}
\caption{NGC 2841 (VLA)}
\includegraphics[scale=0.95]{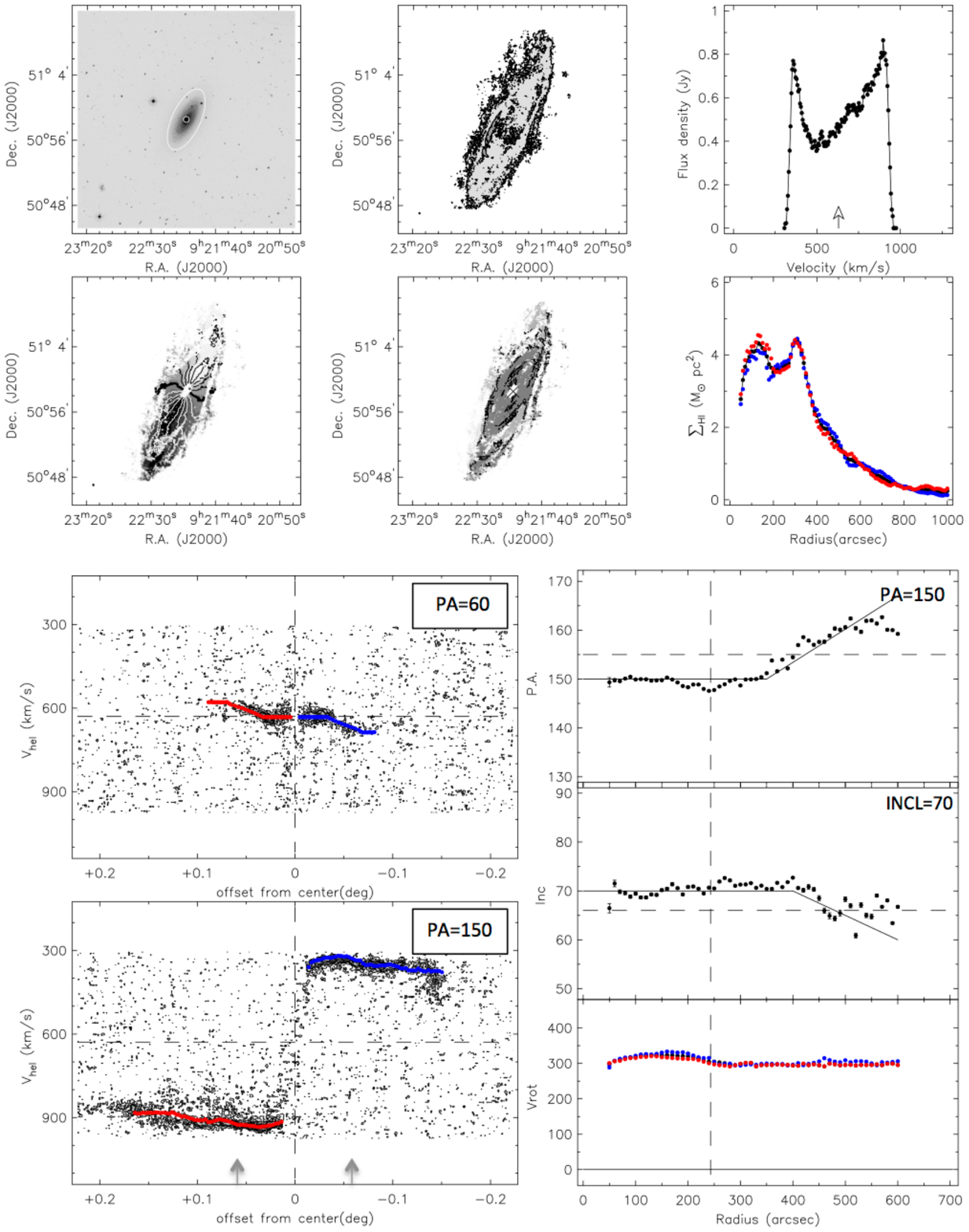}
\end{figure*}
\clearpage

\begin{figure*}
\caption{NGC 2976 (VLA)}
\includegraphics[scale=0.95]{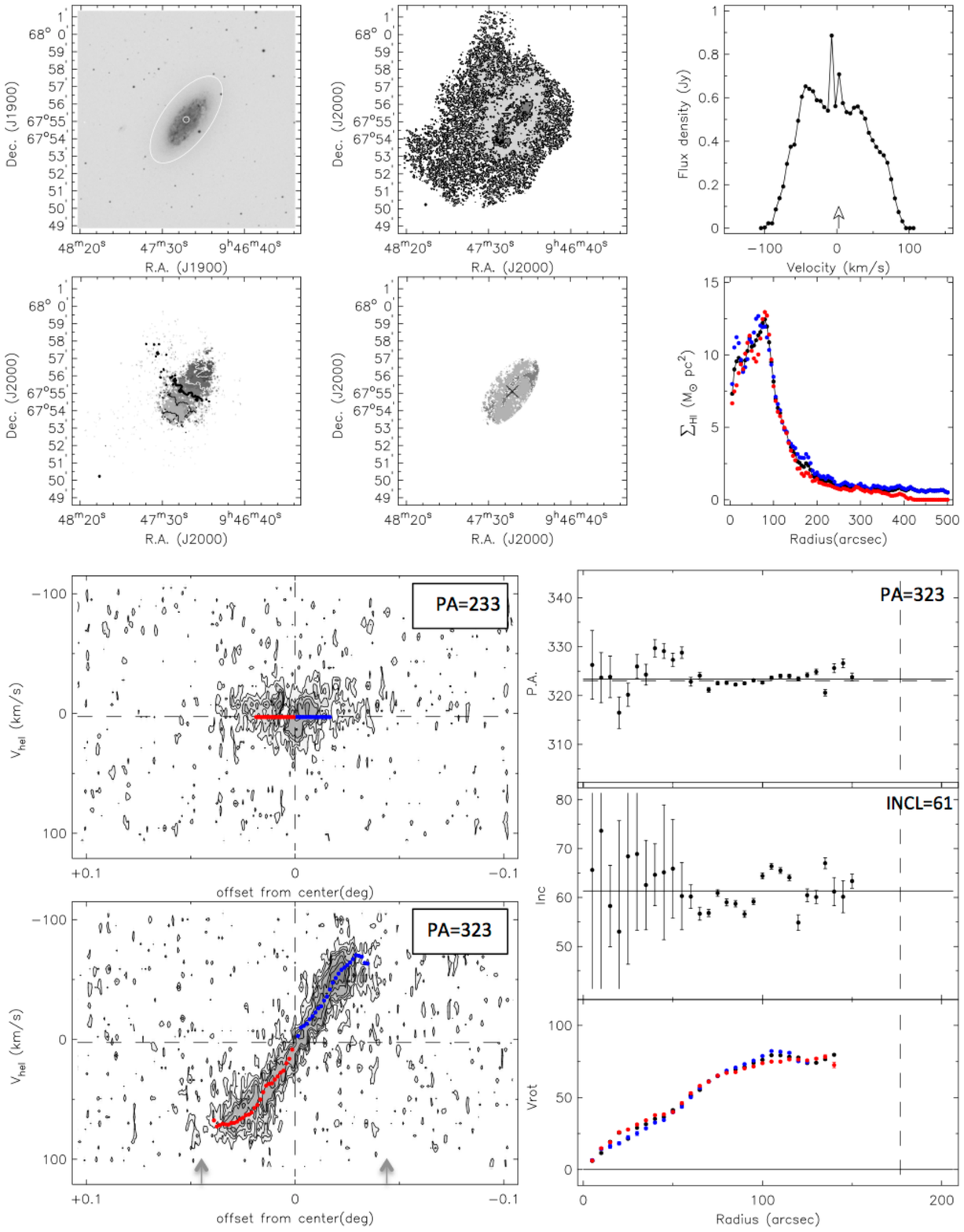}
\end{figure*}
\clearpage

\begin{figure*}
\caption{NGC 3031 (VLA)}
\includegraphics[scale=0.95]{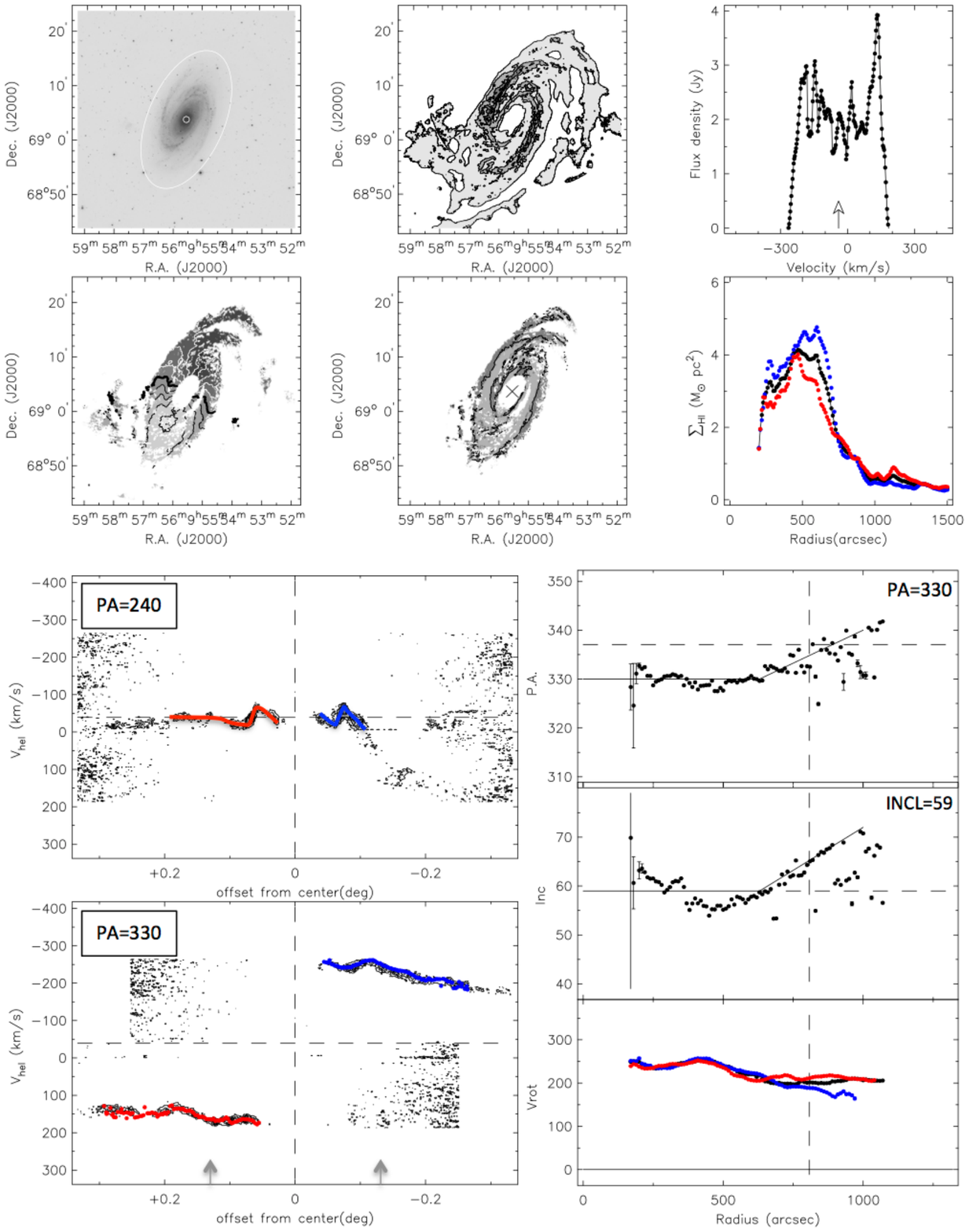}
\end{figure*}
\clearpage

\begin{figure*}
\caption{NGC 3109 (VLA)}
\includegraphics[scale=0.95]{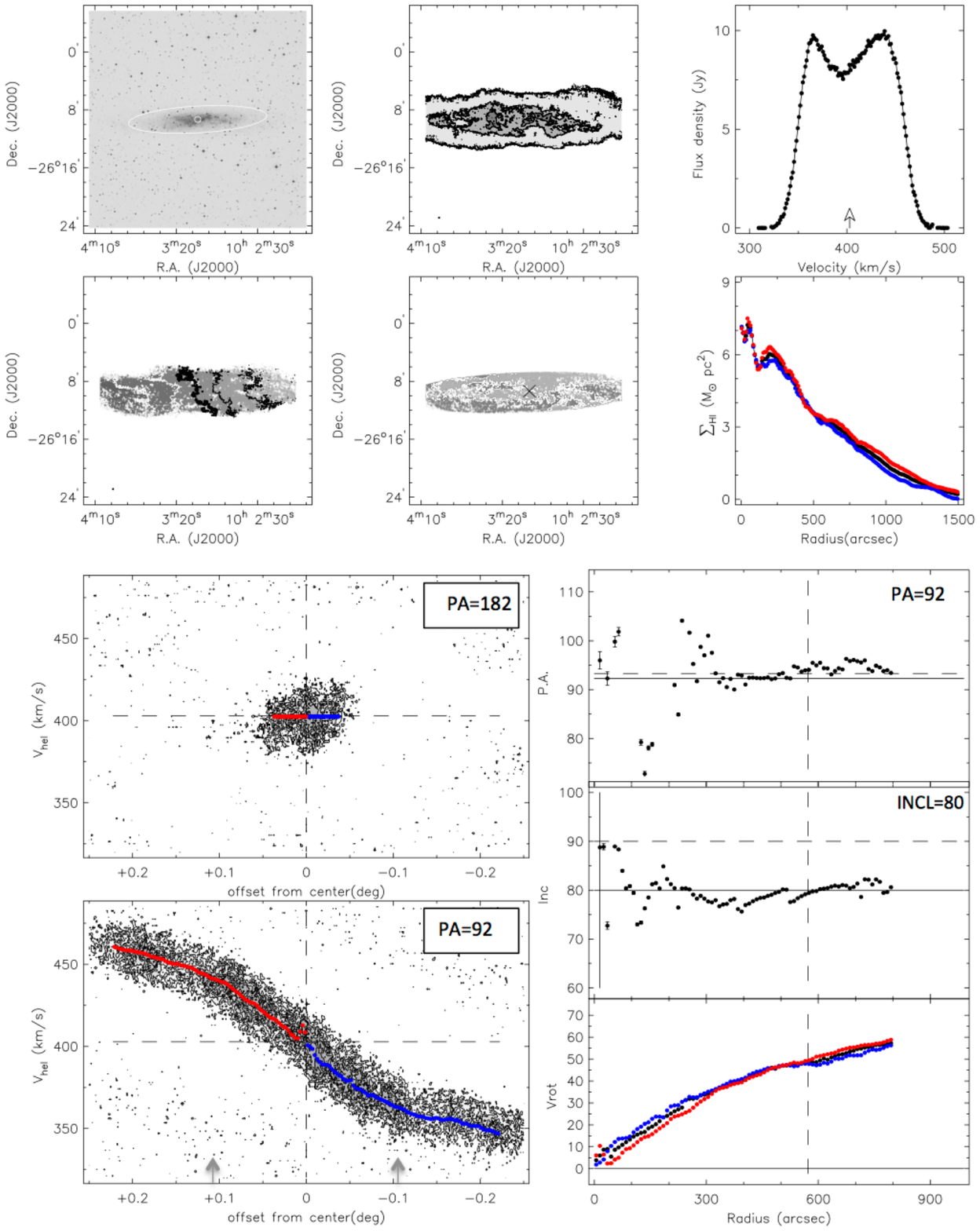}
\end{figure*}
\clearpage

\begin{figure*}
\caption{NGC 3198 (VLA)}
\includegraphics[scale=0.95]{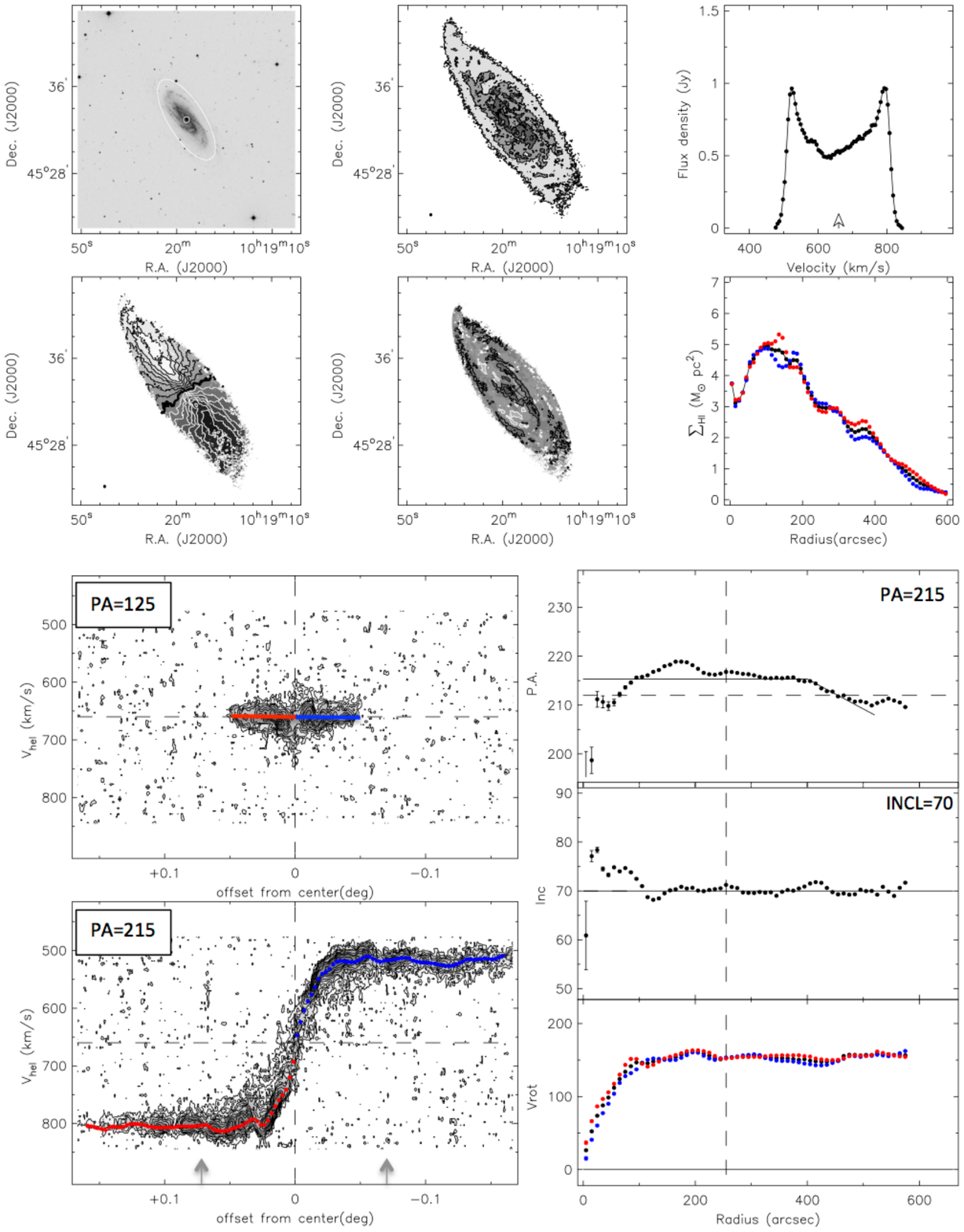}
\end{figure*}
\clearpage

\begin{figure*}
\caption{IC 2574 (VLA)}
\includegraphics[scale=0.95]{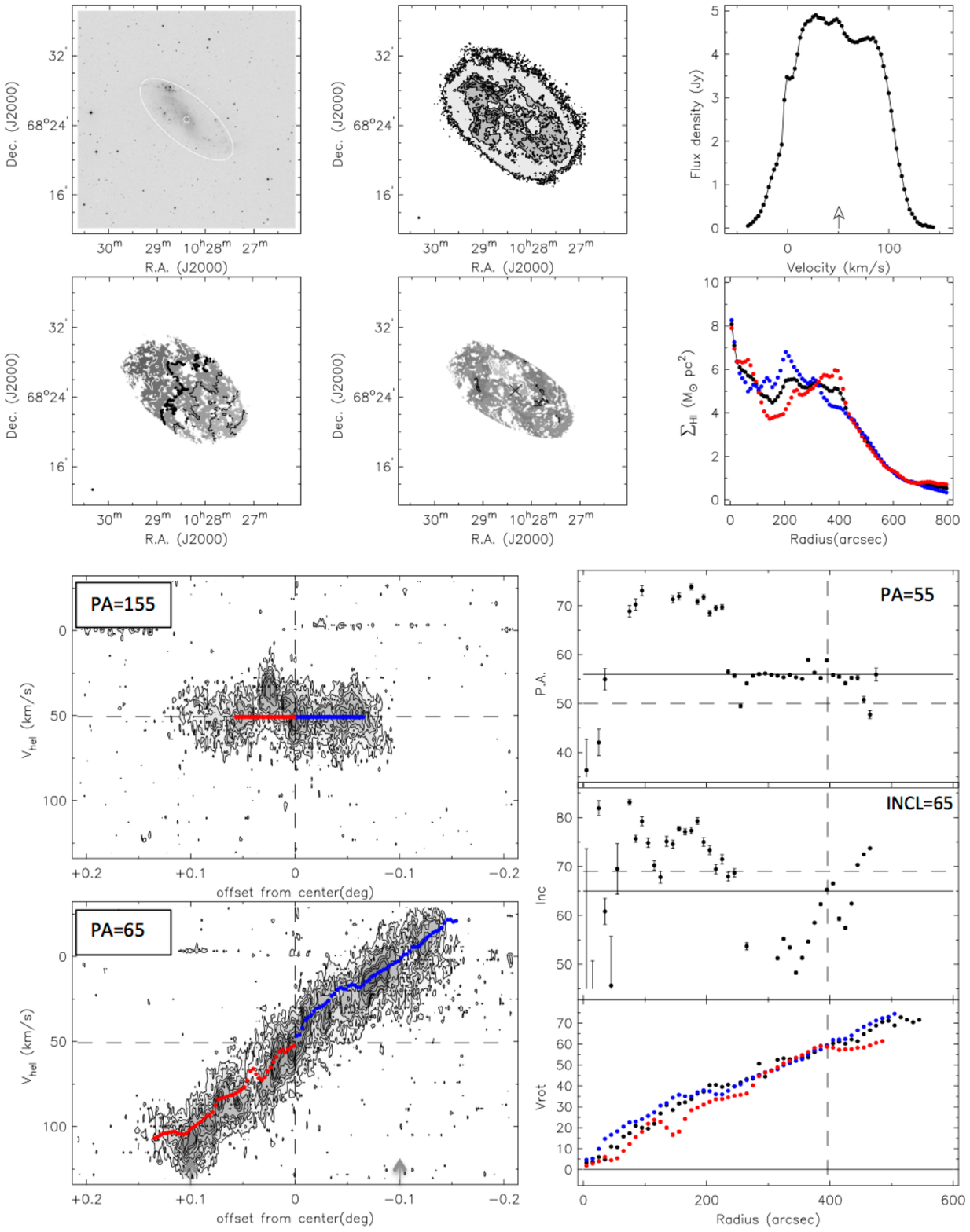}
\end{figure*}
\clearpage

\begin{figure*}
\caption{NGC 3319 (WSRT)}
\includegraphics[scale=0.95]{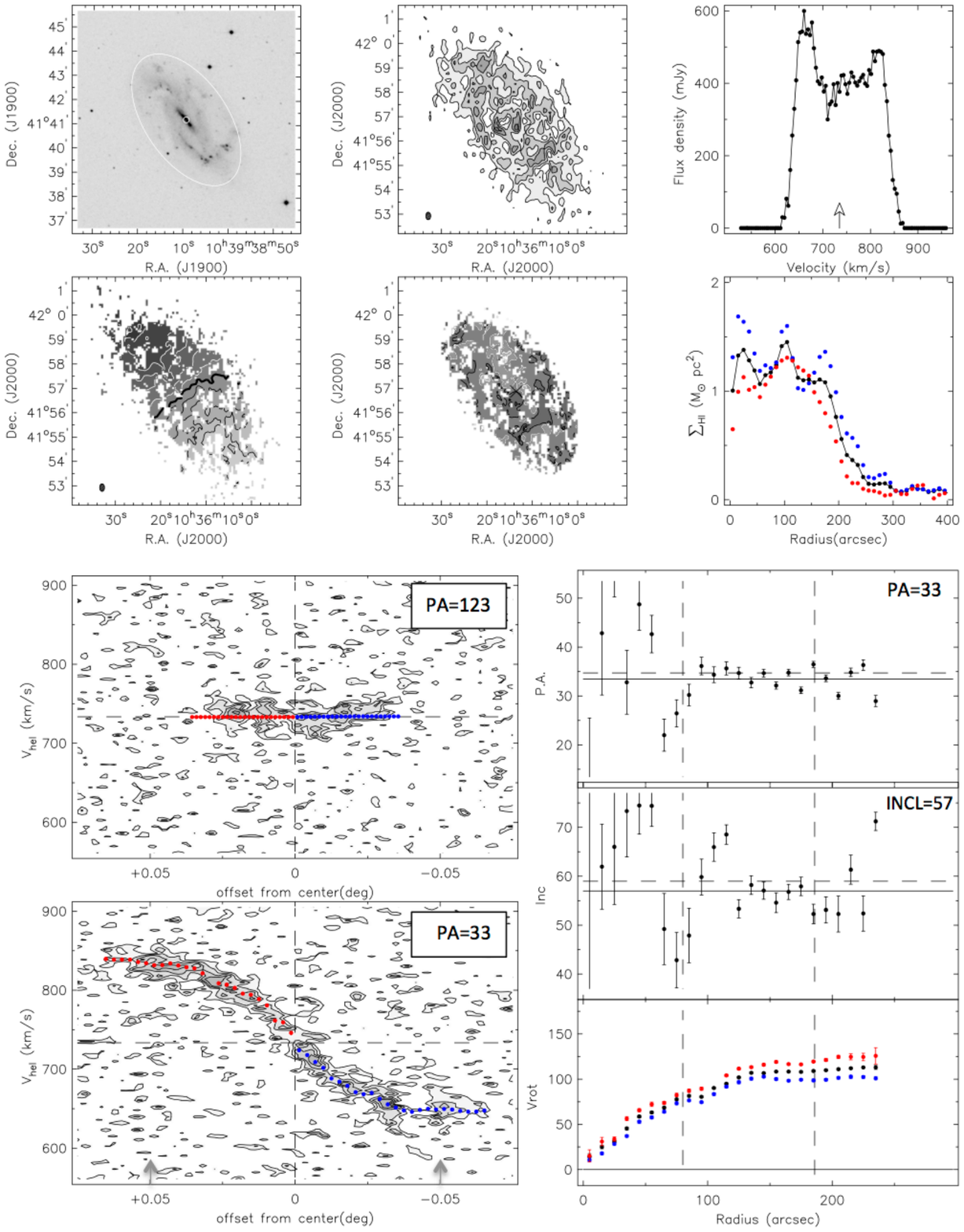}
\end{figure*}
\clearpage

\begin{figure*}
\caption{NGC 3351 (VLA)}
\includegraphics[scale=0.95]{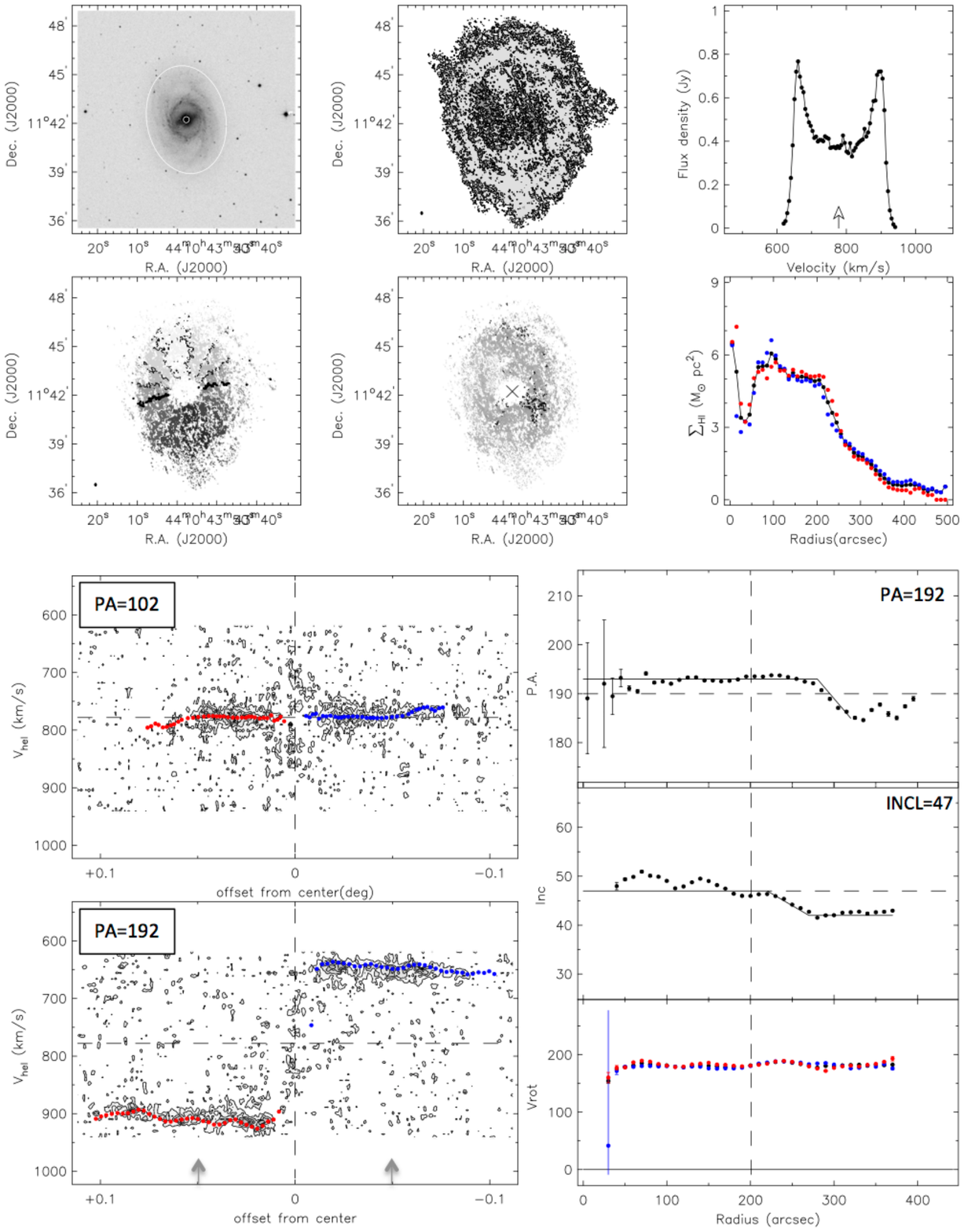}
\end{figure*}
\clearpage

\begin{figure*}
\caption{NGC 3370 (GMRT)}
\includegraphics[scale=0.95]{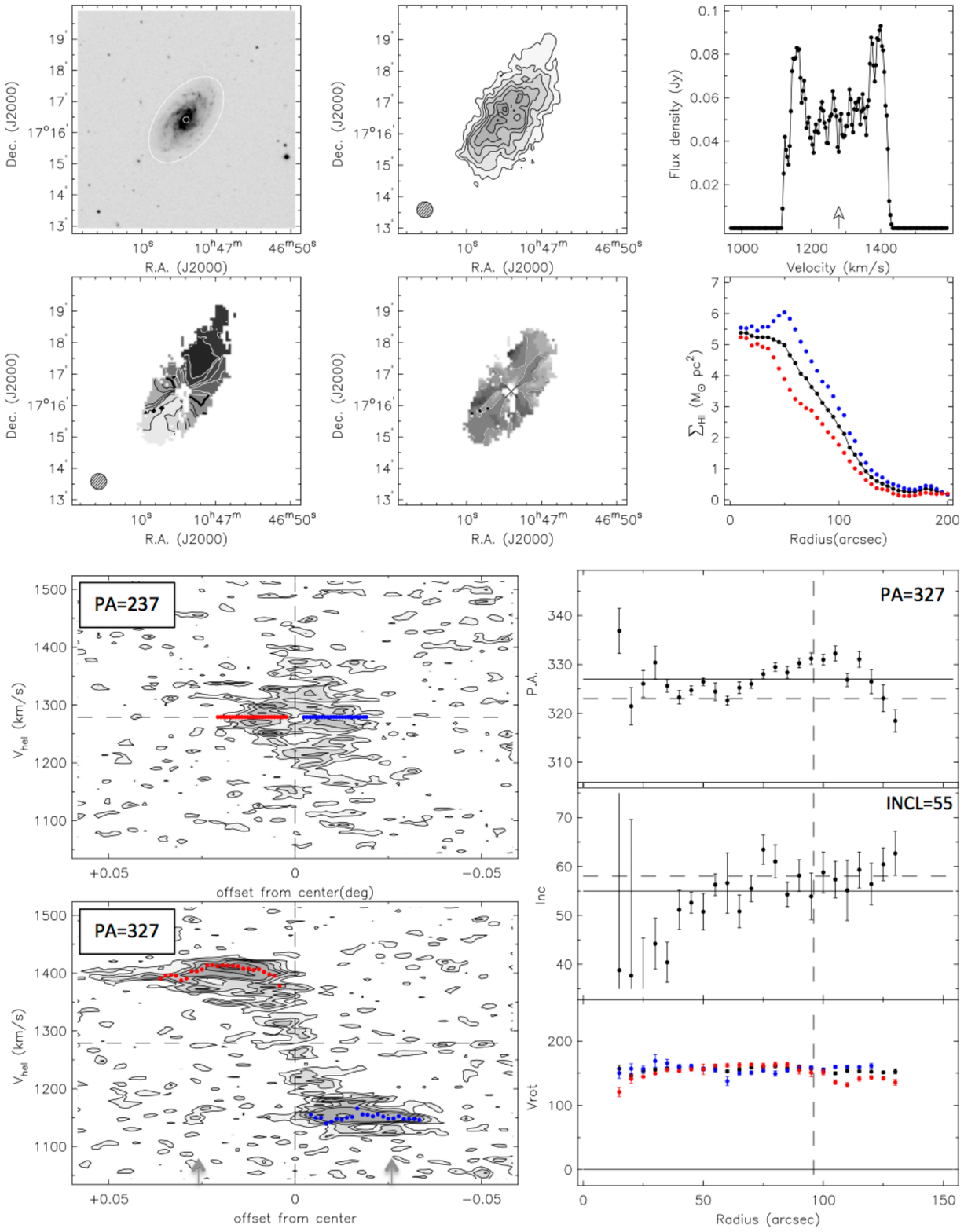}
\end{figure*}
\clearpage

\begin{figure*}
\caption{NGC 3621 (VLA)}
\includegraphics[scale=0.95]{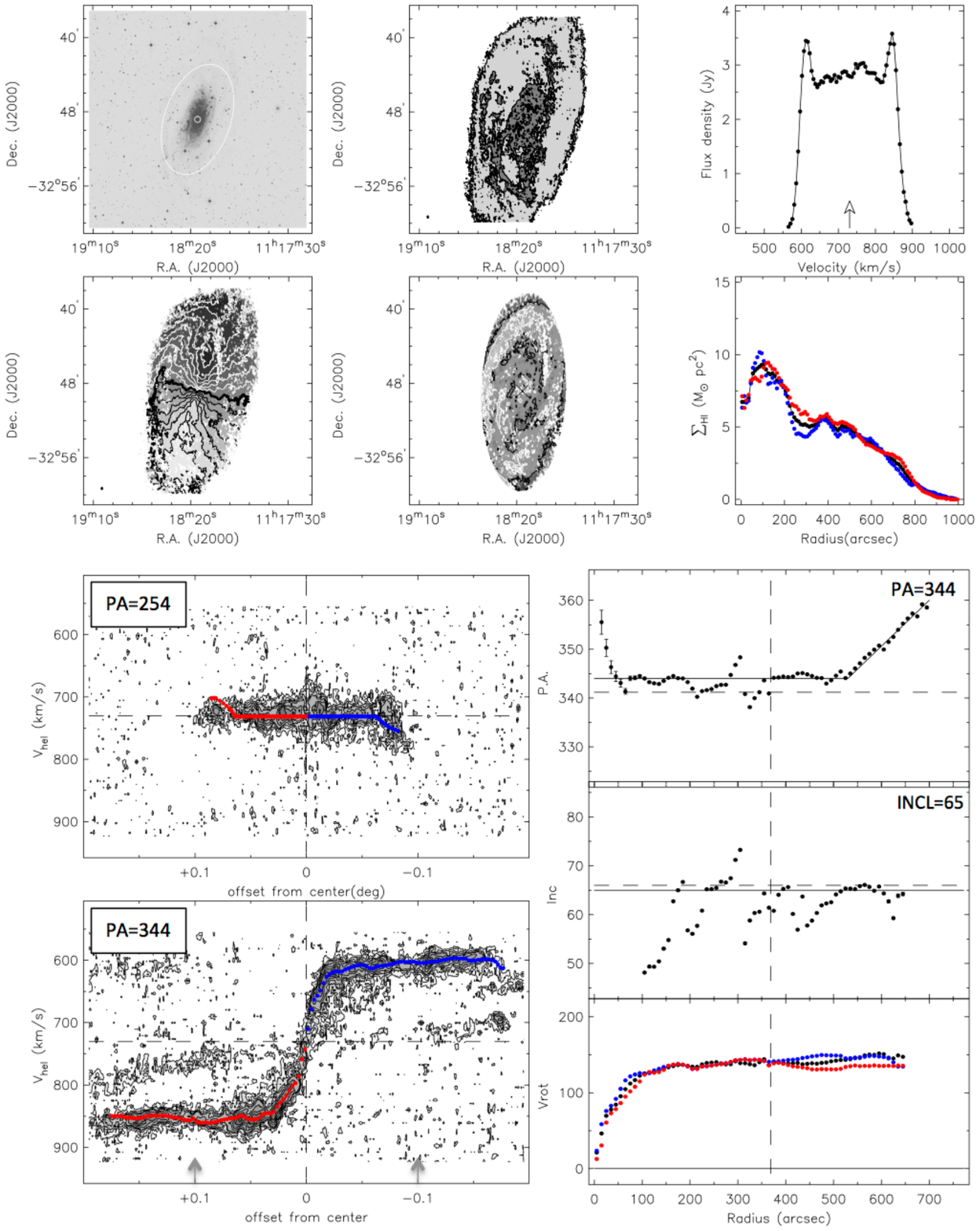}
\end{figure*}
\clearpage

\begin{figure*}
\caption{NGC 3627 (VLA)}
\includegraphics[scale=0.95]{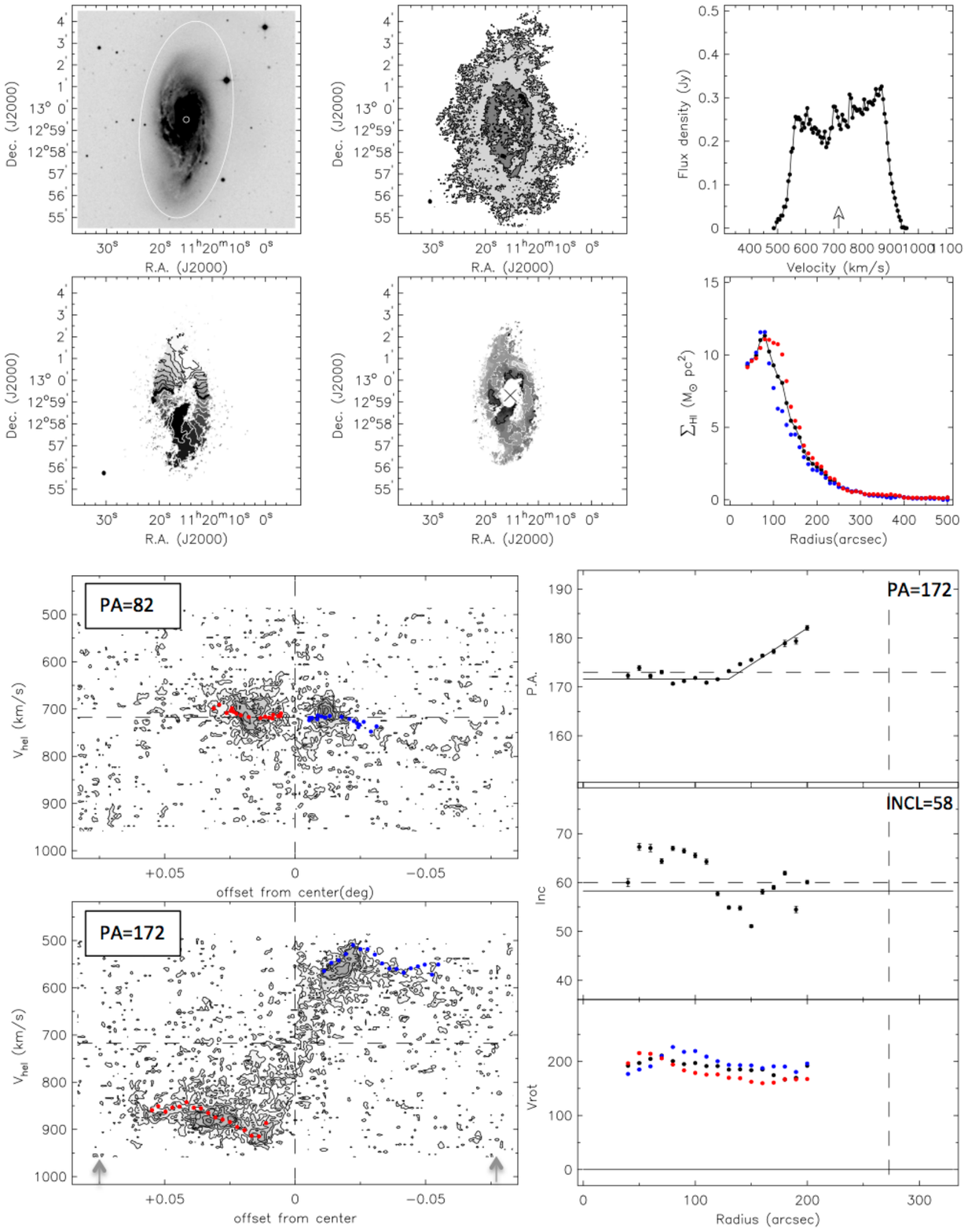}
\end{figure*}
\clearpage

\begin{figure*}
\caption{NGC 4244 (WSRT)}
\includegraphics[scale=0.95]{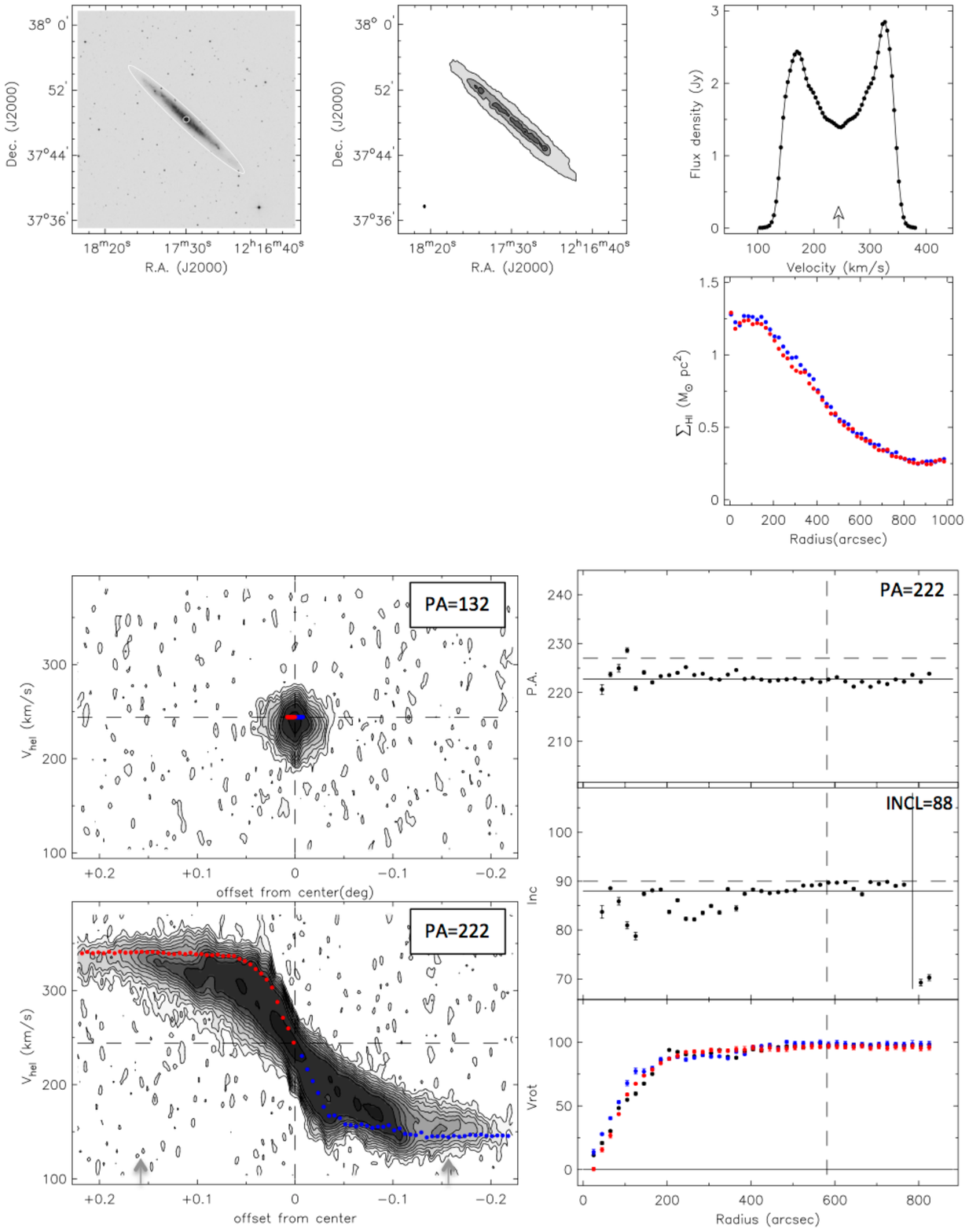}
\end{figure*}
\clearpage

\begin{figure*}
\caption{NGC 4258 (WSRT)}
\includegraphics[scale=0.95]{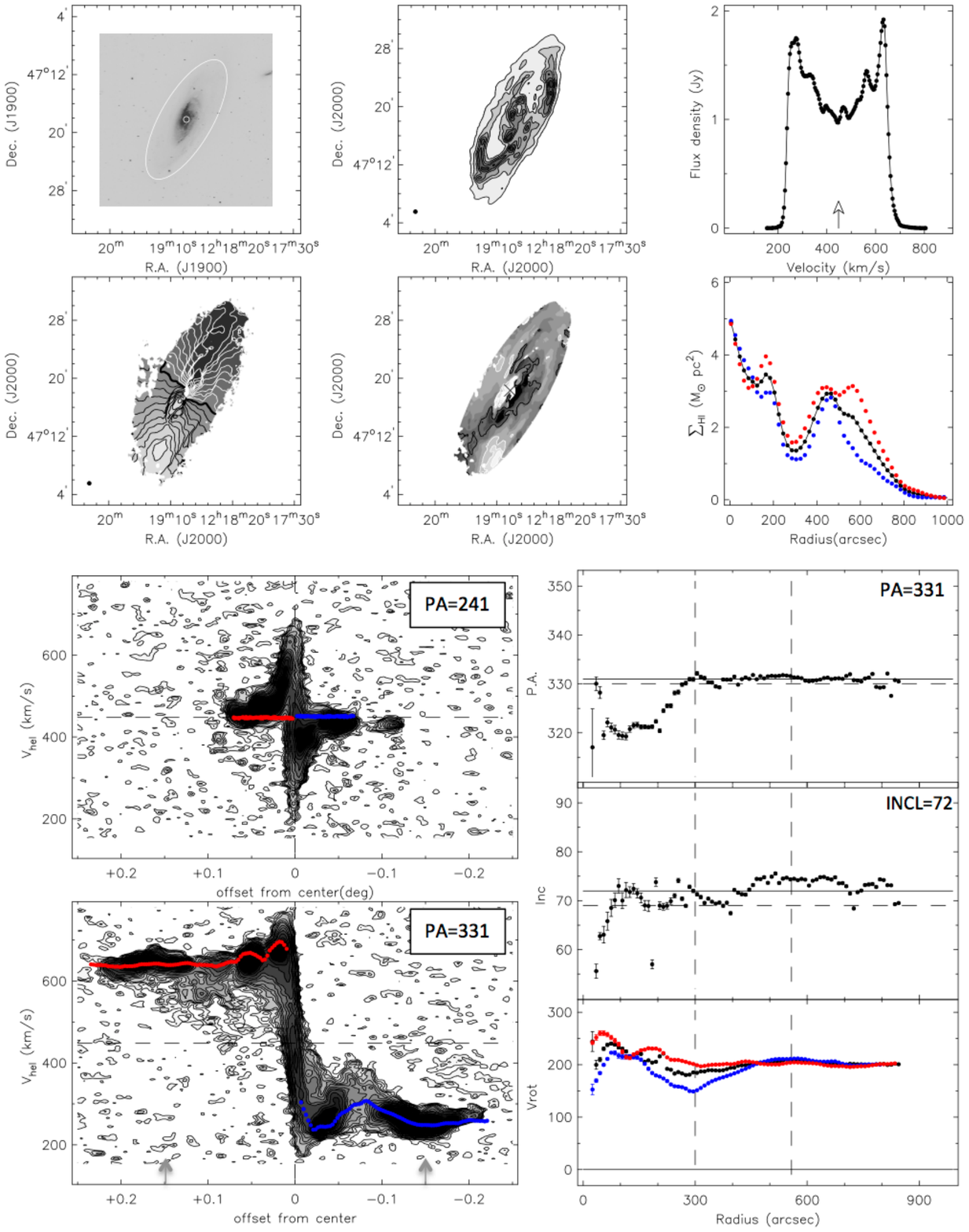}
\end{figure*}
\clearpage

\begin{figure*}
\caption{NGC 4414 (WSRT)}
\includegraphics[scale=0.95]{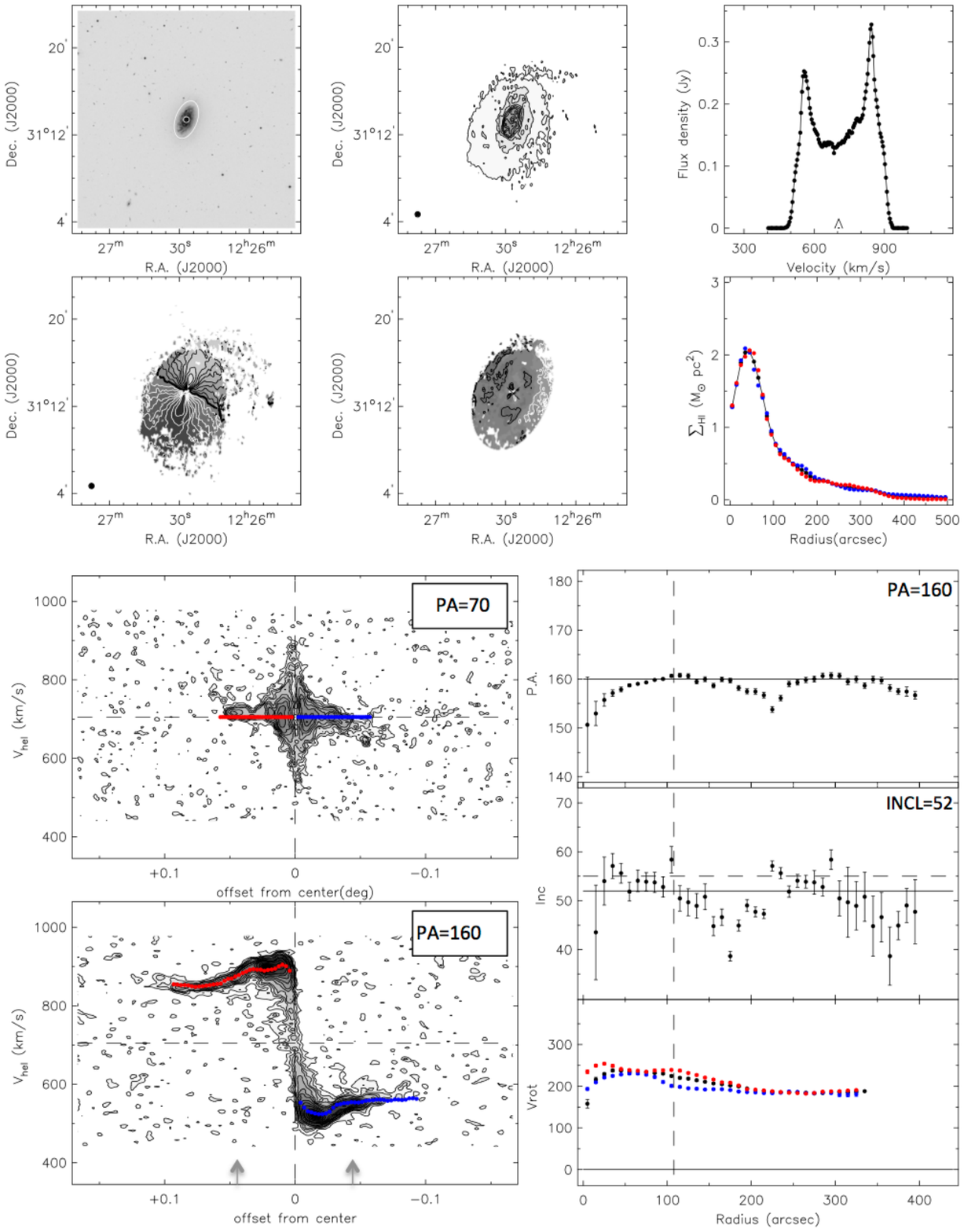}.pdf
\end{figure*}
\clearpage

\begin{figure*}
\caption{NGC 4535 (VLA)}
\includegraphics[scale=0.95]{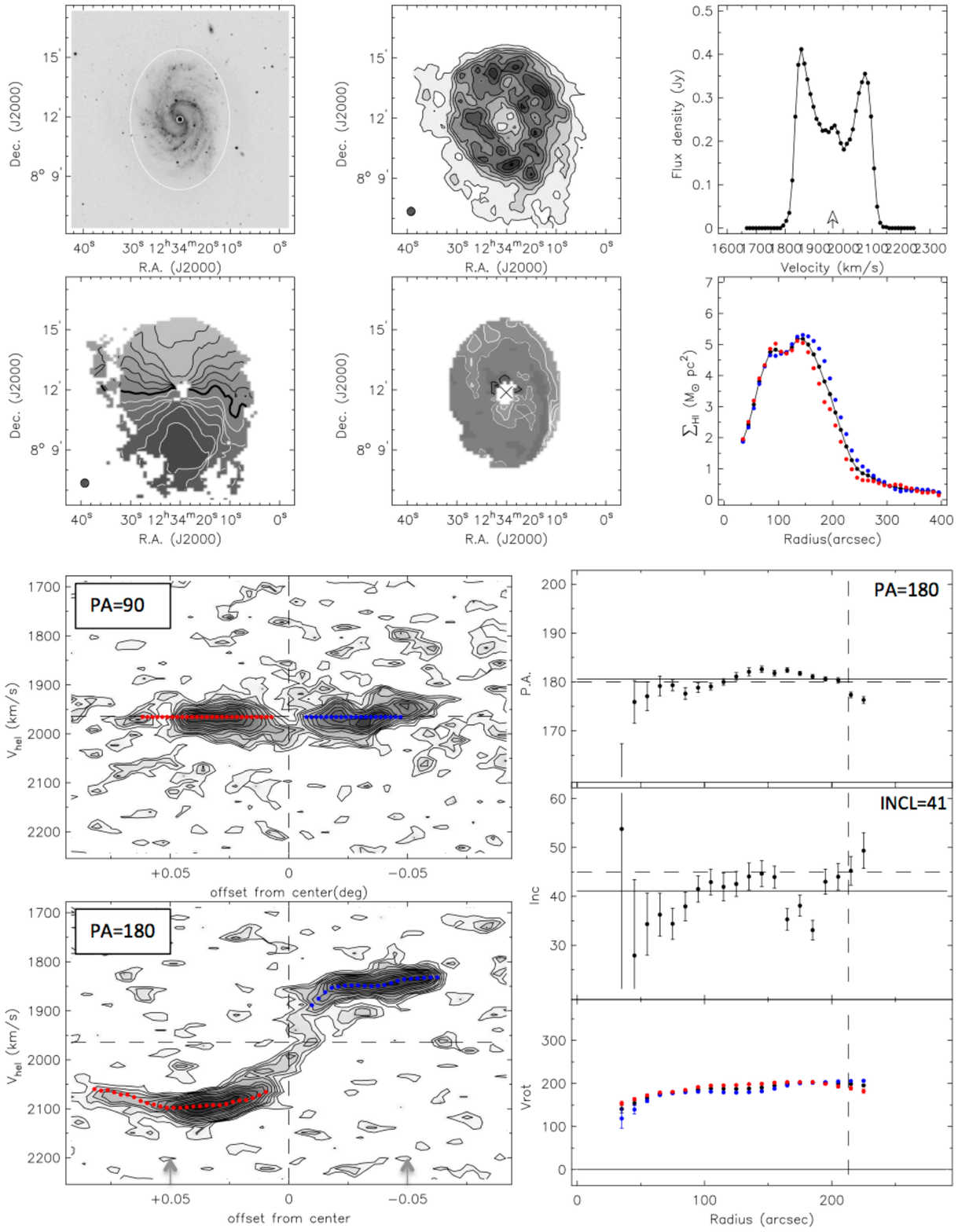}
\end{figure*}
\clearpage

\begin{figure*}
\caption{NGC 4536 (VLA)}
\includegraphics[scale=0.95]{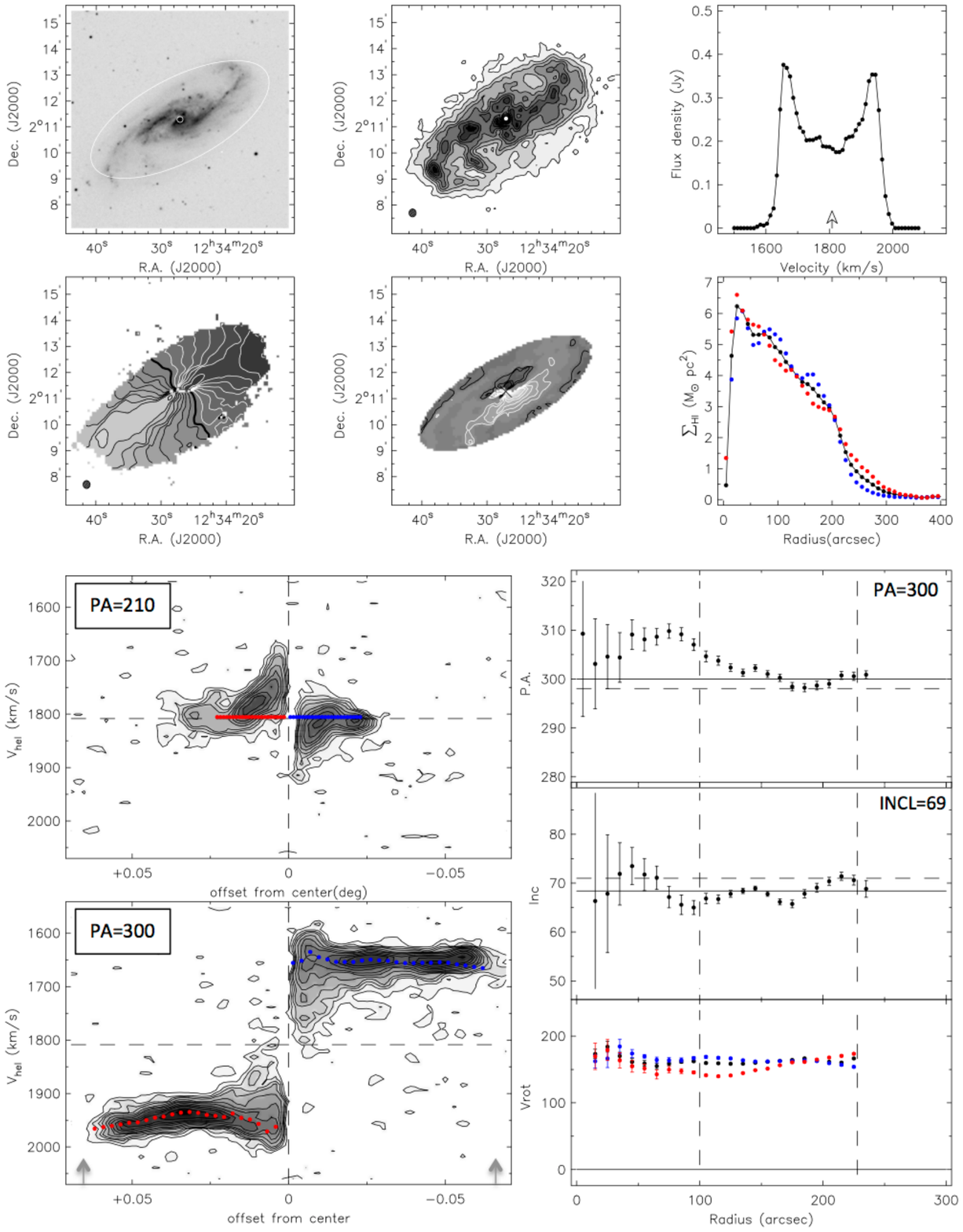}
\end{figure*}
\clearpage

\begin{figure*}
\caption{NGC 4605 (WSRT)}
\includegraphics[scale=0.95]{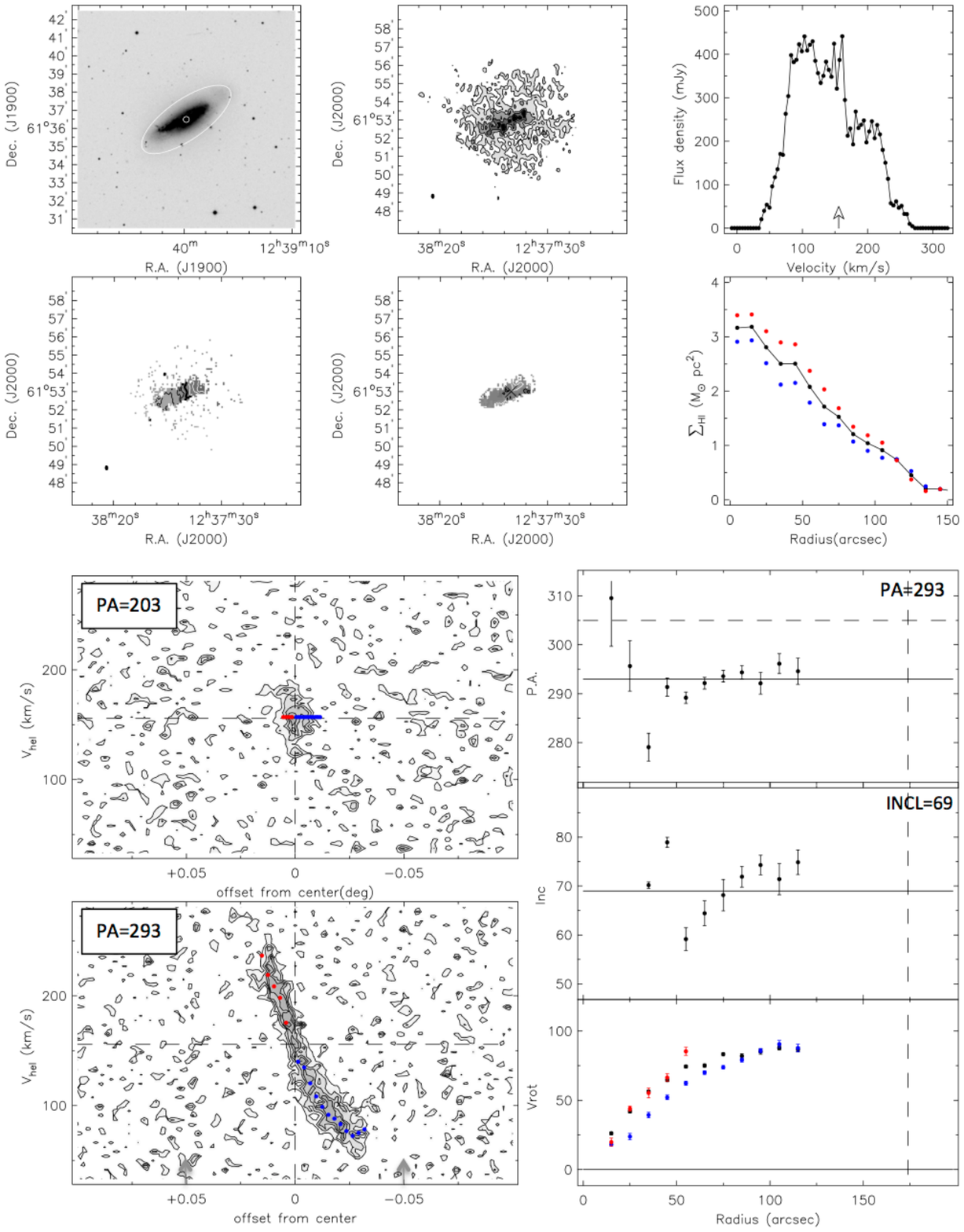}
\end{figure*}
\clearpage

\begin{figure*}
\caption{NGC 4639 (GMRT)}
\includegraphics[scale=0.95]{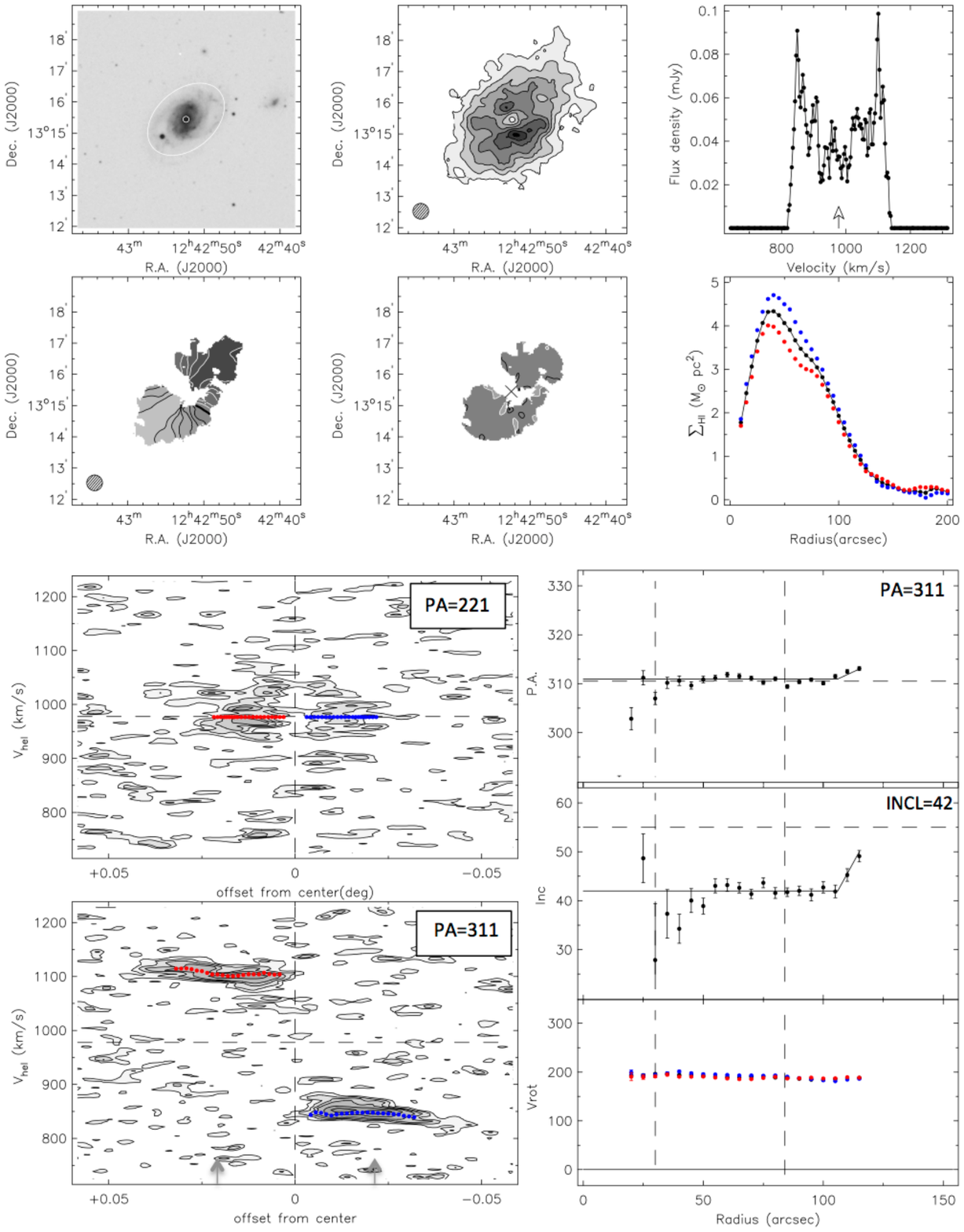}
\end{figure*}
\clearpage

\begin{figure*}
\caption{NGC 4725 (WSRT)}
\includegraphics[scale=0.95]{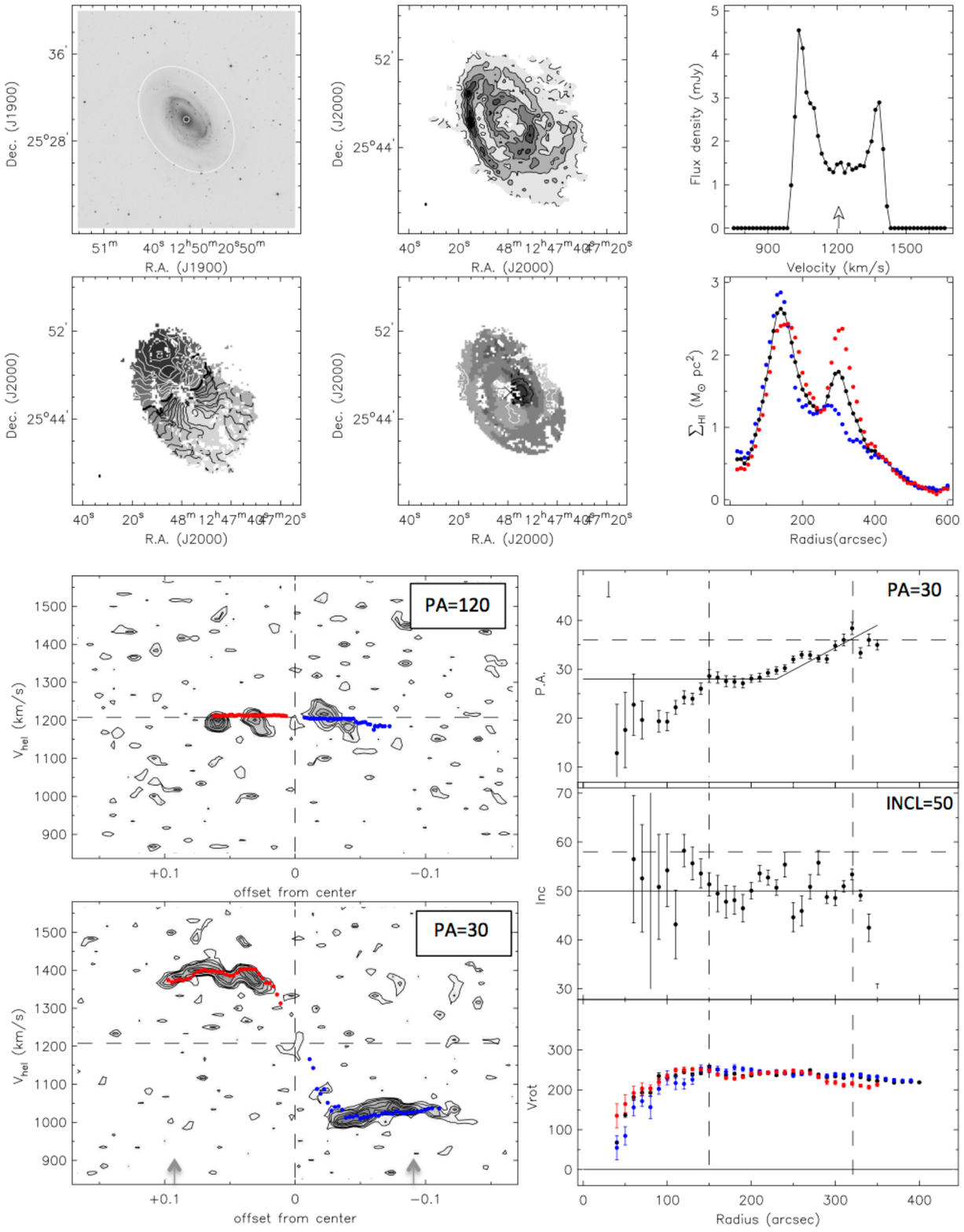}
\end{figure*}
\clearpage

\begin{figure*}
\caption{NGC  5584 (GMRT)}
\includegraphics[scale=0.95]{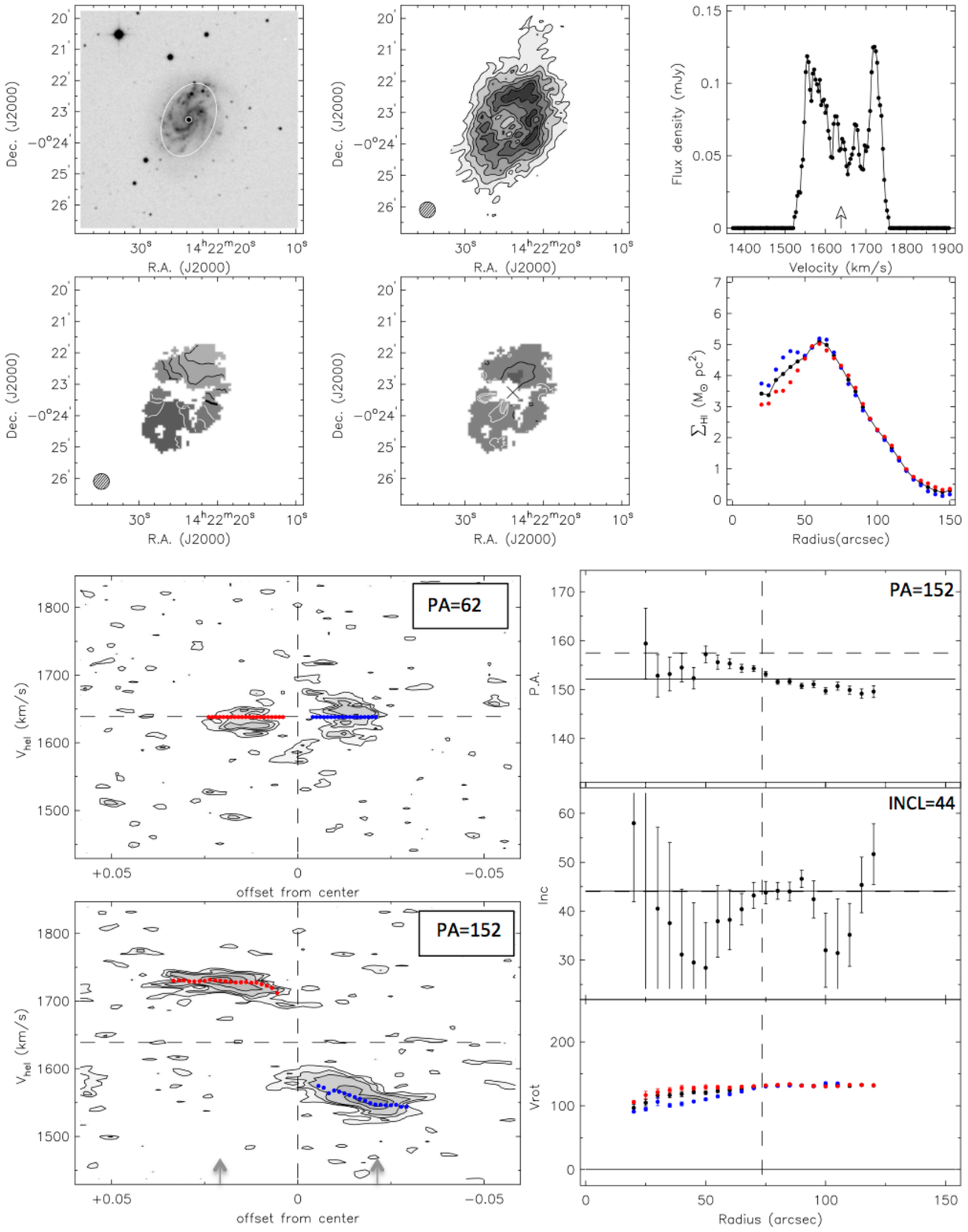}
\end{figure*}
\clearpage

\begin{figure*}
\caption{NGC 7331 (VLA)}
\includegraphics[scale=0.95]{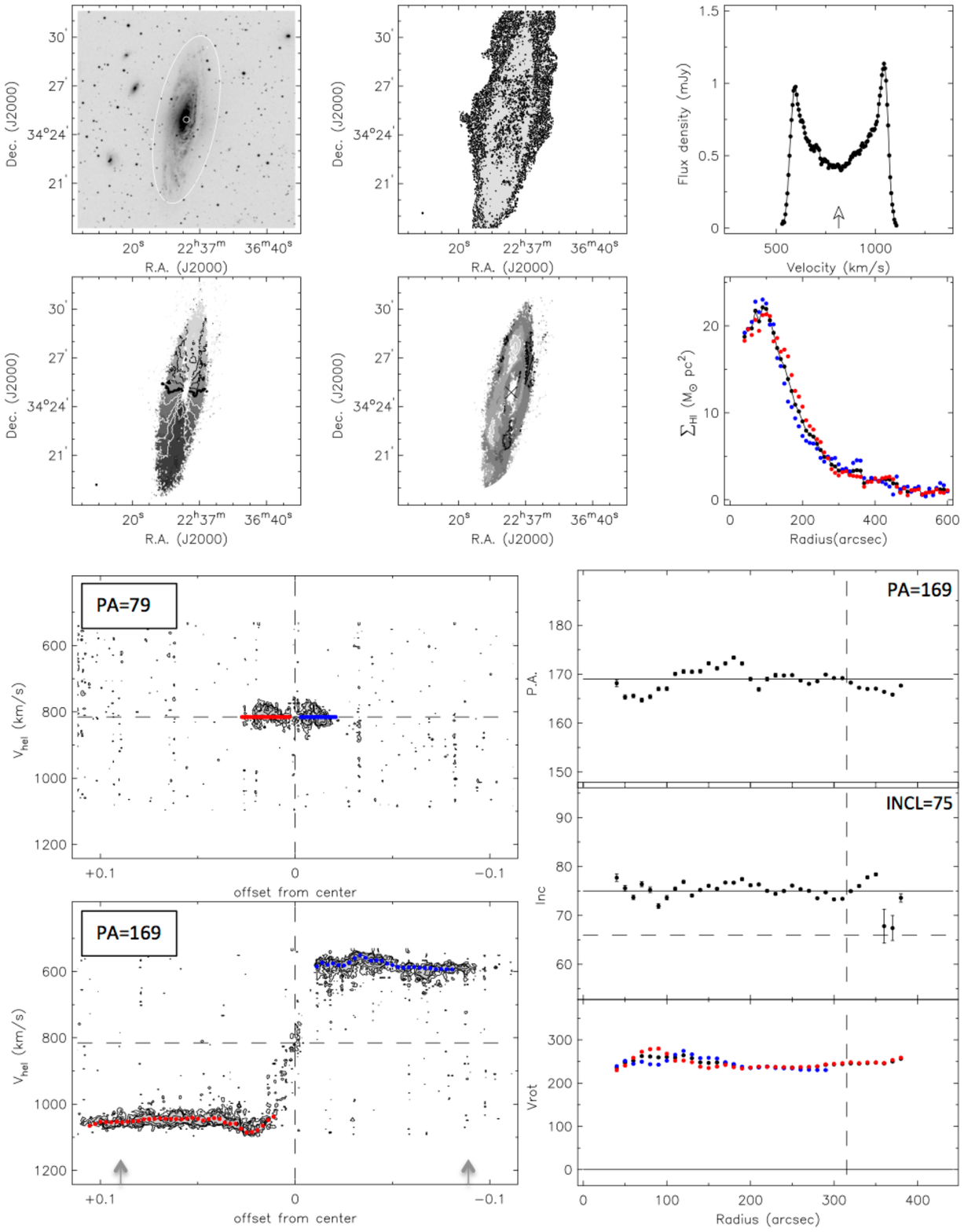}
\end{figure*}
\clearpage

\begin{figure*}
\caption{ NGC 7793 (VLA)}
\includegraphics[scale=0.95]{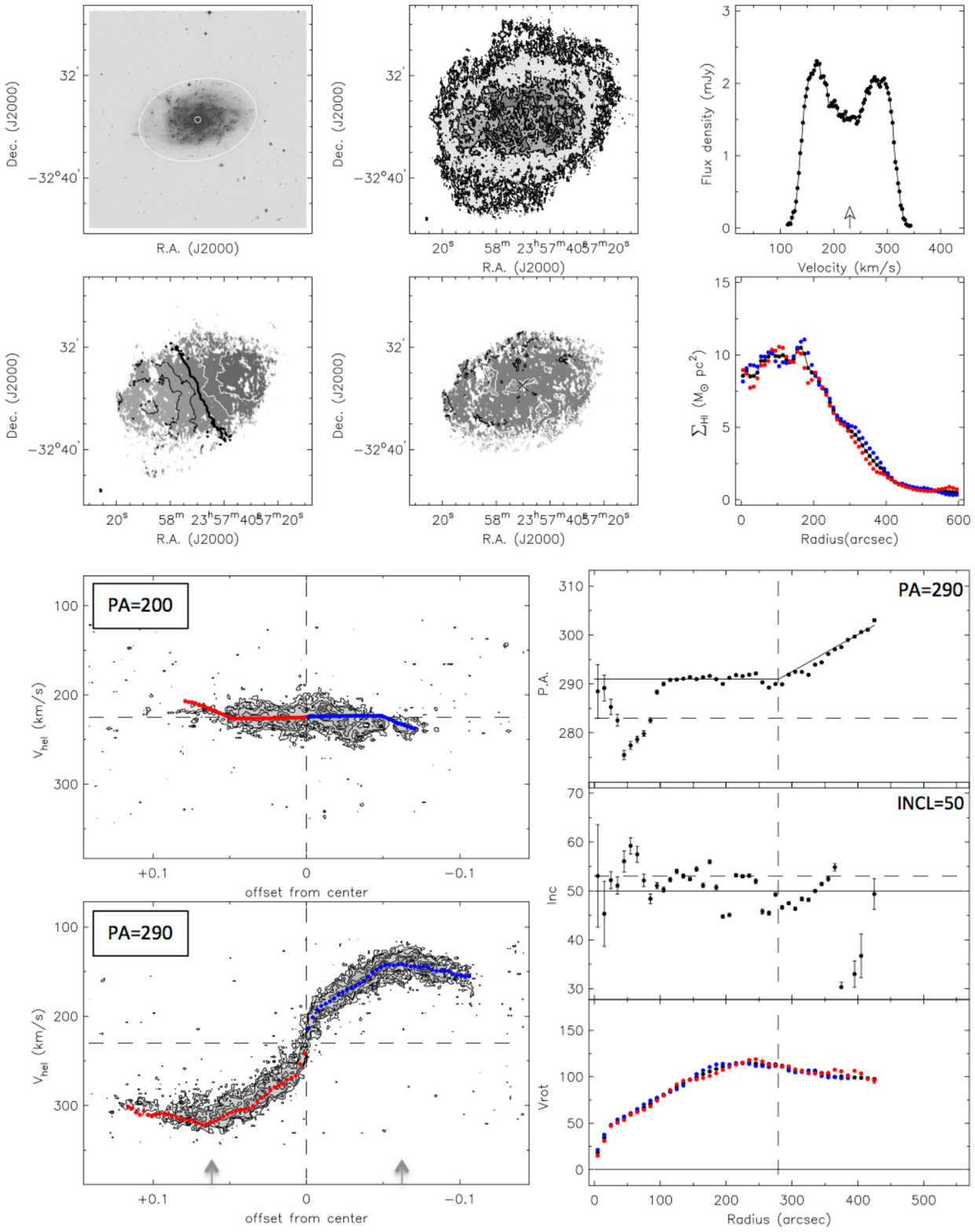}
\end{figure*}
\clearpage

\label{lastpage}

\end{document}